\documentclass{aa}  

\usepackage{graphicx}
\usepackage{txfonts}
\usepackage{xcolor}
\usepackage{natbib}

\usepackage[colorlinks=true,linkcolor=bluea,allcolors=blue]{hyperref}
\begin{document} 
	
	\title{MICONIC: dual AGN, star formation, and ionised gas outflows in NGC\,6240 seen with MIRI/JWST}
	\titlerunning{MICONIC: AGN, star formation, and ionised gas outflows in NGC\,6240 seen with MIRI/JWST}
	
	\author{L. Hermosa Mu{\~n}oz \inst{1}
		\and
		A. Alonso-Herrero \inst{1}
		\and
		A. Labiano\inst{2}
		\and
		P. Guillard\inst{3,4}
		\and
		L. Pantoni\inst{5}
		\and 
		V. Buiten\inst{6}
		\and
		D. Dicken\inst{7}
		\and
		M. Baes\inst{5}
		\and
		T. B{\"o}ker\inst{8}
		\and 
		L. Colina\inst{9}
		\and
		F. Donnan\inst{10}
		\and
		I. Garc{\'i}a-Bernete\inst{1}
		\and
		G. {\"O}stlin\inst{11}
		\and
		P. van der Werf\inst{6}
		\and
		M. J. Ward\inst{12}
		\and
		B. R. Brandl\inst{6,13}
		\and
		F. Walter\inst{14} 
		\and
		G. Wright\inst{7}
		\and
		M. G{\"u}del\inst{15,16} 
		\and
		Th. Henning\inst{17}
		\and 
		P.-O. Lagage\inst{18}
		\and
		T. Ray\inst{19}
	}
	
	\institute{
		1. Centro de Astrobiolog{\'i}a (CAB) CSIC-INTA, Camino Bajo del Castillo s/n, 28692 Villanueva de la Ca{\~n}ada, Madrid, Spain \\
		\email{lhermosa@cab.inta-csic.es}
		\\
		2. Telespazio UK for the European Space Agency (ESA), ESAC, Camino Bajo del Castillo s/n, 28692 Villanueva de la Ca{\~n}ada, Spain \\
		3. Sorbonne Universit{\'e}, CNRS, UMR 7095, Institut d’Astrophysique de Paris, 98bis bd Arago, 75014 Paris, France \\
		4. Institut Universitaire de France, Minist{\`e}re de l’Enseignement Sup{\'e}rieur et de la Recherche, 1 rue Descartes, 75231 Paris Cedex 05, France \\
		5. Sterrenkundig Observatorium, Universiteit Gent, Krijgslaan 281 S9, B-9000 Gent, Belgium \\
		6. Leiden Observatory, Leiden University, PO Box 9513, 2300 RA Leiden, The Netherlands \\
		7. UK Astronomy Technology Centre, Royal Observatory, Blackford Hill Edinburgh, EH9 3HJ, Scotland, UK\\
		8. European Space Agency, c/o Space Telescope Science Institute, 3700 San Martin Drive, Baltimore MD 21218, USA \\
		9. Centro de Astrobiolog{\'i}a (CAB) CSIC-INTA,  Ctra. de Ajalvir km 4, Torrej{\'o}n de Ard{\'o}z, 28850, Madrid, Spain \\
		10. Department of Physics, University of Oxford, Keble Road, Oxford, OX1 3RH, UK \\
		11. Department of Astronomy, Stockholm University, The Oskar Klein Centre, AlbaNova, SE-106 91 Stockholm, Sweden \\
		12. Centre for Extragalactic Astronomy, Durham University, South Road, Durham DH1 3LE, UK \\
		13. Faculty of Aerospace Engineering, Delft University of Technology, Kluyverweg 1, 2629 HS Delft, The Netherlands \\
		14. Max Planck Institute for Astronomy, K{\"o}nigstuhl 17, 69117 Heidelberg\\
		15. Dept. of Astrophysics, University of Vienna, T{\"u}rkenschanzstr. 17, A-1180 Vienna, Austria\\
		16. ETH Z{\"u}rich, Institute for Particle Physics and Astrophysics, Wolfgang-Pauli-Str. 27, 8093 Z{\"u}rich, Switzerland\\
		17. Max Planck Institute for Astronomy, Konigstuhl 17, 69117 Heidelberg, Germany \\
		18. Universit{\'e} Paris-Saclay, Universit{\'e} Paris Cit{\'e}, CEA, CNRS, AIM, F-91191 Gif-sur-Yvette, France\\
		19. School of Cosmic Physics, Dublin Institute for Advanced Studies, 31 Fitzwilliam Place, Dublin, D02 XF86, Ireland
	}
	
	\date{Received M DD, YYYY; accepted M DD, YYYY}
	
	
	\abstract{
		Galaxy mergers are an important and complex phase during the evolution of galaxies. They may trigger nuclear activity and/or strong star forming episodes in the galaxy centres that potentially alter the evolution of the system.}
	{As part of the guaranteed time observations (GTO) program Mid-Infrared Characterization Of Nearby Iconic galaxy Centers (MICONIC), we used the medium-resolution spectrometer (MRS) of the Mid-Infrared Instrument (MIRI) on board of the James Webb Space Telescope (JWST) to study NGC\,6240. We aim to characterise the dual active galactic nuclei (AGN), the ionised gas outflows and the main properties of the interstellar medium (ISM) over a mapped area of $6.6\arcsec \times 7.7\arcsec$.}
	{We obtained integral field spectroscopic (IFS) mid-infrared data (wavelength from 4.9 to 28$\mu$m) of NGC\,6240. We modelled the emission lines through a kinematic decomposition that accounts for the possible existence of various components.}
	{We have resolved both nuclei for the first time in the full 5-28$\mu$m spectral range. The fine-structure lines in the southern (S) nucleus are broader than for the northern (N) nucleus (full width at half maximum of $\geq$1500 vs $\sim$700\,km\,s$^{-1}$ on average, respectively). High excitation lines, such as [Ne\,V], [Ne\,VI], and [Mg\,V], are clearly detected in the N nucleus. In the S nucleus, the same lines can be detected but only after a decomposition of the PAH features in the integrated spectrum, due to a combination of a strong mid-IR continuum, the broad emission lines, and the intense star formation (SF). The SF is distributed all over the mapped FoV of 3.5\,kpc\,$\times$\,4.1\,kpc (projected), with the maximum located around the S nucleus. Both nuclear regions appear to be connected by a bridge region detected with all the emission lines. Based on the observed MRS line ratios and the high velocity dispersion ($\sigma \sim 600$\,km\,s$^{-1}$), shocks are also dominating the emission in this system. We detected the presence of outflows as a bubble north-west from the N nucleus and at the S nucleus. We estimated a ionised mass outflow rate of 1.4$\pm 0.3$\,M$_{\sun}$\,yr$^{-1}$ and $1.8\pm0.2$\,M$_{\sun}$\,yr$^{-1}$, respectively. Given the derived kinetic power of these outflows, both the AGN and the starburst could have triggered them. }
	{}
	
	
	\maketitle
	%
	
	\section{Introduction}
	\label{Sect1:Introduction}
	
	Galaxy mergers play a crucial role in the evolution of galaxies across the Universe. Theoretical simulations suggest that mergers are responsible for the hierarchical growth of structures, contributing significantly to the formation of massive galaxies and shaping their properties \citep[see e.g.][]{Barnes1992,Barnes1996,DiMatteo2005,Conselice2014}. During the merger process, large reservoirs of gas are channelled into the central regions of the system, often leading to intense episodes of star formation, commonly referred to as starbursts, and/or triggering periods of activity in the nucleus, known as active galactic nuclei \citep[AGN, see e.g.][]{Hopkins2006,Hopkins2008,Cox2008,Darg2010a,Ellison2013}. Merging galaxies are observed throughout the Universe in various stages of interaction \citep[see e.g.][]{Toomre1972}, from early encounters where the galaxies have just begun to influence each other gravitationally \citep[see e.g.][]{Darg2010}, to more advanced stages that involve complex dynamical interactions, particularly between the supermassive black holes in the nuclei of the galaxies \citep{Kormendy1995,Veilleux2002,Hopkins2006}. One of the most well known examples for a merger in the local Universe is the system NGC\,6240.
	
	NGC\,6240 is a luminous infrared galaxy (LIRG, see \citealt{PerezTorres2021} for a review of the properties of this class of galaxies) with an infrared luminosity of L$_{\rm IR}$/L$_{\sun}$\,$=$\,10$^{11.93}$ \citep{Kim2013}. It is a galaxy merger located at a distance of 111~Mpc, corresponding to a physical scale of 1\arcsec$=$\,526\,pc (\href{https://ned.ipac.caltech.edu/}{NASA/IPAC Extragalactic Database; NED}). The complex morphology and dynamics of the gas in this system has been studied in the literature in various wavelength bands, such as optical \citep{Veilleux2003,Sharp2010,MullerSanchez2018,Kollatschny2020}, X-rays \citep[][]{Netzer2005,Puccetti2016,Nardini2017,Fabbiano2020,Paggi2022}, sub-millimetre \citep{Greve2009,Feruglio2013,Fyhrie2021,Cicone2018}, radio \citep{Gallimore2004,Hagiwara2011}, and near-infrared \citep{Max2005,Engel2010,Mori2014}. It has two X-ray detected nuclei, north and south \citep[N and S, respectively][]{Scoville2015}, separated by $\sim$740\,pc \citep{Komossa2003}, with 2-10\,keV luminosities of 5.2 and 2.0$\times$10$^{43}$\,erg\,s$^{-1}$, respectively \citep{Puccetti2016}. Recently, \cite{Kollatschny2020} suggested the existence of a third nucleus with optical data, which is not detected in X-rays \citep{Fabbiano2020}. The two X-ray bright nuclei are both classified as AGN \citep{Komossa2003}. Based on optical line ratios, the S nucleus is classified as a type-2 Seyfert and the N nucleus as a low excitation nuclear emission-line region (LINER) with approximately three times lower bolometric luminosity than the S nucleus \citep[8 and 2.6$\times$10$^{44}$\,erg\,s$^{-1}$, respectively;][]{MullerSanchez2018}. Each nucleus hosts a small stellar disc, with an effective radius of $\sim$350\,pc and $\sim$50\,pc for the N and S nuclei, respectively \citep{Medling2014}. The stellar kinematics are decoupled from both the molecular and ionised gas \citep{MullerSanchez2018}. In the region between both nuclei, there is evidence for at least one recent episode of star formation associated with a starburst, and probably more, given the complex structure of the H$\alpha$ emission \citep{Yoshida2016}. This region is responsible for one third of the total K-band luminosity of the source \citep{Tecza2000,Engel2010} and up to 80\% of the total infrared luminosity \citep{Armus2006}, suggesting that neither AGN are the main source of ionisation for the interstellar medium \citep[ISM, see also][]{Lutz2003,Mori2014,Scoville2015}.
	
	The mid-infrared (mid-IR) emission of this system was previously studied in detail with Spitzer/IRS spectroscopic data in \cite{Armus2006}, although both nuclei were contained within the aperture (smallest slit width of 3.6\arcsec). They detected both high-excitation lines [Ne\,V] at 14.32$\mu$m, attributed to the presence of a buried AGN, and low-excitation lines, such as [Ne\,II] at 12.81$\mu$m and [Ne\,III] at 15.55$\mu$m, most likely produced by the central starburst. Indeed NGC\,6240 has been classified as a deeply obscured nuclei in \cite{GarciaBernete2022}. Several other buried nuclei have been observed with the James Webb Space Telescope (JWST; \citealt{Gardner2023}) lately, including LIRGs such as VV\,114 \citep{Rich2023,Donnan2023,Buiten2024,GonzalezAlfonso2024}, II\,Zw9 \citep{GarciaBernete2024}, or NGC\,3256 \citep{PereiraSantaella2024}.
	Ground-based mid-IR imaging observations taken with 8-10m class telescopes with angular resolutions of $\sim$0.3-0.4\arcsec\,resolved the two nuclei of NGC\,6240, as well as emission from the region between them \citep{Egami2006,Asmus2014,AH2014}. Long-slit spectroscopic observations obtained with CanariCam at the Gran Telescopio Canarias also revealed bright 11.3$\mu$m polycyclic aromatic hydrocarbon (PAH) emission peaking at the S nucleus and extending in between the nuclei to the N nucleus \citep{AH2014}.
	
	The spatially-resolved ionised gas morphology in NGC\,6240 traced with optical observations is complex both between the two nuclei \citep[see e.g.][]{Beswick2001} and up to several kpc \citep[see e.g.][]{MullerSanchez2018,Medling2021}. In general, the gas is distributed in a butterfly-like shape (referred to as Butterfly Nebula \citep[see e.g.][]{Medling2021}, with multiple arms and bubbles, some of them associated with superwinds and outflows produced by either the AGN or the nuclear starburst \citep[see also][]{MullerSanchez2018}. NGC\,6240 also hosts massive outflows detected in different gas phases \citep[see e.g.][]{MullerSanchez2018,Cicone2018,Fabbiano2020}. For example, part of the [O\,III] emission is attributed to an outflow produced by the AGN oriented in a conical shape expanding more than 5\,kpc mainly east of the nuclei \citep{MullerSanchez2018,Medling2021}. The CO molecular gas peaks in the region between the two nuclei, with a large-scale outflow extending more than 10\,kpc in the east-west direction \citep{Cicone2018}. The molecular and ionised gas as traced by CO and the H$\alpha$ are expanding in different directions, as the latter tends to expand following the path of least resistance \citep[see][]{Medling2021}. The H$\alpha$ emission extends up to $\sim$90\,kpc \citep{Veilleux2003,Yoshida2016}, and seems correlated with the soft X-ray emission \citep{Nardini2013,Yoshida2016,Paggi2022}, suggesting a common origin for both, as already seen for other Seyfert and LINER galaxies in the literature \citep[see e.g.][]{Veilleux2003,Masegosa2011,HM2022}. 
	
	In this paper, we study NGC\,6240 using the imager and the medium-resolution spectrometer (MRS) of the Mid-Infrared Instrument (MIRI; \citealt{Rieke2015,Wright2015,Wright2023}) on board of the JWST. Similar to other wavelengths, the mid-IR spectroscopic data are very rich and complex. We will mainly focus on the general properties for the two nuclei and their ionised gas emission. The properties of the warm molecular H$_{2}$ lines will be addressed in a future paper (Hermosa Mu{\~n}oz et al. in prep.). 
	
	The paper is organised as follows. Section~\ref{Sect2:Data} describes the observations, data reduction and the line modelling procedure. In Sect.~\ref{Sect3:Results} we present the main results on the global properties of the ionised emission and kinematics. In Sect.~\ref{Sect4:Discussion} we discuss the origin of the ionised gas, the detection of ionised outflows and the presence of high-ionisation lines in both nuclei. In Sect.~\ref{Sect5:Conclusions} we present the summary and main conclusions of this work. 
	
	\section{Data and methodology}
	\label{Sect2:Data}
	
	The observations presented here are part of the guaranteed time observations (GTO) program \textit{Mid-Infrared Characterization Of Nearby Iconic galaxy Centers} (MICONIC) of the MIRI European Consortium. The program includes Mrk\,231 \citep{AH2024}, Arp\,220, NGC\,6240, Centaurus A, and SBS\,0335-052, as well as the region surrounding SgrA* in our Galaxy. The program has been developed in collaboration with the NIRSpec GTO team (for NGC\,6240 see Ceci et al. submitted, and for Arp\,220 see \citealt{Perna2024, Ulivi2024}), providing a complete near- and mid-infrared view into the nuclei of these galaxies. 
	
\begin{figure*}
		\centering
		\includegraphics[trim=0cm 0cm 3.5cm 0cm,clip=True,width=\textwidth]{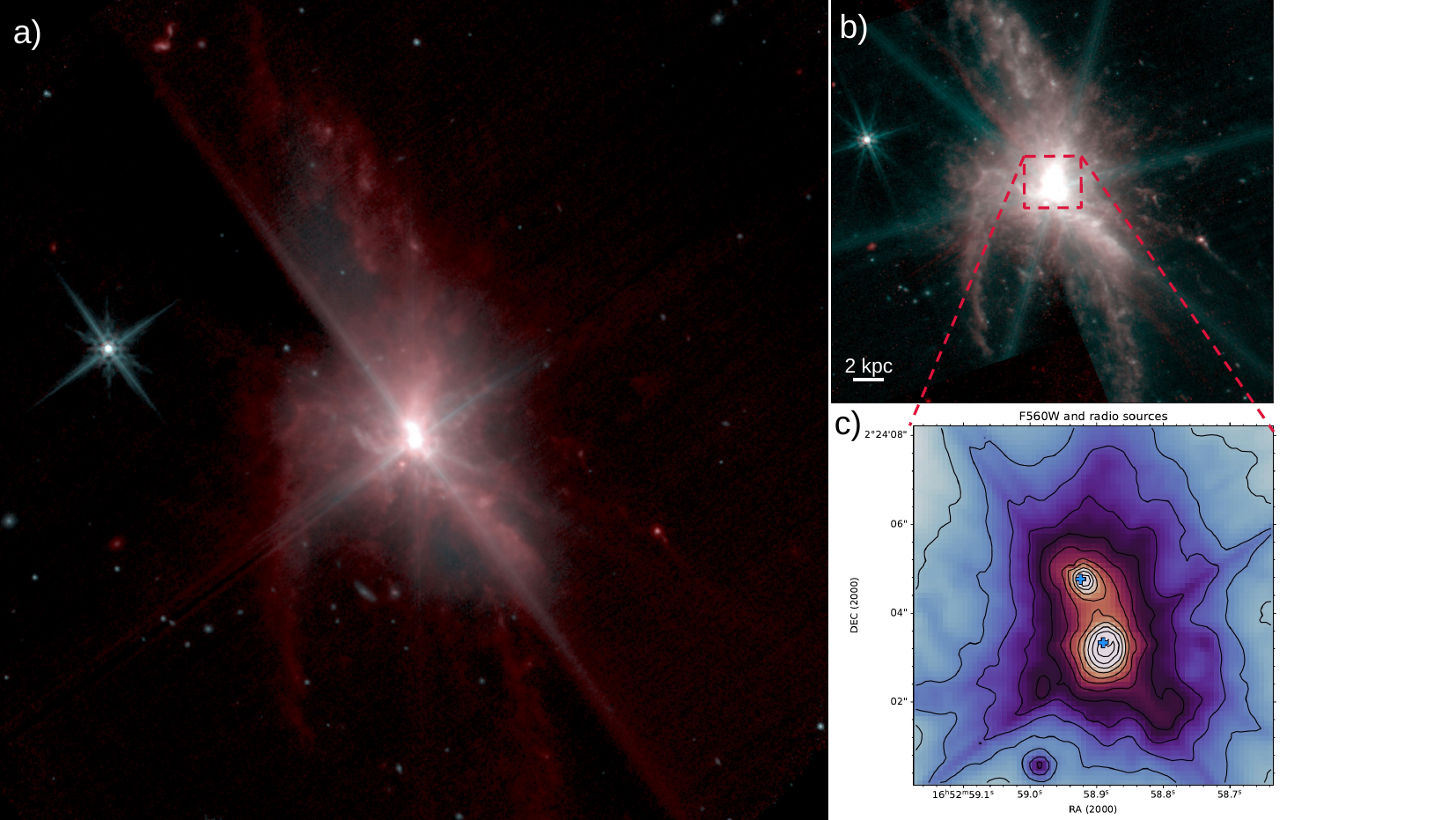}
		\caption{False-colour images of NGC\,6240 produced combining the MIRI images (see Sect.~\ref{Sect2:Data} for the details): a) obtained with the filters F560W and F1130W, within an approximate FoV of 90\arcsec$\times$90\arcsec (1\arcsec$=$526\,pc); b) obtained with the filters F770W and F1130W, within an approximate FoV of 80\arcsec$\times$80\arcsec, and the nuclei saturated to highlight the low level structure; and c) a zoom in (approximate FoV of $7\arcsec\times8\arcsec$) to the F560W-filter image in logarithm scale with the position of the two radio sources in \cite{Gallimore2004} marked as blue crosses. The twelve contour levels in panel c) go from 0.2\% up to 73\% of the flux peak, also in logarithm scale. We note that the diffraction spikes produced by the unresolved component of both nuclei have not been corrected. In all panels, north is up and east to the left.}
		\label{Fig1:Image}      
	\end{figure*}

	The data were observed on the 18th August 2023 with the MRS mode of the MIRI instrument on board of the JWST, and are associated with the GTO program with ID number 1265 (P.I.: A.~Alonso-Herrero). Program 1265 also included direct imaging of NGC6240 (PID 1265 observation 3) as well as parallel imaging with the MRS observations using the filters F560W, F770W, and F1130W. The results of the direct imaging program can be seen in Fig.~\ref{Fig1:Image}, which presents the combination of the F560W and F1130W filters (panel a) and the F770W and F1130 filters (panel b).
	
	\begin{figure*}
		\centering
		\includegraphics[width=\textwidth]{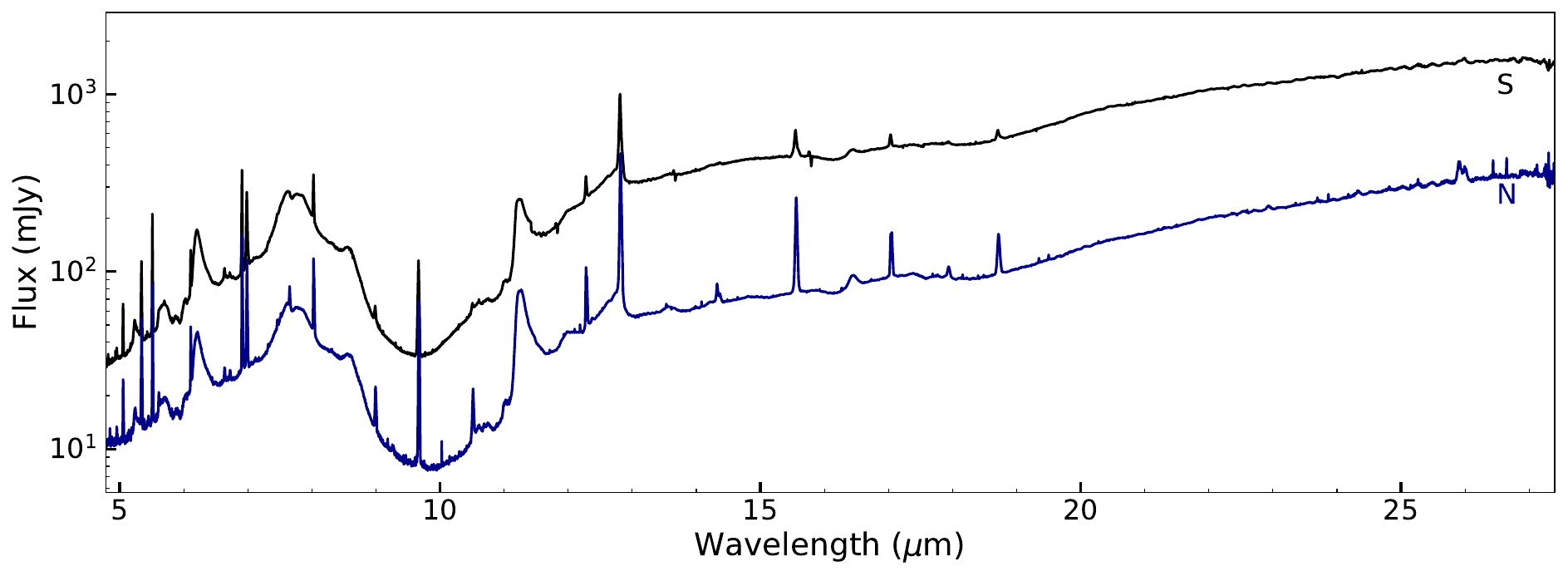} 
		\includegraphics[width=\columnwidth]{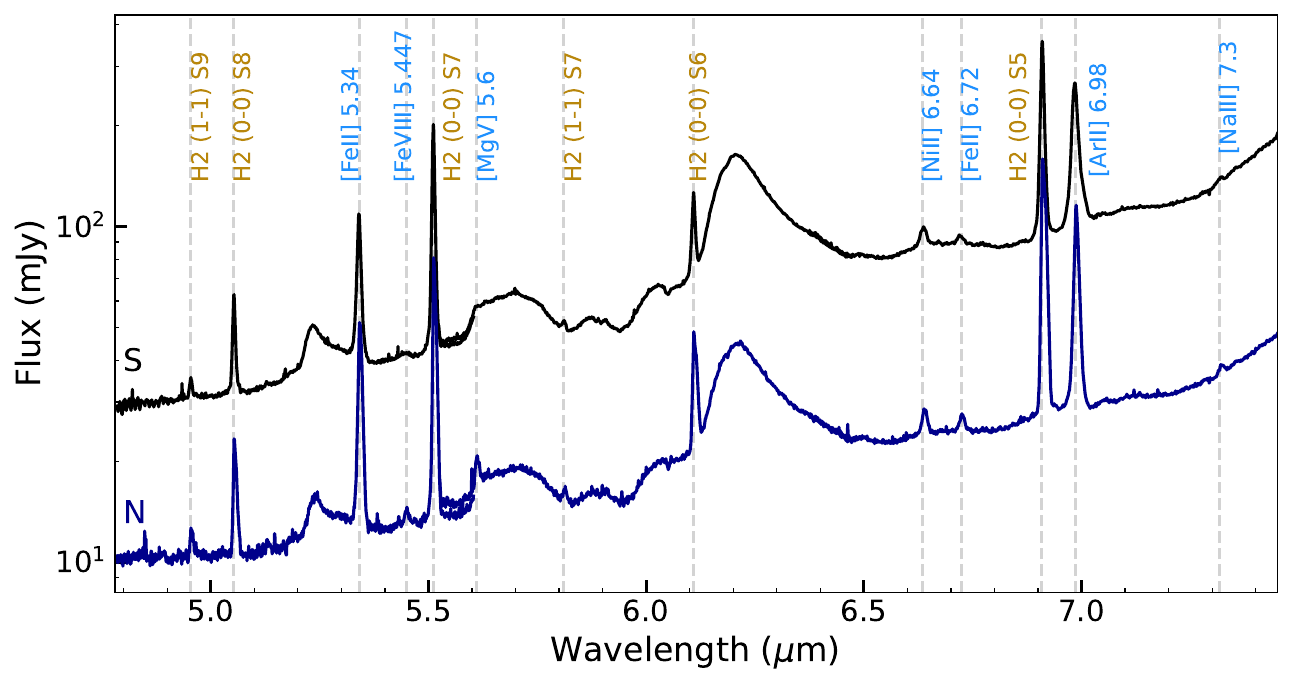} 
		\includegraphics[width=\columnwidth]{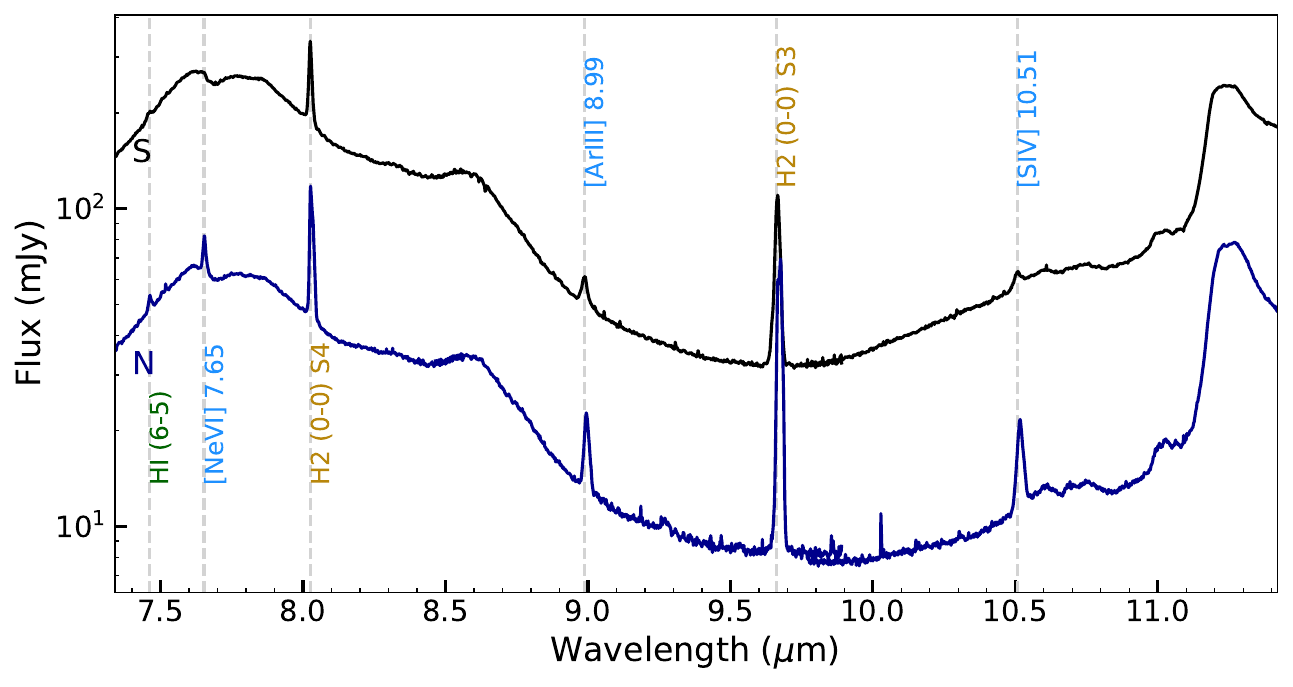} 
		\includegraphics[width=\columnwidth]{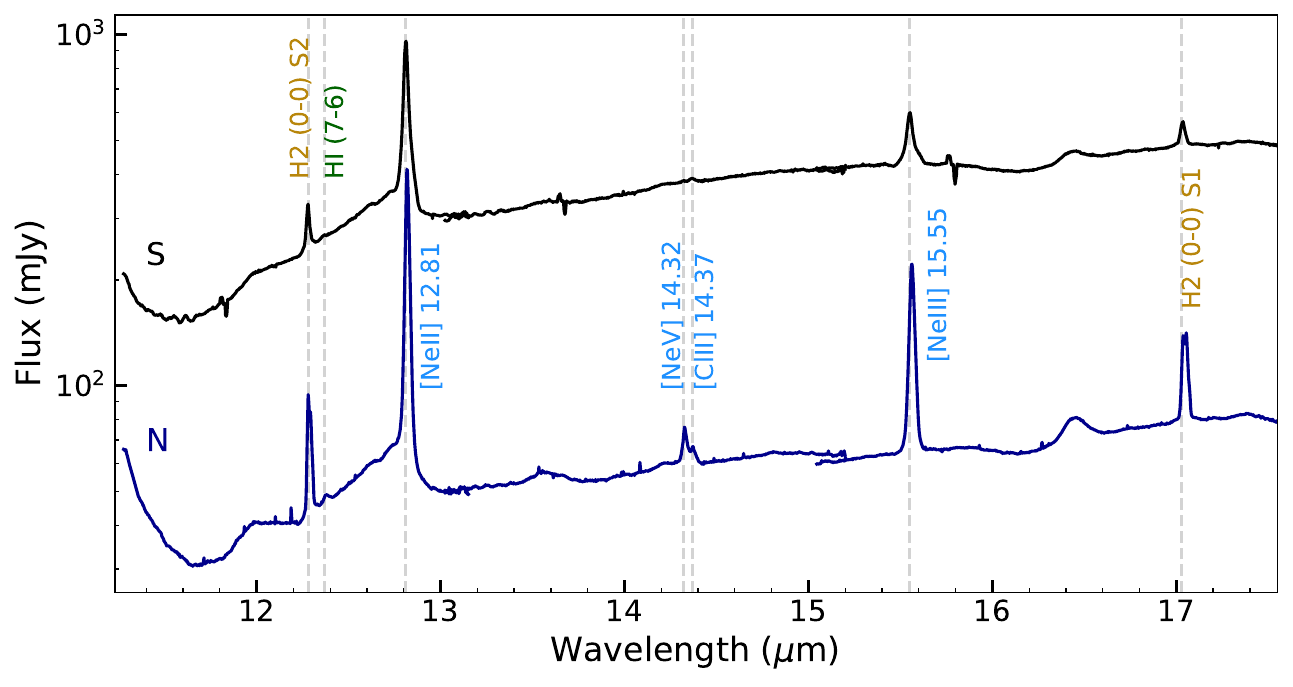} 
		\includegraphics[width=\columnwidth]{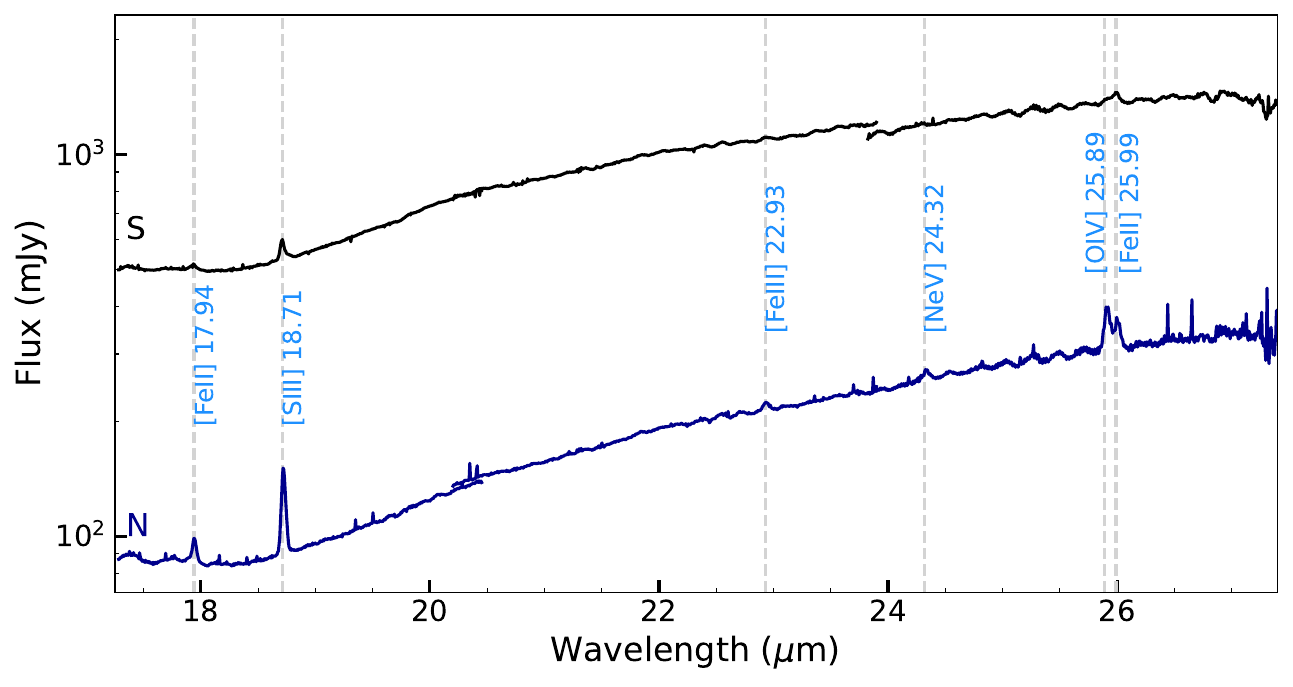} 
		\caption{Total integrated spectra for the northern (dark blue) and southern (black) nuclei of NGC\,6240 obtained with a circular aperture of radius 0.7\arcsec. The top panel shows the full MRS spectral range, after scaling the flux of some individual bands to match between them. The middle and bottom panels show the spectra per channel, with the main ionised gas emission lines indicated in blue, the H$_{2}$ rotational lines in yellow, and the recombination H\,I lines in green. The wavelength is in rest frame assuming a redshift of 0.02448 (see Sect.~\ref{Sect2:Data}). We applied a residual fringe correction for the 1D extracted spectra in ch3 and ch4 \citep[see][]{Argyriou2023}. }
	\label{Fig1:IntSpec}
\end{figure*}

\subsection{Data reduction}
\label{Sect2:Reduction}

The MRS mode of MIRI covers a total wavelength range from $4.9 - 27.9\mu m$, separated in four integral field units (IFUs) referred to as channels, each divided in three bands. The channels cover slightly different field-of-views (FoV), from 3.2\arcsec$\times$3.7\arcsec\,(channel 1) up to 6.6\arcsec$\times$7.7\arcsec\,(channel 4), and have different spatial (from 0.13\arcsec\,in channel 1 to 0.35\arcsec\,in channel 4) and spectral (from $\sim$3700 to $\sim$1500) resolutions \citep[see more details in][]{Labiano2021}. The observations were centred on an intermediate position between the N and S nuclei \citep{Scoville2015}. We used the 4-pt extended source dither pattern, with 50 groups per integration, and one integration per exposure, in FASTR1 readout mode, covering the whole MRS spectral range in three exposures (one per MRS band). This amounts to an on-source integration time of 555s for each MRS band. Using the recommended strategy, we took a background observation with the 2-pt dither, for a total integration time of 227.5 s per band.

The data reduction for the MRS observations was done with the JWST Science Calibration Pipeline \citep[version 1.12.3,][]{Bushouse2023}, with the context 1135 for the Calibration References Data System following the standard procedures \citep[see e.g.][for detailed examples of MRS data reduction and calibration]{Labiano2016, Alvarez2023}. 
Careful examination of the data showed that the stage 1 corrections \citep[][and references therein]{Morrison2023} could be run with default parameters, and did not leave any significant residuals in the data. 
We switched off the background correction in the stage 2 pipeline 
\citep{Argyriou2023, Gasman2023, Patapis2024} and applied the background correction on 1D spectra only when needed. Similarly, we switched off the master background correction and sky matching steps in stage 3 of the pipeline before producing the final fully reconstructed science cubes \citep{Law2023}, as these steps introduced artefacts in the data. Finally, the science cubes were rotated from the observed position (position angle, PA\,$=106^\circ$) to the standard orientation, with north up and east to the left.

We show our combined MIRI image in Fig.~\ref{Fig1:Image}. It was produced combining the images obtained with the broadband filters F560W and F1130W ($\Delta\lambda \sim$1 and 0.7$\mu$m, respectively). The observations were taken in a $1\arcmin\times2\arcmin$ mosaic with total integration times of 166 seconds per filter. These images were processed with the development pipeline version with CRDS version '11.17.26'. Showers correction and tweakreg astrometry correction using GAIA3 were used for this reduction.

The full MRS spectra for each of the two nuclei, extracted by integrating over a circular aperture of r$\sim 0.7$\arcsec\,are shown in Fig.~\ref{Fig1:IntSpec}. This radius has been selected to be larger than the full width at half maximum (FWHM) of the point spread function (PSF) derived from the nuclei in the individual channels, assuming they are point-like sources (see Appendix~\ref{Appendix_figures}). For these spectra we have applied a 1D fringing correction to all the bands from ch3 and ch4 to eliminate the periodic fringes produced by internal reflections in the MIRI detectors and the dichroic optical elements \citep[see more details in][]{Argyriou2020,Argyriou2023}. In Fig.~\ref{Fig:Appendix_ContMaps}, we show the continuum maps from the MIRI/MRS data cubes obtained for all bands in each channel. 

For the analysis of the integrated nuclear spectra we did not apply an aperture correction. The emission from the nuclei is complex, probably containing both resolved and unresolved components. The latter is more noticeable at longest MRS wavelength continuum maps, particularly in the S nucleus (see Fig.~\ref{Fig:Appendix_ContMaps}). Moreover, from the MIRI images (see Fig.~\ref{Fig1:Image}), we detect diffraction spikes coming from both nuclei due to the existence of an unresolved component. The emission lines, which are the focus of this paper, are resolved, thus this correction will not affect the results presented here.

\subsection{Emission line modelling}
\label{Sect2:Methodology}

Once the data cubes were fully reduced, we modelled the emission line profiles to obtain their kinematics and flux maps. The fitting process involved several steps. First, we created an integrated spectrum for each nucleus with a circular aperture of 0.7\arcsec\,(see Fig.~\ref{Fig1:IntSpec}). To fit the emission lines, we used the \textsc{lmfit} library within \textsc{python}, allowing multiple different Gaussian components for each line, following \cite{HM2024}. We used the results from the integrated spectra as initial conditions for fitting each individual line on a spaxel-by-spaxel basis. 
Before the Gaussian fitting, we modelled and subtracted the local continuum. For that we selected a wavelength window on either side of the line that varied depending on the line's width, while carefully avoiding other spectral absorption or emission features. Depending on the steepness of the continuum, we applied either a linear fit or a second-order polynomial function. To avoid spurious detections, we fitted the emission lines only when their signal-to-noise ratio (S/N) was higher than 3. We show some examples of the fitting in Fig.~\ref{Fig:Appendix_FitIntLinesParam}. 

Some lines, such as [S\,IV] at 10.51$\mu$m or [Ar\,III] at 8.99$\mu$m, are embedded within the broad 9.7$\mu$m absorption feature that complicates a correct determination of the continuum and the line flux. For those, we spatially re-binned the corresponding data cube with a box of 2$\times$2 pixels to improve the S/N and obtain a better modelling of the line. These lines are properly indicated throughout the paper. 

For correcting the velocities for the systemic value, we assumed a redshift of z\,$= 0.02448$ \citep{Downes1993}, which coincides with the systemic velocity derived from ALMA data \citep[see e.g.][]{Cicone2018,Treister2020}. Using the velocities derived for the primary component of the warm molecular gas detected with MRS/MIRI for both nuclei (see Hermosa Mu{\~n}oz et al. in prep.) we found a similar value for the redshift. After modelling the Gaussian profiles, we obtained the velocity dispersion and corrected it for instrumental broadening \citep[see][]{Argyriou2023} using $\sigma = \sqrt{\sigma_{\rm obs}^{2} - \sigma_{\rm inst}^{2}}$. 

To obtain the global properties of the integrated emission lines, we additionally used a non-parametric analysis, based on \cite{Harrison2014} (see also \citealt{Speranza2024} and \citealt{HM2024b}). This allows us to measure the parameters v02, v10, v50, v90, and v98, corresponding to the velocities at which the flux is 2\%, 10\%, 50\%, 90\%, and 98\% of the total integrated flux of the line. Similarly to the parametric modelling, we defined the line for spectral elements where the flux is larger than three times the standard deviation from the local continuum. Examples of this modelling for several emission lines for both nuclei at different excitation levels are shown in Figs.~\ref{Fig:Appendix_profilesN} and~\ref{Fig:Appendix_profilesS} (see Appendix~\ref{Appendix1}). Additionally, we estimated the W80 parameter (tabulated in Table~\ref{Table:2_W80}), which is equivalent to $1.088$ times the FWHM for a Gaussian-like profile\footnote{We note that the W80 parameter for the most complicated profiles, such as [Ne\,III] in the S nucleus, should be taken as a first-order approximation to the line width.}. We did not estimate this parameter for the lines that are blended, such as [Ne\,V] at 14.32$\mu$m and [Cl\,II] at 14.37$\mu$m (see Sect.~\ref{Subsect3:Results_IntProperties}). 

In general, we find that the emission lines with similar ionisation potentials (IPs) show similar kinematic properties. Thus, we will focus on a few of the emission lines throughout the paper, and we show the kinematic and flux maps for the rest of the lines in Appendix~\ref{Appendix_figures}. 

\subsection{Modelling and subtraction of PAH features}
\label{Sect2:methodPAH}

In NGC\,6240, the PAH features have a large contribution to the total flux of the spectra, potentially preventing the detection of some of the high excitation emission lines that fall within the same wavelength range (i.e. [Ne\,VI] and the PAH feature at 7.7$\mu$m, see Fig.~\ref{Fig1:IntSpec}). However, for understanding the physical conditions and properties of the gas in this source, it is important to determine if these lines are indeed absent in the spectra, or if they are "buried" due to the strong PAH features. 

In order to answer this question, we use the tool developed by \cite{Donnan2024} which models the near- and mid-IR continuum with different temperatures as well as the line and PAH emission taking into account differential extinction. For the purpose of the following analysis, it is sufficient to only model the mid-IR MRS spectra, as we are only interested in removing the bright PAH emission. The tool allows to model the different PAH features (and emission lines) present in the mid-IR spectra, and returns the properties of each individual component (e.g. central wavelength, equivalent width, flux, etc.). We applied this method only to the integrated spectra of the nuclei (R\,$\sim 0.7$\arcsec), where the high excitation lines are expected to be brighter. 
We combined the individual PAH features to create a total PAH profile and subtract it from the integrated spectra. This allowed us to detect the high excitation lines that were initially buried for the S nucleus. The results from this modelling are presented in Sect.~\ref{Subsect3:HighIonLines}.

\section{Results}
\label{Sect3:Results}

With the MIRI/MRS data we have resolved for the first time the two nuclei in the 5-28$\mu$m spectral range (see continuum maps in Fig.~\ref{Fig:Appendix_ContMaps}). 
We present the integrated properties for both nuclei in Sect.~\ref{Subsect3:Results_IntProperties}, the main morphological properties and features of the ionised emission lines in Sect.~\ref{SubSect3:Results_fluxes}, and the mid-IR line ratios in Sect.~\ref{SubSect3:LineRatios}. We show the kinematical properties of the gas in Sect.~\ref{SubSect3:Results_kin}, and the detection of the high excitation lines in Sect.~\ref{Subsect3:HighIonLines}. The non-parametric modelling for the integrated profiles for both nuclei is described in Appendix~\ref{Appendix1}. 

\subsection{Integrated properties of the nuclei}
\label{Subsect3:Results_IntProperties}

As already noticeable in Fig.~\ref{Fig1:IntSpec}, the two nuclei show different features in their spectra. The S nucleus is more luminous than the N nucleus (see Sect.\ref{Sect1:Introduction}), with stronger emission lines and PAH features, as already seen in \cite{AH2014}. We used the continuum maps of each band (see Fig.~\ref{Fig:Appendix_ContMaps}) to determine the position of the nuclei. They are separated by an average distance of 1.6\arcsec\,(1.7\arcsec\,in channel 4 that has the lowest spatial resolution, see Sect.~\ref{Sect1:Introduction}), and the same separation was measured using the F560W MIRI image (see Fig.~\ref{Fig1:Image}). These positions are approximately coincident with the radio sources N1 and S, where the two AGN are believed to be located. The two radio sources are separated 1.5\arcsec\,as reported by \cite{Gallimore2004}, and coincident with the X-ray detected sources \citep{Max2007}. A more detailed discussion on the position of the nuclei based on NIRSpec observations is presented in Ceci et al. (submitted).

\begin{figure*}
	\centering
	\includegraphics[width=.67\columnwidth]{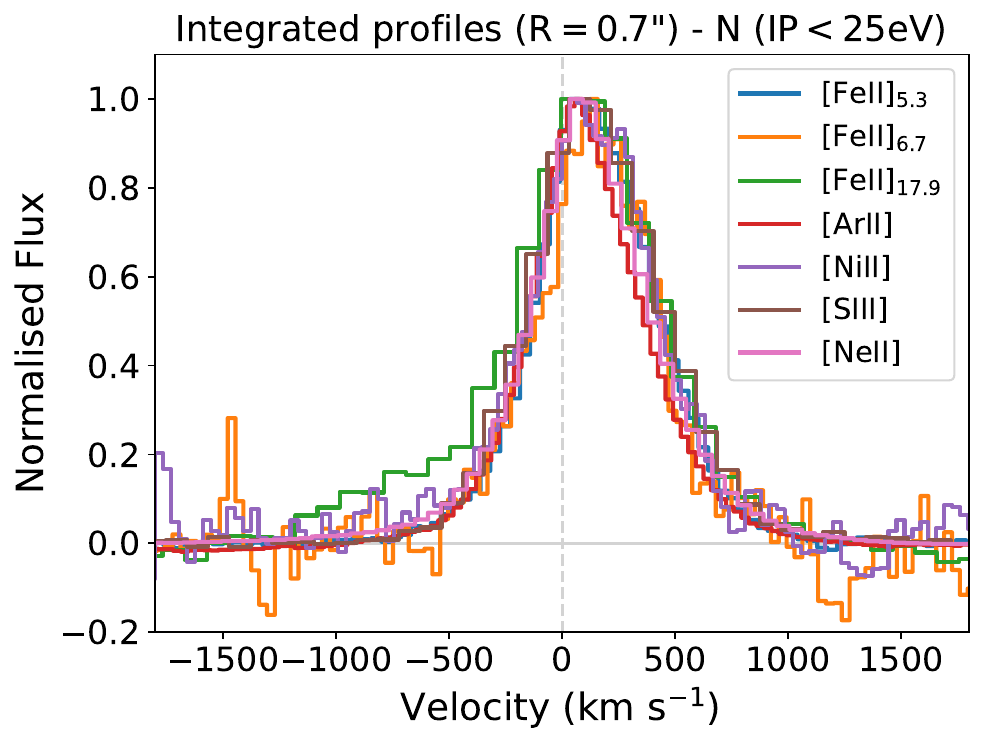} 
	\includegraphics[width=.68\columnwidth]{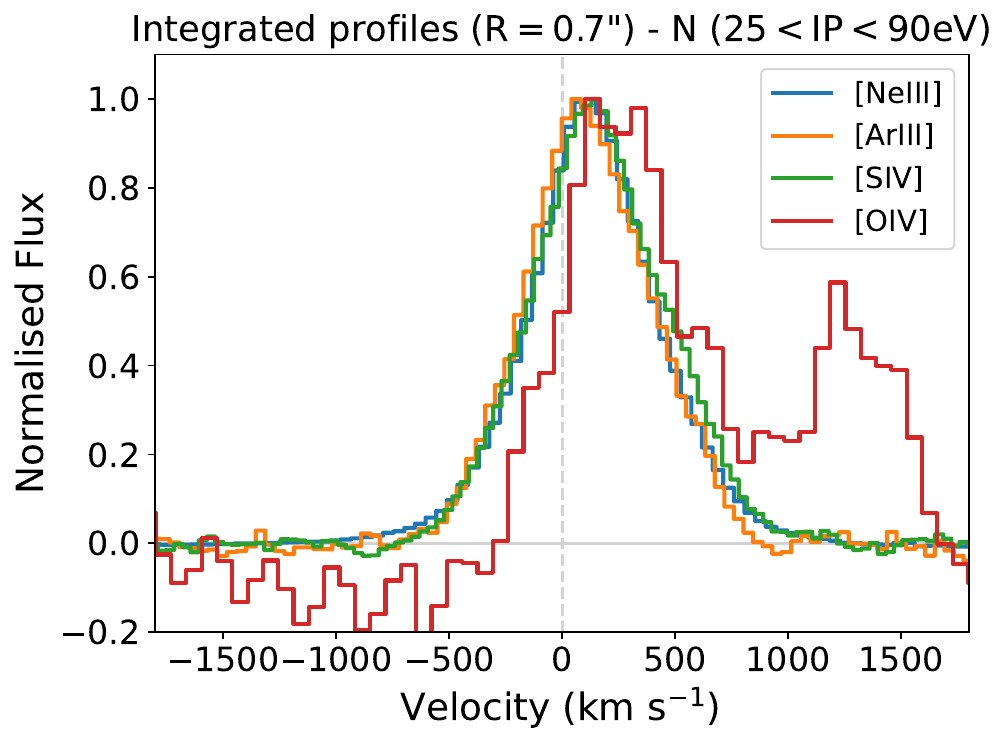} 
	\includegraphics[width=.67\columnwidth]{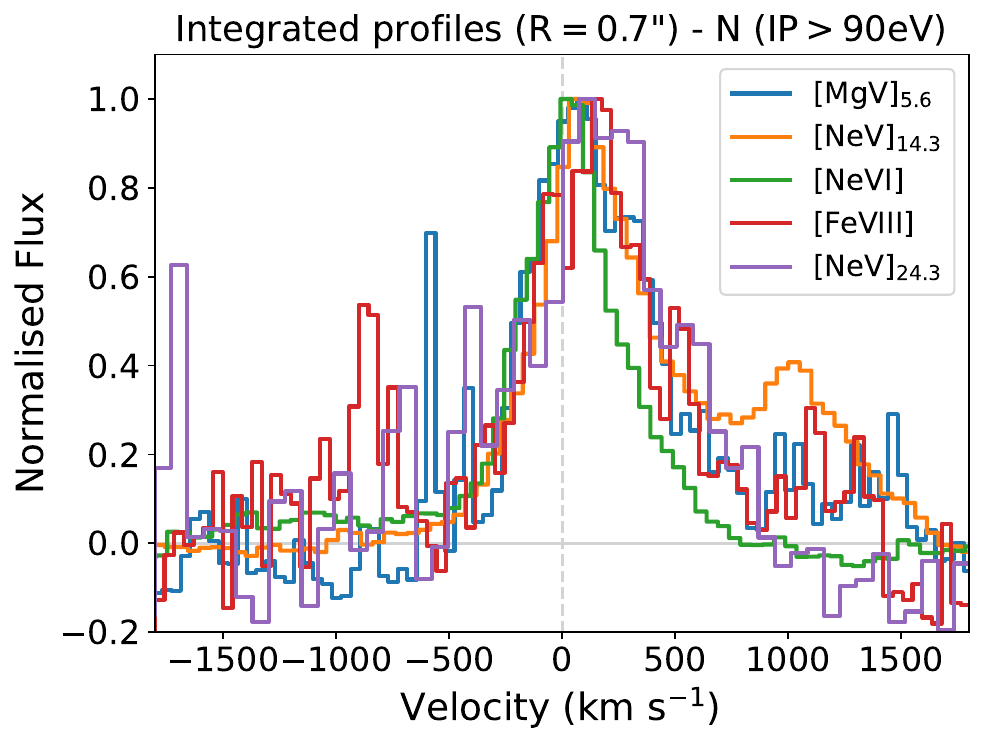} 
	\includegraphics[width=.67\columnwidth]{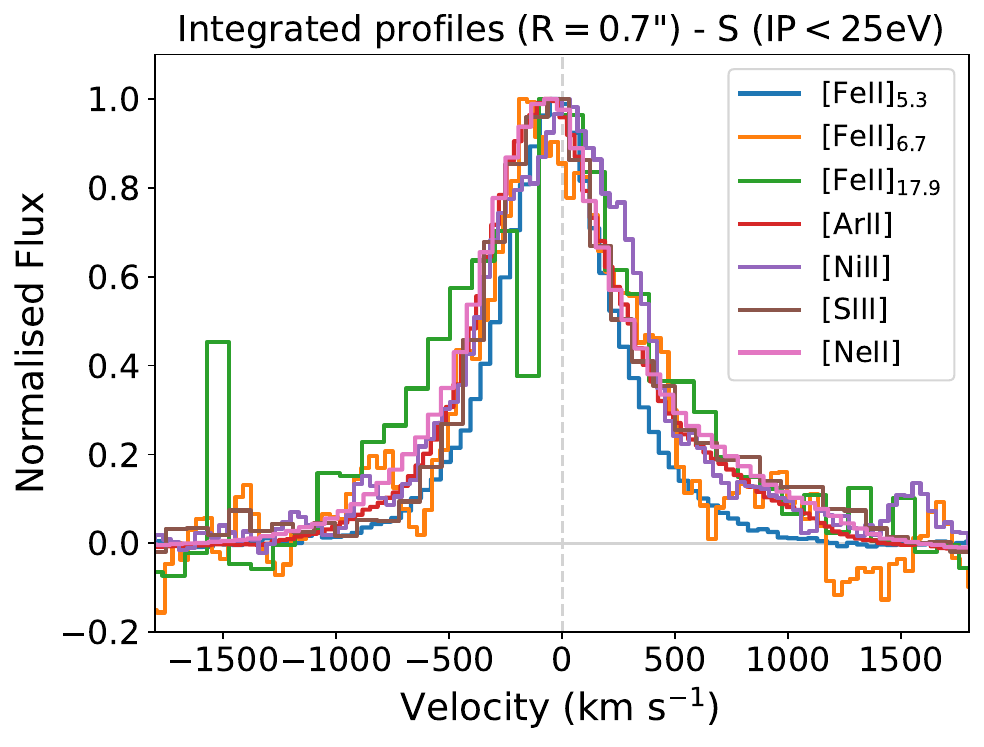} 
	\includegraphics[width=.68\columnwidth]{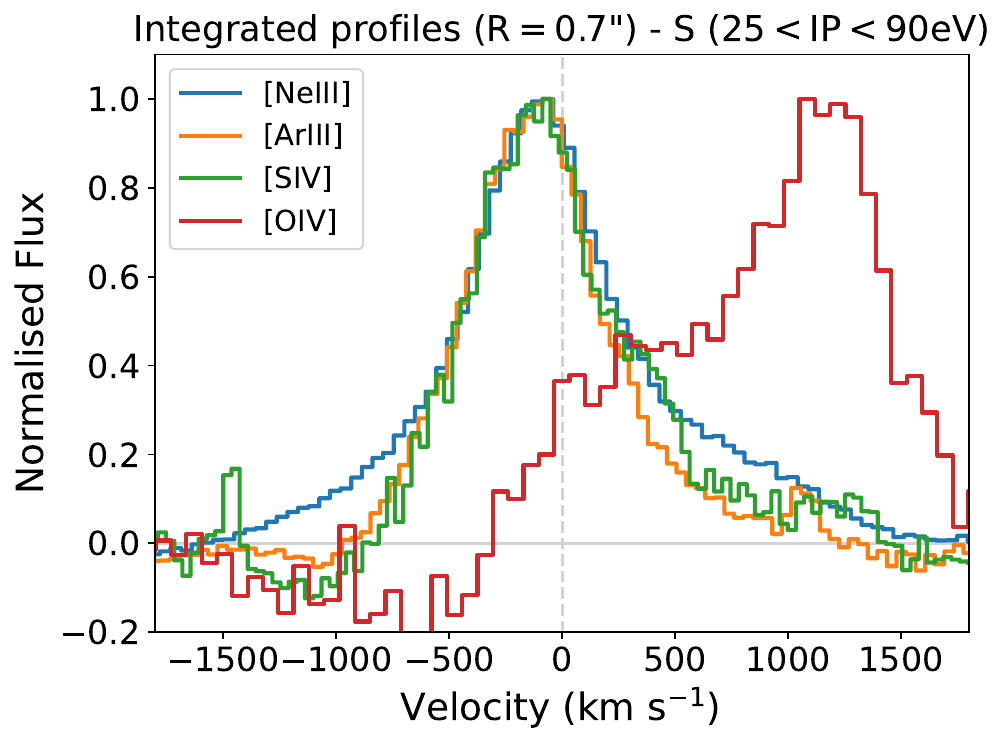} 
	\includegraphics[width=.67\columnwidth]{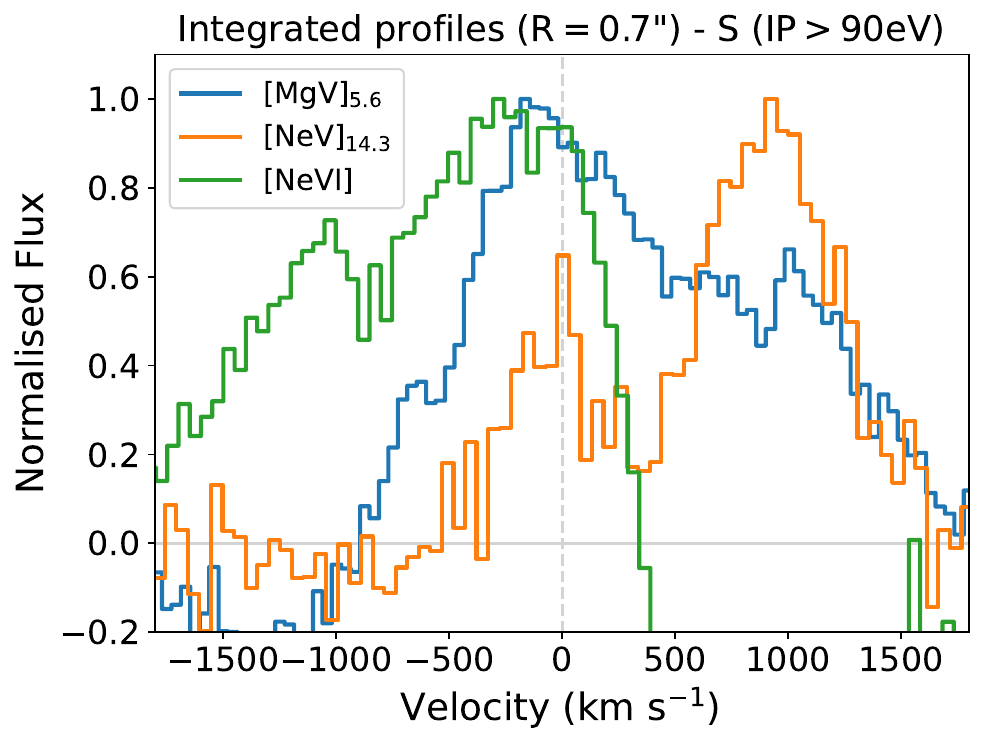} 
	\caption{Integrated profiles (r\,$= 0.7$\arcsec) of all the emission lines detected in the northern (upper panels) and southern (bottom panels) nucleus, after applying the fringing correction (see Sect.~\ref{Sect2:Reduction}), normalised to the maximum flux. The lines are separated by IP, left IP\,$<$\,25\,eV, middle 25\,$<$\,IP\,$<$\,90\,eV, and right IP\,$>$\,90\,eV. The rest frame velocity corresponding to a z\,$=$\,0.02448 is indicated with a dashed, grey, vertical line. We note that [O\,IV] is blended with [Fe\,II] at 25.99$\mu$m and [Ne\,V] with [Cl\,II] at 14.37$\mu$m (middle and right panels, see Sect.~\ref{Subsect3:Results_IntProperties}). We note that in the lower right panel [Ne\,VI] and [Mg\,V] are embedded into PAH features, and thus they are not detected in the original integrated spectrum (see Sects.~\ref{Sect3:Results} and~\ref{SubSect4:Disc_HighIonLines}). }
	\label{Fig:LineProfiles}
\end{figure*}

We have detected a total of 20 different fine-structure emission lines with IPs ranging from 7.6 to 126 eV (see Table~\ref{Table:1}), two hydrogen recombination lines, namely Pf$\alpha$ and Hu$\alpha$ (i.e. H\,I\,(6-5) at 7.46$\mu$m and H\,I\,(7-6) at 12.37$\mu$m, respectively), and 10 warm molecular (H$_{2}$) emission lines (Hermosa Mu{\~n}oz et al. in prep.). All these lines are indicated in Fig.~\ref{Fig1:IntSpec}. In general, the lines for both nuclei have very complex, non-symmetrical profiles, especially for the S nucleus. We have measured widths of at least 1000\,km\,s$^{-1}$ (FWHM), and even higher for some lines such as [Ne\,II] at 12.81$\mu$m and [Ne\,III] at 15.55$\mu$m in the S nucleus (see Fig.~\ref{Fig:LineProfiles}). As a result, lines close in wavelength, such as [Ne\,V] at 14.32$\mu$m and [Cl\,II] at 14.37$\mu$m, and [O\,IV] at 25.89$\mu$m and [Fe\,II] at 25.99$\mu$m, are blended in the spectra (see Fig.~\ref{Fig:LineProfiles}). 

Despite the S nucleus being $\sim$3 times brighter than the N nucleus, the latter shows strong high excitation emission lines (IP\,$> 90$\,eV; e.g. [Ne\,VI] at 7.65$\mu$m) that are barely detected for the S nucleus (see Fig.~\ref{Fig1:IntSpec} and Sect.~\ref{Subsect3:HighIonLines}). These emission lines (e.g. [Ne\,V], [Ne\,VI], and [Mg\,V] at 5.6$\mu$m\footnote{Unless specified otherwise, from here [Ne\,V] and [Mg\,V] refer to the lines at 14.32$\mu$m and at 5.6$\mu$m, respectively.}) can only be explained by AGN photoionisation \citep[see e.g.][]{Genzel1998,Sturm2002,PereiraSantaella2022}, whereas for lines such as [S\,IV] or [O\,IV] with intermediate IPs, and especially for lower ionisation potentials (see Table~\ref{Table:1}), star formation processes contribute the most to the emission \citep[see e.g.][]{Pereira2010, AH2012}. 

We present in Table~\ref{Table:1} the measurements for the fluxes in the integrated spectra (r\,$=$\,0.7\arcsec) for both nuclei. These fluxes are measured by modelling the profiles of the lines with a single Gaussian component. This is a simplification, given that for some lines the profiles are complex and a single Gaussian is not enough to fully reproduce their shape. However, we compared the line ratios obtained with these flux measurements and a simple integration of the line profile, and the values were similar (differences $<$\,5\%). We note that for the blended pairs [Ne\,V] + [Cl\,II], and [O\,IV] + [Fe\,II] at 25.99\,$\mu$m we cannot integrate the profiles, so their fluxes were obtained with a two Gaussian modelling of the lines. From the flux measurements, the strongest ionised gas lines for both nuclei are [Ne\,II], [Ne\,III], [Ar\,II] at 6.98$\mu$m, and [Fe\,II] at 5.34$\mu$m. 

The most relevant low excitation lines for this work (namely [Ne\,II], [Ne\,III], [Fe\,II] at 5.34$\mu$m, and Pf$\alpha$) are shown in Fig.~\ref{Fig:KinMapsLowExcit}, and the maps for the remaining lines are in Appendix~\ref{Appendix_figures}. We show in Fig.~\ref{Fig:KinMapsHighIon} the kinematic and flux maps obtained with a single Gaussian for the detected high-ionisation emission lines (namely [O\,IV], [Ne\,V], [Ne\,VI], and [Mg\,V]).

\subsection{Morphological properties}
\label{SubSect3:Results_fluxes}

In the MIRI images (see Fig.~\ref{Fig1:Image}) we detect ionised gas emission extending $\sim$50\,kpc in the north-south direction. In the inner parts of the image we detect the "Butterfly Nebula" (see Sect.~\ref{Sect1:Introduction}), extending $\sim$5\,kpc east-west, with a bubble-like structure towards the north-west part and several ionised gas filaments to the east and south \citep[see also][]{Medling2021}.

\begin{figure*}
	\centering
	\includegraphics[width=.85\textwidth]{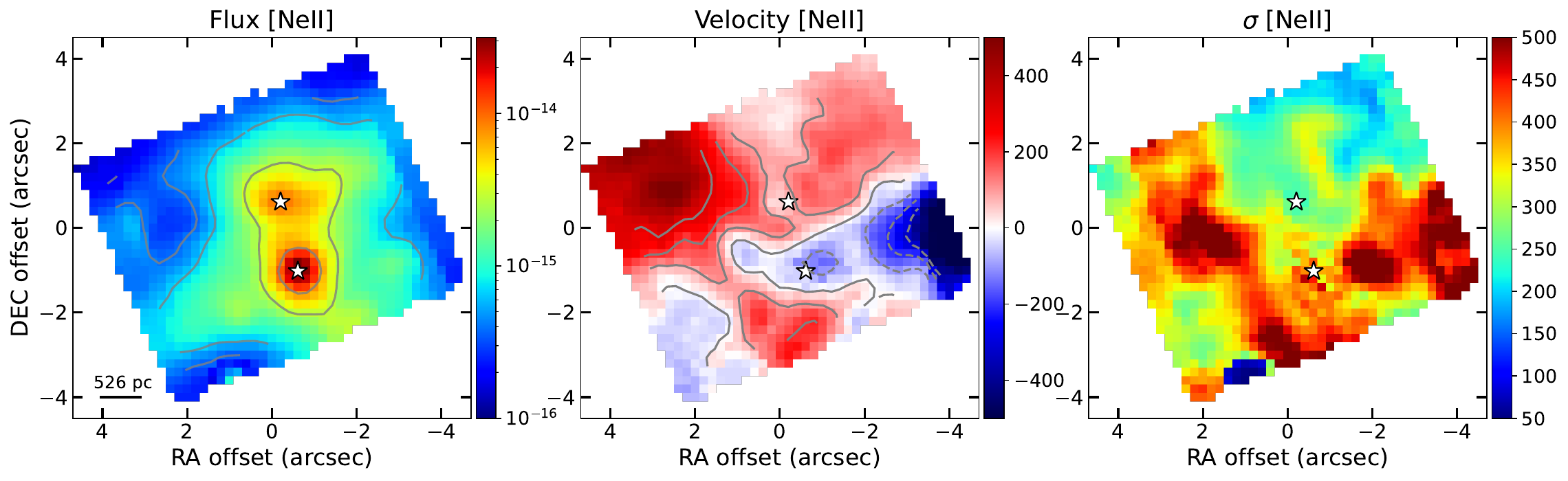} 
	\includegraphics[width=.85\textwidth]{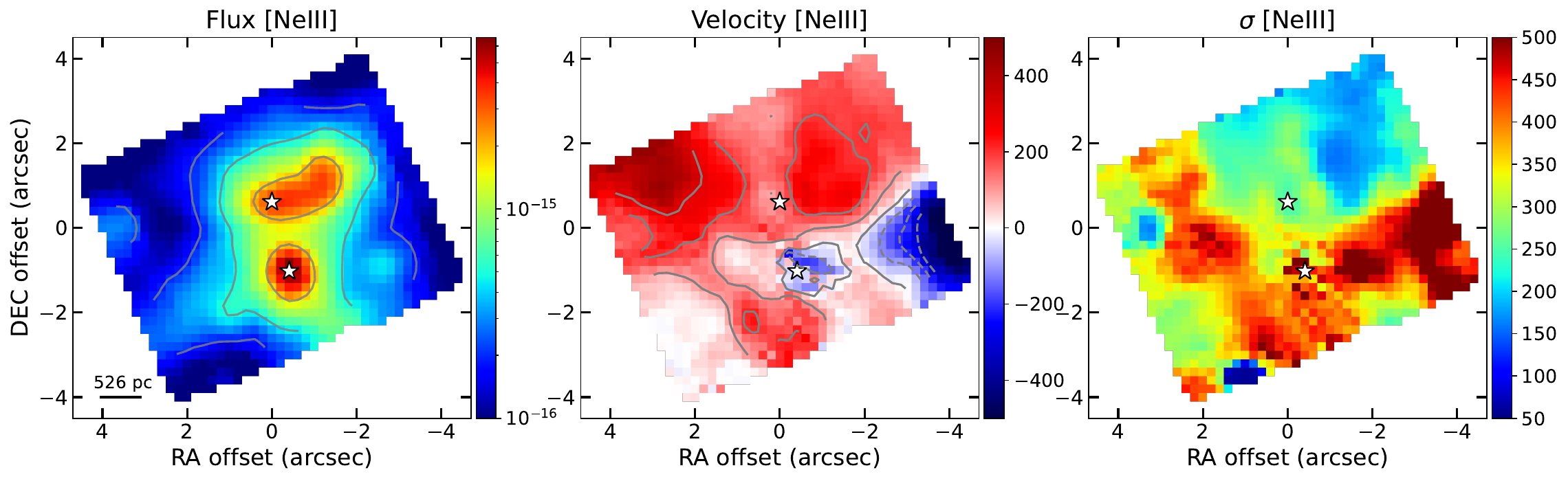}
	\includegraphics[width=.85\textwidth]{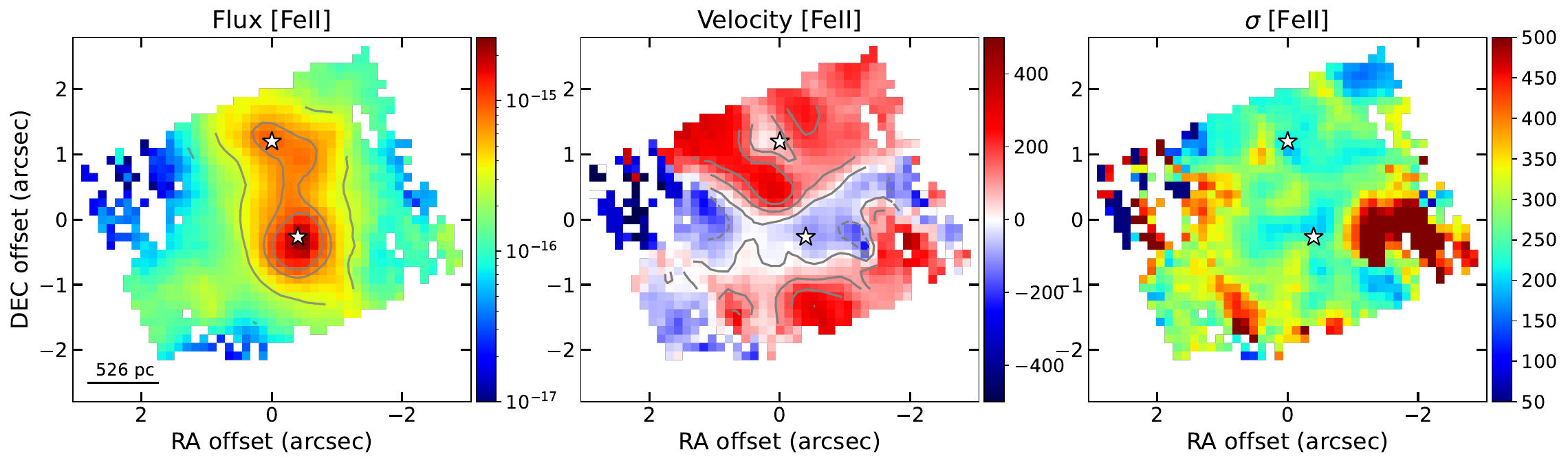} 
	\includegraphics[width=.85\textwidth]{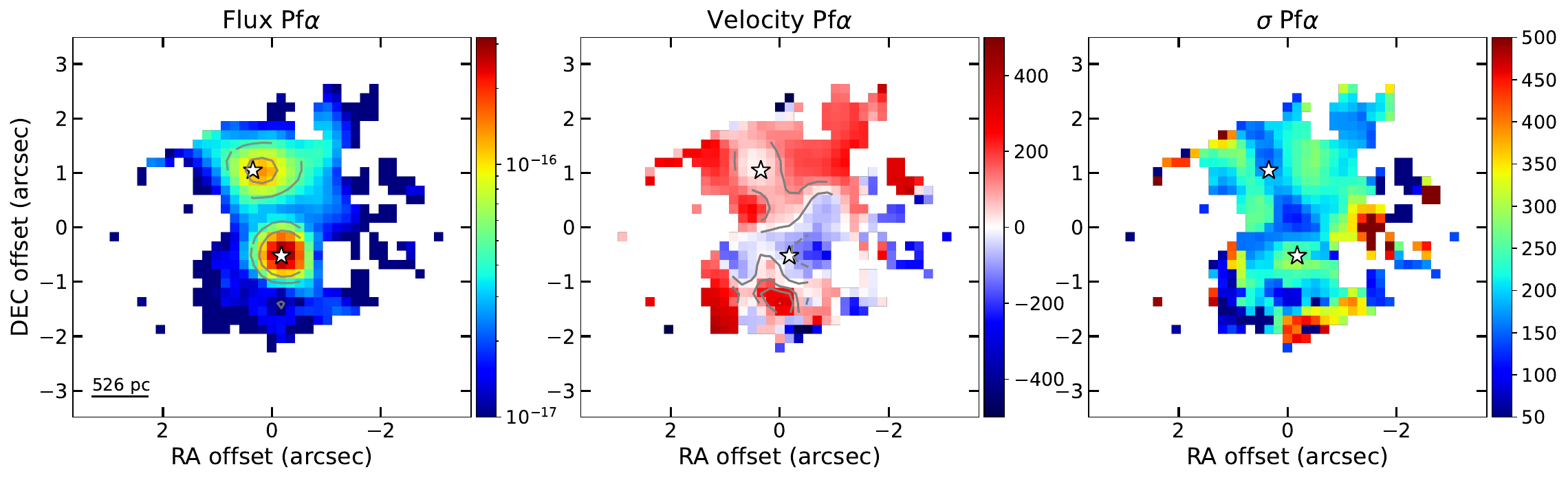} 
	\caption{Kinematic maps obtained with the modelling of a single Gaussian for the [Ne\,II], [Ne\,III], [Fe\,II] at 5.34$\mu$m, and Pf$\alpha$ lines (from top to bottom, respectively). The latter line kinematics was obtained after re-binning the cube with a 2$\times$2 box (see Sect.~\ref{Sect2:Methodology}). From left to right: flux in erg\,s$^{-1}$\,cm$^{-2}$, velocity in km\,s$^{-1}$, and velocity dispersion in km\,s$^{-1}$. The contours in the velocity maps go from $-$300 to 300\,km\,s$^{-1}$ (increments of 100\,km\,s$^{-1}$). The contours in the flux maps go from 10$^{-16}$ to 10$^{-14}$\,erg\,s$^{-1}$\,cm$^{-2}$ (divided in 5 contours), except for Pf$\alpha$ (from $10^{-17}$ to $10^{-15}$\,erg\,s$^{-1}$\,cm$^{-2}$). White stars indicate the photometric centre for both nuclei in their corresponding sub-channels, and the lower-left line indicate the 1\arcsec\,physical scale. The (0,0) point in each panel marks the centre of the FoV at each channel. For all maps, north is up and east to the left.}
	\label{Fig:KinMapsLowExcit}
\end{figure*}

\begin{figure*}
	\centering
	\includegraphics[width=.85\textwidth]{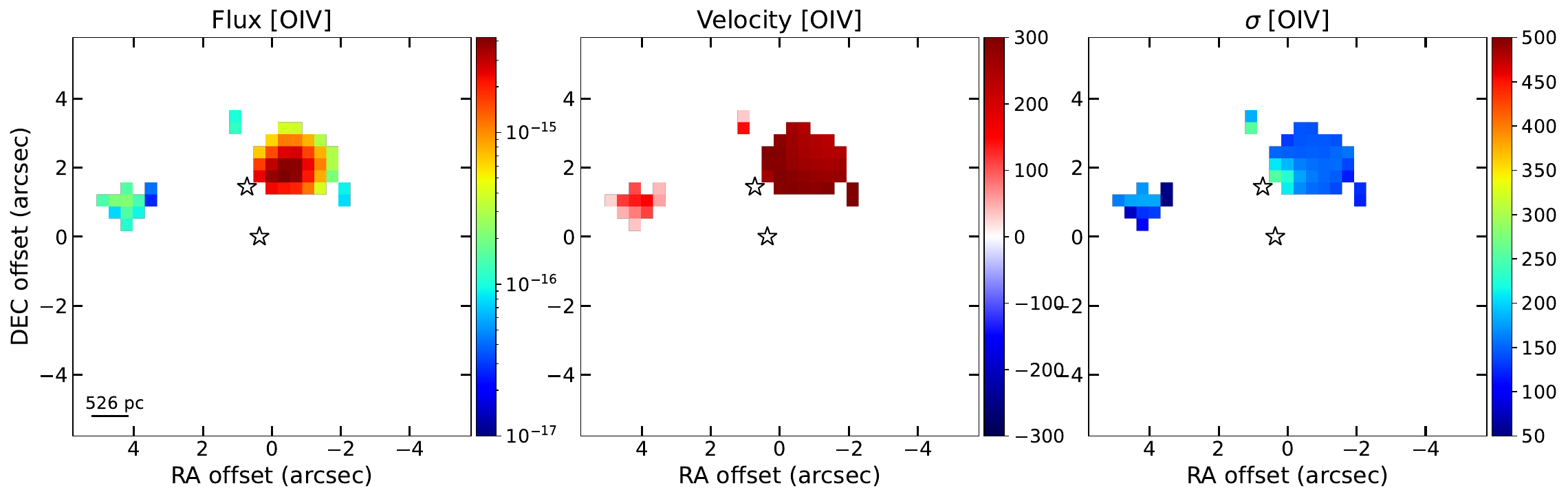}
	\includegraphics[width=.85\textwidth]{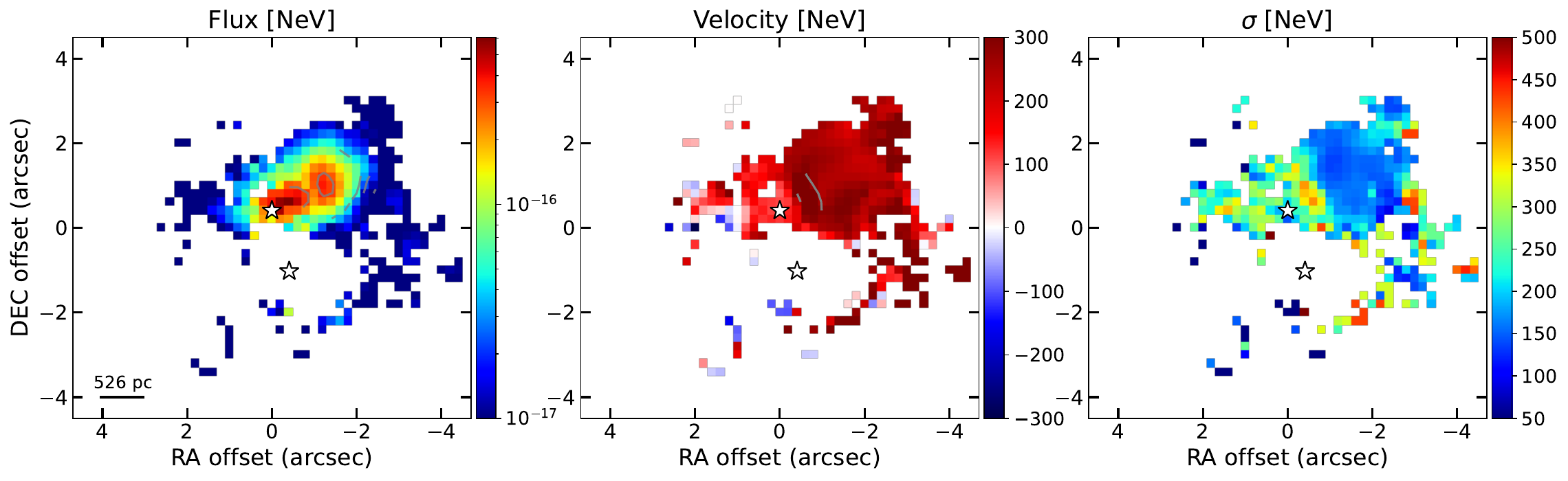}
	\includegraphics[width=.85\textwidth]{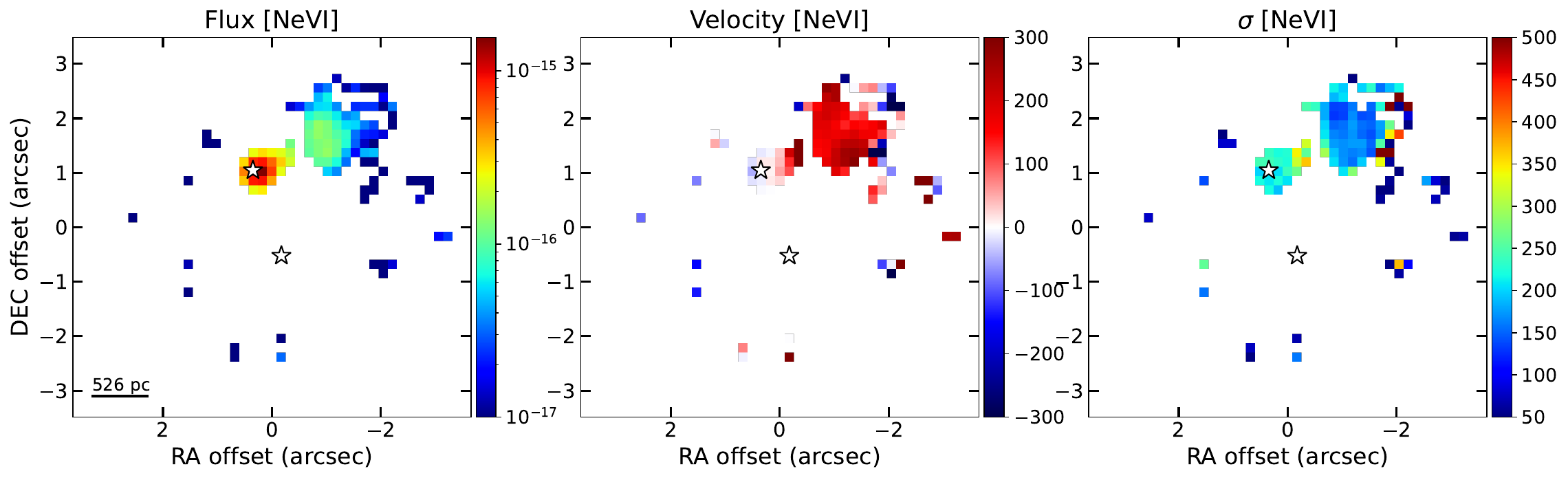} 
	\includegraphics[width=.85\textwidth]{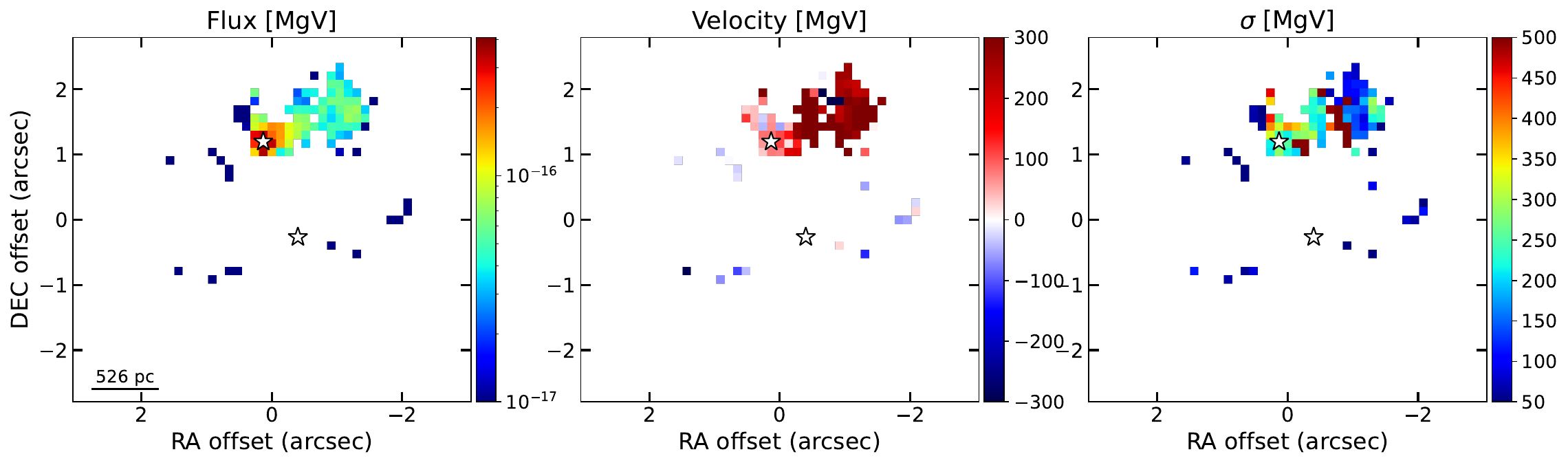} 
	\caption{Kinematic maps obtained with the modelling of a single Gaussian for the [O\,IV], [Ne\,V], [Ne\,VI], and [Mg\,V] lines, top to bottom respectively. Full description in Fig.~\ref{Fig:KinMapsLowExcit}.}
	\label{Fig:KinMapsHighIon}
\end{figure*}

\begin{figure}
	\centering
	\includegraphics[width=.8\columnwidth]{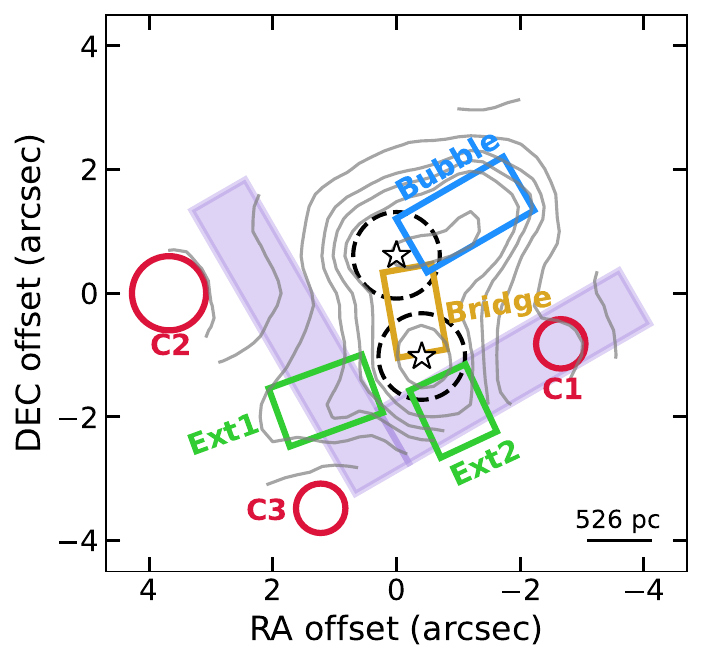} 
	\caption{Schematic figure of the main features seen in the flux and kinematic maps of the emission lines superimposed to the [Ne\,III] flux map contours (see Fig.~\ref{Fig:KinMapsLowExcit}). In red we mark the position of the different detected clumps (namely C1, C2, and C3) and in green the extended emission regions (namely Ext1 and Ext2, see Sect.~\ref{SubSect3:Results_fluxes}). The bubble-like structure mainly detected with the high-excitation lines is in blue (extending up to $\sim5.2\arcsec$, i.e. 2.74\,kpc, see Fig.~\ref{Fig:KinMapsHighIon} and Sect.~\ref{SubSect4:Disc_Outflow}), and the bridge detected between both nuclei (separated $\sim 1.6\arcsec$, i.e. $\sim 840$\,pc, see Sect.~\ref{Subsect3:Results_IntProperties}) is marked in yellow. The purple regions indicate the "V"-shaped structure detected in the velocity dispersion maps. Additionally, we mark with black, dashed circles the region where we integrated the spectra for both nuclei, as well as their position with the white stars.}
	\label{Fig:SummaryRegions}
\end{figure}

\begin{figure}
	\includegraphics[width=\columnwidth]{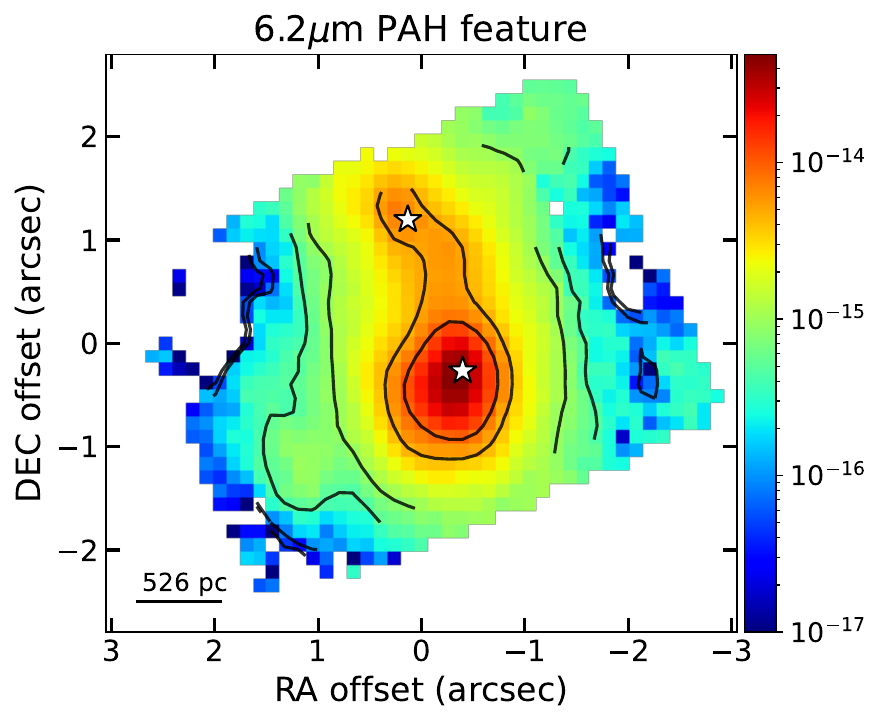} 
	\caption{Flux map of the PAH feature at 6.2$\mu$m in log scale, estimated after subtracting a local continuum at both sides of the feature. The contour levels go from $10^{-16}$ to 10$^{-14}$\,erg\,s$^{-1}$\,cm$^{-2}$.}
	\label{Fig:PAHmap}
\end{figure}

Focusing on the inner regions studied with the MRS data, in the left panels of Figs.~\ref{Fig:KinMapsLowExcit} and~\ref{Fig:KinMapsHighIon} we see that both low and high excitation lines show distinct morphological features. The ionised gas traced by the low excitation lines is detected across the entire FoV (see Fig.~\ref{Fig:KinMapsLowExcit}), with most of the emission concentrated around the two nuclei. The main features detected in the flux maps are indicated in Fig.~\ref{Fig:SummaryRegions} and described below. For all lines, the flux peaks at the S nucleus, although [Ne\,II] shows a bridge connecting both nuclei \citep[see also][]{AH2014}, which is not as prominent for [Ne\,III] (see also the [Fe\,II] map in Fig.~\ref{FigAp:KinMaps_extralines} and the PAH map in Fig.~\ref{Fig:PAHmap}). There is, however, significant extended emission detected both to the north-west of the N nucleus and to the south-east (Ext1 in Fig.~\ref{Fig:SummaryRegions}) and south of the S nucleus, seen most clearly in the [Ne\,II] and [Ne\,III] flux maps (namely Ext2 in Fig.~\ref{Fig:SummaryRegions}, see also left panels in Fig.~\ref{Fig:KinMapsLowExcit}). Both regions roughly coincide with the molecular outflow and one of the ionisation cones detected with the NIRSpec data in Ceci et al. (submitted, see their Fig.~10). This Ext1 region coincides with the NES extension defined in Fig.~5 in \cite{Paggi2022} using soft X-rays. A few additional clumps are apparent in the flux maps, labelled C1 (at $\sim$2\arcsec\, or projected distance $\sim$1\,kpc west from the S nucleus), C2 (at $\sim$3.8\arcsec\, or projected distance $\sim$2\,kpc east of the N nucleus), and C3 (at $\sim$3\arcsec\, or projected distance $\sim$1.6\,kpc southeast of the S nucleus). The C3 clump is seen directly in the MIRI images (see Fig.~\ref{Fig1:Image}), but for the MRS it is only detected in ch3 and ch4 due to their larger FoVs (see Sect.~\ref{Sect2:Data}). The region around C2 coincides with the "Outflow ridge" defined with soft X-ray data \citep[see Fig.~5 in][]{Paggi2022}. In general, there is little correspondence between the morphological features in the flux maps and those in the kinematic maps (see Sect.~\ref{SubSect3:Results_kin}), although C1 coincides with a high velocity dispersion region ($\sigma \sim$\,500\,km\,s$^{-1}$) for both neon lines. In general, we found that the morphology of the ionised gas (see Figs.~\ref{Fig:KinMapsLowExcit} and~\ref{FigAp:KinMaps_extralines}) closely resembles that of the $0.3 - 3$\,keV X-ray emission (see Fig.~2 in \citealt{Paggi2022}), as already noted in several other studies \citep{Nardini2013,Yoshida2016,Paggi2022}.

In Fig.~\ref{Fig:PAHmap} we show the flux map of the PAH feature at 6.2$\mu$m, obtained by integrating over its total wavelength range (6.15$\mu$m to 6.5$\mu$m) after subtracting a linearly interpolated continuum estimated from points at either sides of the feature. As for the low excitation lines, the peak of the flux is located in the S nucleus, and we clearly detect the bridge between both nuclei. The extended emission towards the south-east from the S nucleus (Ext1 in Fig.~\ref{Fig:SummaryRegions}) is also detected. 

The extended emission detected to the north-west of the N nucleus is even more evident in the high excitation line emission (see Fig.~\ref{Fig:KinMapsHighIon}). These high excitation lines (IP\,$>$\,50\,eV) are only detected in and around the N nucleus in the pixel-by-pixel modelling. They expand mainly in a bubble-like shape towards the north-west region at a PA of $\sim 25^{\circ}$ (measured west to north from the N nucleus), with similar morphologies for [Ne\,VI], [Mg\,V], and [Ne\,V]. The [O\,IV], which is observed in the larger FoV ch4-long channel (see Sect.~\ref{Sect2:Reduction}), reveals some emission to the south-east at $\sim$\,4\arcsec\,(projected distance $\sim$2.1\,kpc) from the N nucleus (namely C2, see Fig.~\ref{Fig:SummaryRegions}) at a similar PA that the NW emission. This emission is coincident with the beginning of the east arm detected with [O\,III] in \cite{MullerSanchez2018} \citep[see also][]{Medling2021}.

\begin{table}
	\caption{Flux measurements of the emission lines of NGC\,6240 in the integrated spectra (r$\sim$0.7\arcsec) of both nuclei, ordered by ionisation potential. }
	\label{Table:1}
	\centering          
	\begin{tabular}{lrrcc}
		\hline\hline
		Line & $\lambda$ & IP & Flux N & Flux S \\ 
		& ($\mu$m) & (eV) & \multicolumn{2}{c}{($\times 10^{-16}$\,erg\,cm$^{-2}$\,s$^{-1}$)} \\
		\hline           
		Pf$\alpha$      & 7.46  & --   & $21.3\pm0.3$    & $65.4\pm1.2$     \\
		$[\rm Ni\,II]$  & 6.64  & 7.6  & $45.7\pm0.1$    & $142.5\pm5.7$    \\ 
		$[\rm Fe\,II]$  & 5.34  & 7.9  & $471.7\pm4.0$   & $835.6\pm14.9$   \\ 
		$[\rm Fe\,II]$  & 6.72  & 7.9  & $26.2\pm0.8$    & $54.6\pm4.4$     \\ 
		$[\rm Fe\,II]$  & 17.94 & 7.9  & $57.1\pm5.1$    & $87.3\pm33.1$    \\ 
		$[\rm Fe\,II]$  & 25.99 & 7.9  & $158.4\pm10.6$  & $289.6\pm20.7$   \\ 
		$[\rm Cl\,II]$  & 14.37 & 13.0 & $36.4\pm2.1$    & $47.1\pm3.3$     \\ 
		$[\rm Ar\,II]$  & 6.98  & 15.8 & $767.3\pm21.9$  & $2034.6\pm211.6$ \\ 
		$[\rm Fe\,III]$ & 22.93 & 16.2 & $38.0\pm0.1$    & -- \\ 
		$[\rm Ne\,II]$  & 12.81 & 21.6 & $1855.1\pm10.3$ & $4492.9\pm342.2$ \\
		$[\rm S\,III]$  & 18.71 & 23.3 & $238.9\pm4.6$   & $301.7\pm23.0$   \\
		$[\rm Ar\,III]$ & 8.99  & 27.6 & $73.3\pm2.8$    & $93.4\pm9.7$     \\ 
		$[\rm S\,IV]$   & 10.51 & 34.9 & $70.1\pm1.9$    & $43.9\pm7.5$     \\
		$[\rm Ne\,III]$ & 15.55 & 41.0 & $711.8\pm4.1$   & $1110.9\pm73.2$  \\
		$[\rm Na\,III]$ & 7.32  & 47.3 & $14.0\pm2.3$    & $12.4\pm3.5$     \\
		$[\rm O\,IV]$   & 25.89 & 54.9 & $250.9\pm10.5$  & $162.9\pm21.5$   \\
		$[\rm Ne\,V]$   & 14.32 & 97.2 & $59.8\pm1.6$    & $16.4\pm2.6$     \\
		$[\rm Ne\,V]$   & 24.32 & 97.2 & $47.6\pm13.3$   & -- \\
		$[\rm Mg\,V]^{*}$  & 5.60 & 109 & $43.4\pm6.5$   & $13.1\pm0.5$     \\ 
		$[\rm Fe\,VIII]$   & 5.45 & 125 & $14.3\pm1.4$   & -- \\
		$[\rm Ne\,VI]^{*}$ & 7.65 & 126 & $124.8\pm8.9$  & $24.1\pm3.1$     \\ 
		\hline        
	\end{tabular}\\
	\tablefoot{"--" indicates that the line is not detected with S/N\,$> 3$. $^{*}$ indicates that for those lines the fluxes of the S nucleus have been measured in the PAH subtracted spectrum (see Sect.~\ref{SubSect4:Disc_HighIonLines}).}
\end{table}

\subsection{Mid-IR line ratios}
\label{SubSect3:LineRatios}

In general, the low excitation lines such as [Ne\,II] are typically ionised by star formation processes and/or shocks, whereas the high excitation lines (IP\,$>$\,90\,eV) such as [Ne\,V] can only be produced by AGN photoionisation. As mentioned in Sect.~\ref{Sect3:Results}, the emission lines with intermediate IPs (IP\,$>$\,25\,eV) can be produced due to ionisation from both stars and AGN, thus they tend to show more complex kinematics and flux properties \citep[see also][]{Armus2023,Dasyra2024,HM2024b}. We use the mid-IR line ratios obtained by \cite{Pereira2010}, in particular [Ne\,V]/[Ne\,II], [Ne\,III]/[Ne\,II], [O\,IV]/[Ne\,II], and [O\,IV]/[Ne\,V], to disentangle between the main ionising source of the different regions in NGC\,6240.
First, we obtained the overall properties of the nuclei with the integrated line ratios (r$\sim$0.7\arcsec), using both high and low excitation lines. We found that for the N nucleus, all ratios using the Neon and [O\,IV] lines are consistent with those found in LINERs and/or star-forming regions \citep{Pereira2010}: [O\,IV]/[Ne\,V]\,$=$\,4.20 (0.62 in log); [O\,IV]/[Ne\,II]\,$=$\,0.14 ($-$0.86 in log); [Ne\,III]/[Ne\,V]\,=\,11.9 (1.08 in log); and [Ne\,V]/[Ne\,II]\,$=$\,0.03 ($-$1.49 in log). As for the S nucleus, all the line ratios are consistent with star-forming regions \citep{Pereira2010}: [O\,IV]/[Ne\,V]\,$=$\,9.9 (1.0 in log); [O\,IV]/[Ne\,II]\,$=$\,0.04 ($-$1.44 in log); [Ne\,III]/[Ne\,V]\,=\,67.7 (1.8 in log); and [Ne\,V]/[Ne\,II]\,$=$\,0.004 ($-$2.438 in log). However, using the [O\,IV] ratios, if we put both nuclei in Fig.~6 in \cite{Hernandez2023}, the S nucleus lies between the starburst-dominated systems, whereas the N nucleus falls in the AGN-dominated region.   

\begin{figure}
	\centering
	\includegraphics[width=.9\columnwidth]{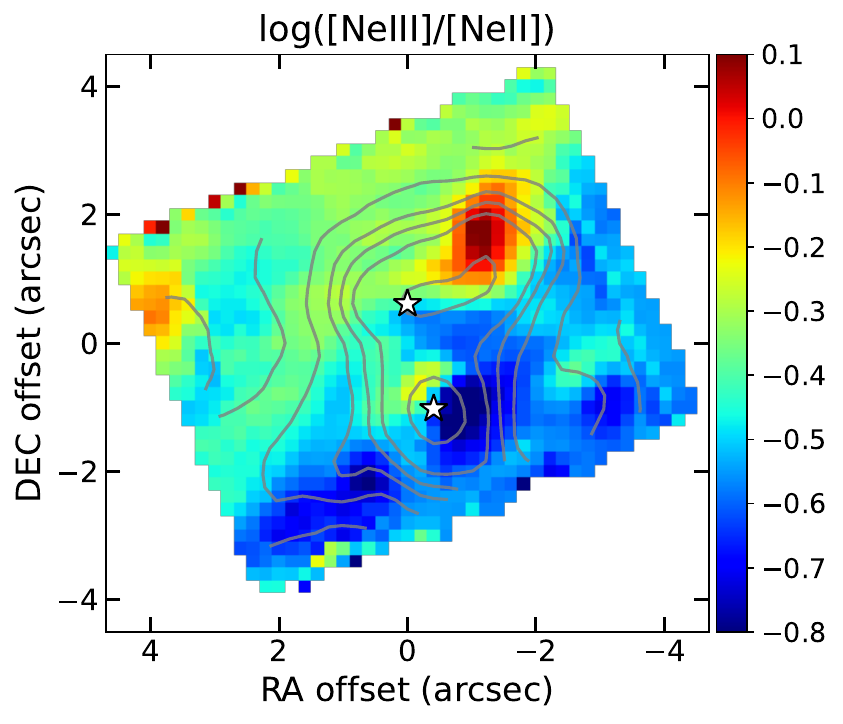} 
	\includegraphics[width=.9\columnwidth]{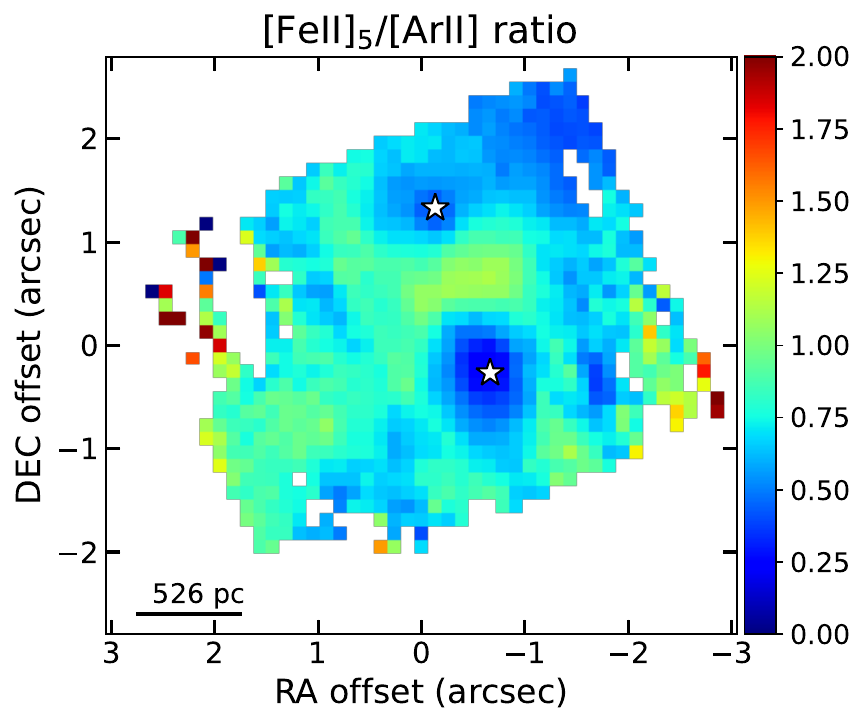} 
	\caption{Line ratios between [Ne\,III] and [Ne\,II] in log scale (upper panel) and [Fe\,II] over [Ar\,II] in linear scale (bottom panel). In the top panel we superimposed the contours from the [Ne\,III] flux map (see Fig.~\ref{Fig:KinMapsLowExcit}). The white stars indicate the position of both nuclei.}
	\label{Fig:RatioLines}
\end{figure}

Additionally, we created a [Ne\,III]/[Ne\,II] line ratio map in Fig.~\ref{Fig:RatioLines} (top panel), as [Ne\,III] is the line with the highest IP (41\,eV) detected in single-spaxel spectra across the entire FoV, including the S nucleus. In general, the ratio is low (median of $-0.44$ and standard deviation of 0.16 in logarithm scale), and consistent with those found in LINERs in \cite{Pereira2010}.
The various regions identified in the flux maps (see Fig.~\ref{Fig:SummaryRegions}) also show clear differences in their line ratios. The highest (lowest) ratios are located towards the northern (southern) part of the FoV. We found the minimum around the S nucleus (average value of $-0.82\pm0.08$ in log) and towards the extended emission Ext1 (see Fig.~\ref{Fig:SummaryRegions}). The region of the bubble, located west of the N nucleus, and C2, located $\sim$3.8\arcsec\,east of it, have the highest ratios ($>-0.1$ in log). Following again \cite{Pereira2010}, these ratios are closer to those expected for regions photoionised by a Seyfert nucleus. Also, the extended region west of the N nucleus, C2, is oriented similarly to the bubble-like structure clearly seen in the high excitation lines. 
We also created a [Fe\,II] at 5.34$\mu$m over [Ar\,II] map, shown in the bottom panel of Fig.~\ref{Fig:RatioLines}. This ratio could be useful for disentangle between shocks (using [Fe\,II]) and SF (traced by [Ar\,II]). The minimum values of this ratio ($\sim 0.5$) are located in both nuclei and in the bubble region. The maximum values ($\sim 1.2$) are however located in the bridge between both nuclei, which could be tracing shocks produced by the interaction between the original galaxies, and in the extended region Ext1. 
We further discuss these ratios and their comparison to optical diagnostics in Sect.~\ref{SubSect4:Disc_LineRatios}. 

\subsection{Kinematical properties}
\label{SubSect3:Results_kin}

\noindent We represent in Fig.~\ref{Fig:LineProfiles} the profiles of the emission lines detected in the integrated spectra (see Fig.~\ref{Fig1:IntSpec}). When simply integrating, the high excitation lines in the S nucleus are not clearly detected ([Mg\,V] and [Ne\,VI] are inside a PAH feature, as shown in lower-right panel of Fig.~\ref{Fig:LineProfiles}). From the emission line profiles, it is clearly seen that the two nuclei have different rest-frame velocities. In the N nuclei, the maximum is shifted by $\sim 100$\,km\,s$^{-1}$ after correcting for the redshift (see Sect.~\ref{Sect1:Introduction}), but for the S nucleus the peak is shifted by $-100$\,km\,s$^{-1}$ for all the detected lines (see v50 parameter in Table~\ref{Table:2_W80}). For some lines, such as [Ne\,II] or [Ne\,III], there is a prominent redshifted wing reaching velocities $> 1000$\,km\,s$^{-1}$ for the S nucleus, that was similarly detected for the molecular gas \citep[see e.g.][]{Cicone2018}. 

\subsubsection{Non-parametric modelling}
\label{Subsubsect3:Result_Kin_NonParamModel}

From the non-parametric analysis (see Sect.~\ref{Sect2:Methodology}), we found that the average value for the W80 parameter is $\sim$600\,km\,s$^{-1}$ for the N nucleus, while the S nucleus presents systematically larger widths, up to $\sim 1400$\,km\,s$^{-1}$. From Table~\ref{Table:2_W80}, it is clear that the low excitation lines tend to have lower widths than the high excitation lines. 
\cite{Dasyra2011} suggested that, if the narrow line region of the AGN is stratified, then the lines with higher IPs would have larger FWHM that those with lower IPs. This would be due to the fact that they originate closer to the AGN and thus are more affected by the strong motions taking place in the inner regions. These trends have been seen for other objects \citep{Armus2023}, such as the type-2 Seyfert NGC\,7172 (see \citealt{HM2024b}). We have plotted the W80 parameter versus the IP for the N nucleus, where we detect lines at all IPs, and we only detected a trend with moderate significance ($\rho \sim 0.07$ and p-value $\sim 0.79$). We did not detect a trend of the derived velocities with the IP of the lines. In terms of the largest velocities characterised with the v02 and v98 parameters, overall, the velocity structure of the lines spans more than 800\,km\,s$^{-1}$ and up to $\sim 1300$\,km\,s$^{-1}$ (N) and $\sim 2000$\,km\,s$^{-1}$ (S) for lines such as [Ne\,III]. Although the rotational motions contribute to the line widths, these high velocities and the highly asymmetric line profiles are typically attributed to the presence of non-rotational motions such as outflows (see discussion in Sect.~\ref{SubSect4:Disc_Outflow}). 

\subsubsection{Parametric modelling}
\label{Subsubsect3:Result_Kin_ParamModel}

As mentioned in Sect.~\ref{Subsect3:Results_IntProperties}, the line profiles are complex and this makes it challenging a proper decomposition and ordering of all the components within the FoV. We tried modelling the ionised gas emission lines on a spaxel-by-spaxel basis using a multi-Gaussian approach. At least two Gaussians are found across most of the FoV, but there are additional broad components and visible double peaks particularly towards the south-west and south-east regions. In addition, for the two nuclei, the different components blend together, generating very broad profiles that cannot be easily disentangled, especially at the longest wavelengths. All of this makes it difficult to correctly follow the individual kinematic components across the FoV, and associate them consistently with a particular physical process. Thus, we decided to just discuss the overall kinematic properties using a one Gaussian fitting, but we show some examples of the multi-Gaussian modelling of the integrated line profiles for both nuclei in Fig.~\ref{Fig:Appendix_FitIntLinesParam}. We discuss and compare the non-parametric approximation and the parametric results in Appendix~\ref{Appendix_figures}. 

We have estimated the kinematic and flux maps for all lines with a single Gaussian component, which are equivalent to moment 0 and moment 1 maps (see examples of the maps for the main lines in Fig.~\ref{Fig:KinMapsLowExcit},~\ref{Fig:KinMapsHighIon} and~\ref{FigAp:KinMaps_extralines}). 
As mentioned in Sect.~\ref{SubSect3:Results_fluxes}, for all the high-ionisation emission lines we detect the peak of the emission at the N nucleus, and some extended emission up to 2\arcsec\, towards the north-west. Additionally, we detect some extended [O\,IV] emission $\sim$4\arcsec\,east of the nucleus. 
For all the lines the global kinematic properties (i.e. velocity and velocity dispersion, $\sigma$) are very similar. The nuclear region shows higher values of the velocity dispersion ($\sigma \sim$300\,km\,s$^{-1}$) than the north-west, extended region ($\sigma \sim$150\,km\,s$^{-1}$). The lines are completely redshifted with velocities of up to 300\,km\,s$^{-1}$ except at the nucleus, where the lines are at almost rest-frame values. However, the low and intermediate ionised emission line kinematics show a more complex picture. 

In Fig.~\ref{Fig:KinMapsLowExcit} we present flux maps of the [Ne\,II] and [Ne\,III] lines, modelled with a single Gaussian. These are representative for other lines with similar IPs while covering a larger field of view (see also Fig.~\ref{FigAp:KinMaps_extralines}). The velocity and velocity dispersion maps are highly disturbed, i.e. do not follow a classical rotation pattern. As mentioned in Sect.~\ref{SubSect3:Results_fluxes}, this makes it difficult to associate their morphology with any particular kinematic feature(s). In general, the kinematic maps for both lines are similar, with the redshifted velocities (up to $500$\,km\,s$^{-1}$) located towards the northern and eastern part of the FoV and the blueshifted (down to $-600$\,km\,s$^{-1}$) mainly towards the west from the S nucleus. The region around the N nucleus has almost rest-frame velocities and a velocity dispersion of $\sim$300\,km\,s$^{-1}$ for both [Ne\,II] and [Ne\,III] lines. However, the region around the S nucleus is blueshifted for both lines (v$\sim -100$\,km\,s$^{-1}$) with larger velocity dispersion ($\sigma \sim$400\,km\,s$^{-1}$). This velocity difference was already hinted at in the analysis of the integrated spectra of the nuclear regions (see Table~\ref{Table:2_W80}). 

The velocity dispersion is highly disturbed, with the highest values ($\sigma >$\,400\,km\,s$^{-1}$) distributed in a "V"-like shape extending from the north-east part towards the south-west part, passing through the S nucleus. It is particularly enhanced towards the south-west region (maximum $\sigma \sim 590$\,km\,s$^{-1}$ for [Ne\,III], see bottom left panel in Fig.~\ref{Fig:KinMapsLowExcit}), where the more blueshifted velocities are found. 
In contrast, the bubble region has redshifted velocities (up to $\sim$200\,km\,s$^{-1}$) and low velocity dispersion ($\sim$\,250\,km\,s$^{-1}$ for [Ne\,II] and $\sim$\,200\,km\,s$^{-1}$ for [Ne\,III]), consistently to the kinematic properties derived from the high excitation line (see Fig.~\ref{Fig:KinMapsHighIon}).

There is a region at the limits of the FoV to the southeast from the S nucleus (C3 in Fig.~\ref{Fig:SummaryRegions}), with almost rest frame velocities and very low velocity dispersion ($\sim$50\,km\,s$^{-1}$), that is clearly seen as a blob in [S\,III] (see Fig.~\ref{FigAp:KinMaps_extralines}). This region is also detected in the [Ar\,II] map and in the F560W filter MIRI image, located at 3\arcsec\,SE (61$^{\circ}$ east to south) from the S nucleus. It is likely a SF region corresponding to one of the clusters previously detected in this system \citep[see][]{Pollack2007}.

\begin{table*}
	\caption{Results from the non parametric method (see Sect.~\ref{Sect2:Data}) for all the ionised gas emission lines (r$\sim$0.7\arcsec).}
	\label{Table:2_W80}
	\centering          
	\begin{tabular}{lcccccccc}
		\hline\hline
		Line & W80$_{N}$ & W80$_{S}$ & v02$_{N}$ & v02$_{S}$ & v50$_{N}$ & v50$_{S}$ & v98$_{N}$ & v98$_{S}$ \\ 
		& (km s$^{-1}$) & (km s$^{-1}$) & (km s$^{-1}$) & (km s$^{-1}$) & (km s$^{-1}$) & (km s$^{-1}$) & (km s$^{-1}$) & (km s$^{-1}$) \\ 
		\hline           
		$[\rm Ni\,II]$      & 635  & 1023 & $-$355 & $-$814  & 104  & $-$2   & 562 & 1233 \\ 
		$[\rm Fe\,II]_{5}$  & 658  & 745  & $-$449 & $-$756  & 121  & $-$54  & 735 & 647  \\ 
		$[\rm Fe\,II]_{6}$  & 627  & 662  & $-$365 & $-$469  & 123  & $-$51  & 610 & 436  \\ 
		$[\rm Fe\,II]_{17}$ & 1077 & 1273 & $-$984 & $-$984  & $-$5 & $-$103 & 680 & 1071 \\ 
		$[\rm Cl\,II]$      & --   & --  & --     & --   & --   & --  & --  & --  \\ 
		$[\rm Ar\,II]$      & 704  & 1005 & $-$481 & $-$816  & 89   & $-$45  & 726 & 961  \\  
		$[\rm Ne\,II]$      & 799  & 1142 & $-$707 & $-$936  & 92   & $-$79  & 834 & 1063 \\ 
		$[\rm S\,III]$      & 751  & 1032 & $-$534 & $-$628  & 123  & $-$65  & 780 & 968  \\  
		$[\rm Ar\,III]$     & 719  & 889  & $-$424 & $-$721  & 83   & $-$128 & 633 & 972  \\  
		$[\rm S\,IV]$       & 760  & 796  & $-$449 & $-$630  & 130  & $-$123 & 745 & 456  \\
		$[\rm Ne\,III]$     & 753  & 1317 & $-$557 & $-$1121 & 102  & $-$86  & 760 & 1184 \\
		$[\rm O\,IV]$       & --   &  --  & --     & --      & --   & --     & --  & --   \\ 
		$[\rm Ne\,V]_{14}$  & --   &  --  & --     & --      & --   & --     & --  & --   \\ 
		$[\rm Ne\,V]_{24}$  & 866  &  --  & $-$792 & --      & 75   & --     & 580 & --   \\ 
		$[\rm Mg\,V]_{5}$   & 668  &  --  & $-$601 & --      & 25   & --     & 400 & --   \\ 
		$[\rm Fe\,VIII]$    & 688  &  --  & $-$470 & --      & 132  & --     & 648 & --   \\ 
		$[\rm Ne\,VI]$      & 597  &  --  & $-$405 & --      & $-$7 & --     & 490 & --   \\ 
		\hline        
	\end{tabular}\\
	\tablefoot{"--" indicates that the line is not detected with S/N\,$> 3$, except for [Ne\,V]+[Cl\,II] and [O\,IV]+[Fe\,II], where the modelling was not applied as they are blended (see Fig.~\ref{Fig:LineProfiles}).}
\end{table*}

\subsection{High ionisation emission lines}
\label{Subsect3:HighIonLines}

As mentioned in Sect.~\ref{SubSect3:Results_fluxes}, evaluation of the high excitation lines on a spaxel-by-spaxel basis is only possible around the N nucleus, likely because the S nucleus emits a much brighter MIR continuum \citep[see e.g.][]{Puccetti2016}. The largest difference for the continuum emission between nuclei is found at the longest wavelengths (factor $\sim$\,4; see Fig.~\ref{Fig1:IntSpec} and also Fig.~\ref{Fig:Appendix_ContMaps}). This could impact the detection of high ionisation emission lines both in the spatially-resolved (see Fig.~\ref{Fig:KinMapsHighIon}) and the integrated analysis of the S nucleus (see Fig.~\ref{Fig:LineProfiles}). 

From the spectra in Fig.~\ref{Fig1:IntSpec}, it is clear that in both nuclei of the NGC\,6240 system, the emission from PAH features is intense and complex, thus hampering the detection and characterisation of weak emission lines. This is particularly true for [Ne\,VI] and [Mg\,V] lines, that lie inside a PAH feature at 7.7$\mu$m and 6.2$\mu$m, respectively. In fact, there is PAH emission all over the mapped FoV (see the 6.2$\mu$m PAH feature map in Fig.~\ref{Fig:PAHmap}, and see also \citealt{AH2014} for the PAH feature at 11.3$\mu$m), which is particularly prominent around the S nucleus and slightly fainter in the N nucleus and in the bridge connecting the two nuclei, similarly to [Fe\,II] (see Sect.~\ref{SubSect3:Results_kin} and Fig.~\ref{FigAp:KinMaps_extralines}). 

For these reasons, to properly detect and characterise the high excitation lines in both nuclei, we model and subtract the PAH emission from the integrated spectra using the tool developed by \cite{Donnan2024} (see Sect.~\ref{Sect2:methodPAH}). 

\begin{figure*}
	\centering
	\includegraphics[width=\textwidth]{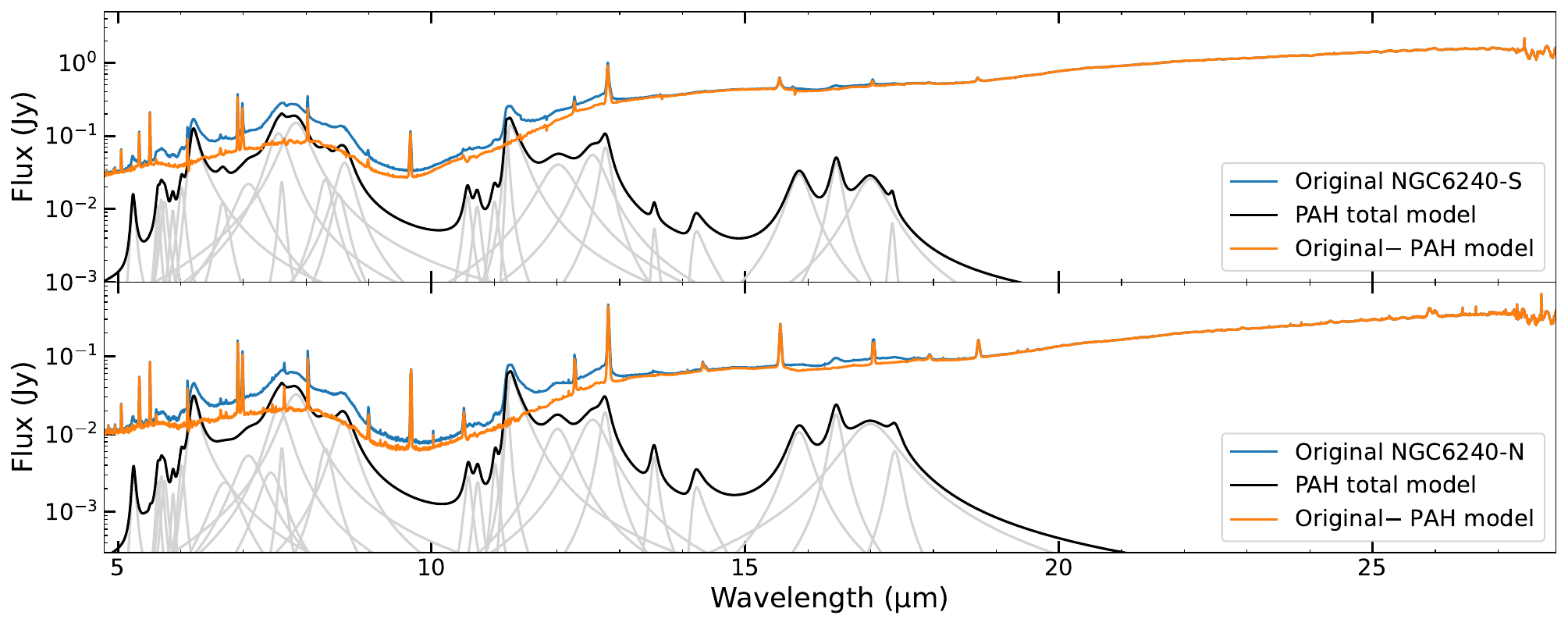} 
	\includegraphics[width=\textwidth]{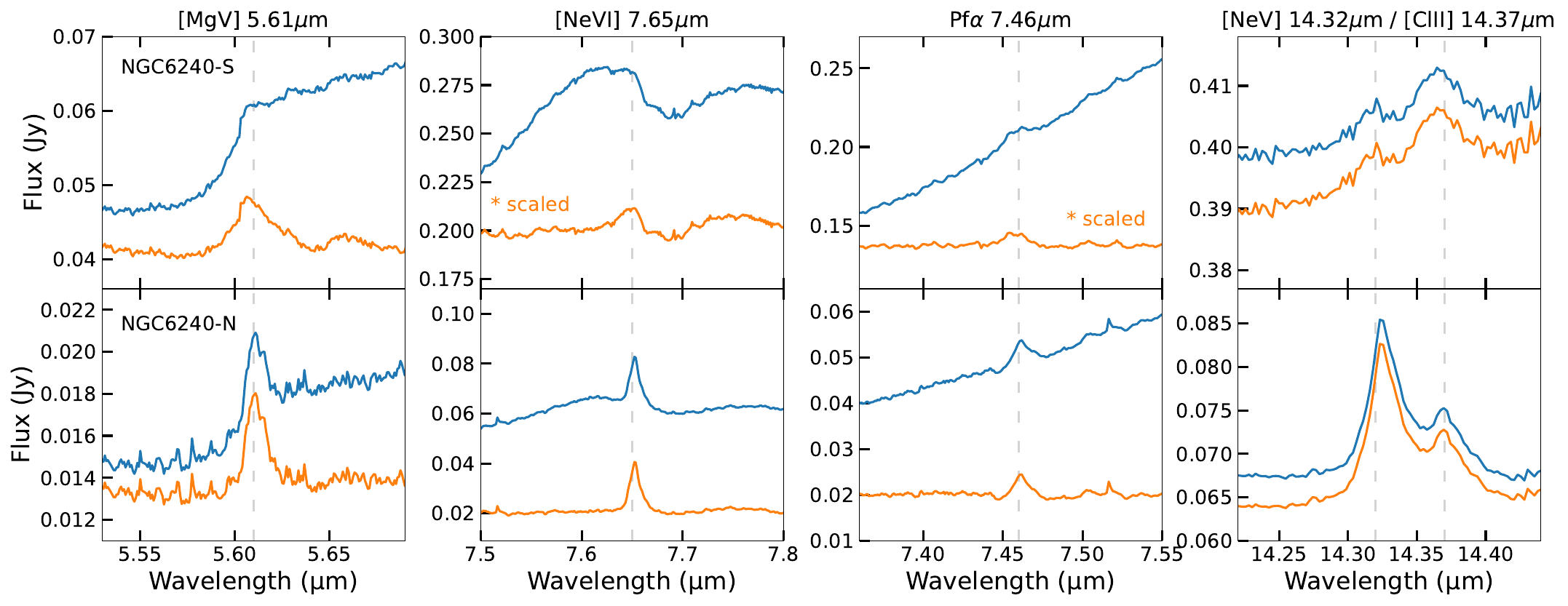} 
	\caption{Full integrated (r$\sim$0.7\arcsec) spectrum of the two nuclei in rest-frame after modelling the dust features and continuum (see details in Sect.~\ref{SubSect4:Disc_HighIonLines}). The upper figures show the global fit to the S (top) and N (bottom) nucleus before (blue) and after (orange) subtracting the total PAH emission (black). The modelled contribution of the individual PAH features are shown in grey. The bottom panels show some insets near the region of [Mg\,V], Pf$\alpha$, [Ne\,VI], and [Ne\,V] lines for both nuclei. The rest-frame wavelengths are indicated with a dashed, grey, vertical line. The Pf$\alpha$ and [Ne\,VI] lines from the S nucleus have been scaled for visualisation purposes. }
	\label{Fig:fitPAH}
\end{figure*}

We show in Fig.~\ref{Fig:fitPAH} the results for the PAH-subtracted spectra of both nuclei, highlighting the region around the [Mg\,V], [Ne\,VI], Pf$\alpha$, and [Ne\,V] emission lines. As already mentioned, the first three overlap with a PAH feature, whereas the [Ne\,V] is in a wavelength range less affected by the PAH features \citep{Chown2024}. The main result is that when subtracting the PAH model from the observed spectrum, the high excitation lines [Mg\,V] and [Ne\,VI], as well as Pf$\alpha$, are now detected in the S nucleus. For the particular case of [Ne\,VI], the line is still contaminated by the presence of the absorption due to the ice CH$_{4}$ band at $\sim$7.7\,$\mu$m. This demonstrates that in the S nucleus, these lines are indeed \textit{buried} by the strong continuum and dust features, similarly to what has been proposed for Arp\,220 \citep{Perna2024}, II\,Zw96 \citep{GarciaBernete2024}, or for Mrk\,231, where the weak X-ray nature of its AGN likely also contributes to the absence of high-excitation lines \citep{AH2024}. 
In Table~\ref{Table:1} we report the flux measurements for [Mg\,V] and [Ne\,VI] after the PAH subtraction for both nuclei. These lines are broader in the S nucleus than in the N nucleus ($\sigma_{\rm [Ne\,VI]}^{\rm S} = 357\pm36$\,km\,s$^{-1}$ vs $\sigma_{\rm [Ne\,VI]}^{\rm N} = 222\pm5$\,km\,s$^{-1}$, and $\sigma_{\rm [Mg\,V]}^{\rm S} = 526\pm17$\,km\,s$^{-1}$ vs $\sigma_{\rm [Mg\,V]}^{\rm N} = 279\pm9$\,km\,s$^{-1}$), similarly to what we found for the rest of the emission lines (see Table~\ref{Table:2_W80}). 

Even after PAH modelling and subtraction, the [Ne\,V] line for the S nucleus is still not detected with good S/N in the data ([Ne\,V] at 24.32$\mu$m is not detected), although it is clearly detected in the N nucleus. This line is blended with [Cl\,II] and both are broad, like other lines with similar IPs. This would dilute the line and hamper a good detection given the strong continuum in the S nucleus. Moreover, [Cl\,II] is stronger in the S nucleus than in the N nucleus with respect to [Ne\,V], which suggests that the relative contribution of the star formation is larger in the S nucleus. The strong contribution from the continuum is likely the cause for the non-detection for other lines, such as the [Ar\,V] at 7.9$\mu$m and at 13.1$\mu$m (IP of 59.6\,eV), that is detected in the region of the bubble structure seen in the spatially-resolved maps (see Figs.~\ref{Fig:KinMapsHighIon} and~\ref{Fig:SummaryRegions}), but not in the nuclear apertures (see Fig.~\ref{Fig1:IntSpec}). We will further discuss about the detection of the high excitation lines in Sect.~\ref{SubSect4:Disc_HighIonLines}.

\section{Discussion}
\label{Sect4:Discussion}

NGC\,6240 is an interacting system in a merging phase \citep{vanderWerf1993,Engel2010,Fyhrie2021}. From the results presented in Sect.~\ref{Sect3:Results} and from previous works (see Sect.~\ref{Sect1:Introduction}), there are several processes occurring simultaneously in this galaxy, affecting the gas kinematics and distribution. The stellar disks of the two galaxies were clearly detected using near-IR observations \citep[][see also Ceci et al. submitted]{Medling2014}. However, based on our results as well as in optical observations \citep{MullerSanchez2018}, the ionised gas is decoupled from the stellar emission. The velocity maps do not show a regular rotating pattern anywhere along the FoV, although there is a clear shift between the velocities from both nuclei (see Sect.~\ref{SubSect3:Results_kin}), previously detected also with the stellar component \citep{Kollatschny2020}. The existence of multiple intensity peaks and kinematic components in the emission lines points towards a highly perturbed and shocked gas as a result of the interaction between the initial galaxies and the triggering of both the AGN and the starburst activity. All these perturbations are likely producing the apparent lack of correlation detected between the morphologies and kinematic properties of the emission lines (see Figs.~\ref{Fig:KinMapsLowExcit},~\ref{Fig:KinMapsHighIon}, and~\ref{FigAp:KinMaps_extralines}).

In Sect.~\ref{SubSect4:Disc_LineRatios} we discuss the origin of the ionised gas based on our mid-IR line ratios. We explore in Sect.~\ref{SubSect4:Disc_HighIonLines} about the presence and detection of the high excitation lines for this source and other AGN observed with the JWST. Finally, we give evidence about the presence of outflows in the system and their properties as seen in the mid-IR in Sect.~\ref{SubSect4:Disc_Outflow}.

\subsection{Origin of the ionised gas}
\label{SubSect4:Disc_LineRatios}

From optical line ratios, the N nucleus was previously associated with LINER-like photoionisation, and the S nucleus with Seyfert-like photoionisation \citep[e.g.][]{MullerSanchez2018}. However, other works in the literature have detected the presence of multiple shocked regions throughout the system \citep[see e.g.][]{Fosbury1979,vanderWerf1993,Nardini2013,Medling2021}. This would explain the LINER-like ratios around the N nucleus, given the degeneration in optical diagnostic diagrams between shock models and LINERs \citep{Heckman1980,Baldwin1981,Dopita1995}. In general, the [Ne\,III]/[Ne\,II] ratios (see Fig.~\ref{Fig:RatioLines}) are consistent with observational measurements derived with Spitzer for galaxies classified as H\,II/LINER \citep{Pereira2010}. They are also consistent with the presence of shocks except for the region north-west of the N nucleus (bubble in Fig.~\ref{Fig:SummaryRegions}), that has similar ratios to those observed in Seyfert galaxies \citep[see][]{Pereira2010}. Moreover, the detection of high excitation lines in the N nucleus, such as [Mg\,V], [Ne\,VI], and [Fe\,VIII] provides evidence that it is indeed an AGN. 

\subsubsection{Nuclear regions}
\label{Subsubsect4:Disc_LineRat_Nucleus}

\begin{figure}
	\centering
	\includegraphics[width=\columnwidth]{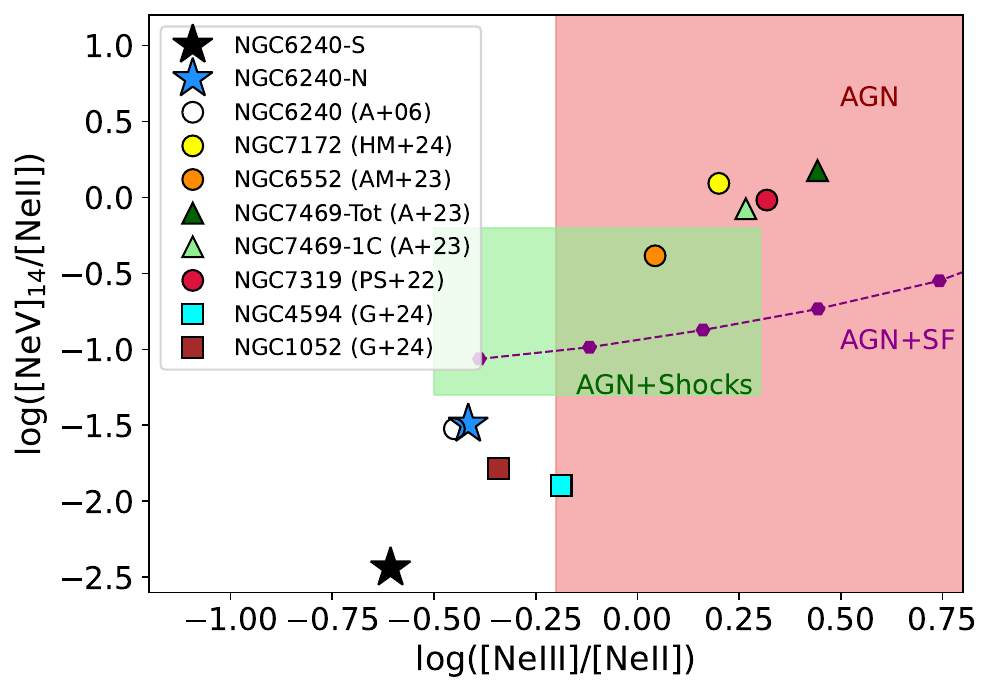} 
	\caption{Mid-IR diagnostic diagram representing log([Ne\,V]/[Ne\,II]) vs log([Ne\,III]/[Ne\,II]) to divide between AGN (red area), shocks (green area) and SF (purple line) ionisation based on Fig.~5 in \cite{Feltre2023}. The shaded areas schematically represent part of the grid of models in \cite{Feltre2023}, and are used here just as a reference. We represent with a blue and a black star the position of the N and S nucleus, respectively (see Sect.~\ref{SubSect3:LineRatios}). For comparison, we show the integrated value derived by \cite{Armus2006} for NGC\,6240 as a white circle, and the nuclear values obtained with similar JWST data for other local type-2 Seyfert galaxies: yellow circle for NGC\,7172 \citep{HM2024b}, orange circle for NGC\,6552 \citep{Alvarez2023}, dark (light) green triangles for the total (primary component) flux of the line for the type-1.5 Seyfert NGC\,7469 \citep{Armus2023}, and red circle for NGC\,7319 \citep{PereiraSantaella2022}. Additionally, we show the ratios for two LINERs, namely NGC\,4594 and NGC\,1052, as a blue and brown squares, respectively \citep{Goold2024}.}
	\label{Fig:Feltre23}
\end{figure}

The majority of the central region of NGC\,6240 is consistent with being dominated by shocks and by SF ionisation, as previously suggested. In Fig.~\ref{Fig:Feltre23} we show a simplified version of the diagnostic diagrams from Fig.~5 in \cite{Feltre2023}, comparing the ratios of different neon transitions. We found that the N nucleus has log([Ne\,III]/[Ne\,II])\,$= -0.42\pm0.01$ and log([Ne\,V]/[Ne\,II])\,$= -1.49\pm0.02$ (measured with r\,$\sim 0.7\arcsec$, see Table~\ref{Table:1}), which are consistent with AGN+shock ionisation according to the diagram. For the S nucleus, on the other hand, we measured a ratio log([Ne\,III]/[Ne\,II])\,$= -0.61\pm0.04$ and log([Ne\,V]/[Ne\,II])\, $=-2.44\pm0.08$. If we compare with other models in \cite{Feltre2023} \citep[see also][]{Allen2008,Sutherland2017,Alaire2019}, the shocked regions reach the lower-left part of the diagram, indicating that the ratios for both nuclei are consistent with shocks. With Spitzer/IRS data, \cite{Armus2006} also derived the line ratios encompassing both nuclei, finding similar values to our determination for the N nucleus (log([Ne\,III]/[Ne\,II])\,$\sim -0.45$ and log([Ne\,V]/[Ne\,II])\,$\sim -1.52$, uncorrected for extinction, see Fig.~\ref{Fig:Feltre23}). This is expected, as the contribution from the [Ne\,V] emitted in the N nucleus is larger than for the S nucleus (see Table~\ref{Table:1} and Fig.~\ref{Fig1:IntSpec}). They stated that these ratios are consistent with being excited by the presence of a nuclear starburst. 

Additionally, for comparison we show in Fig.~\ref{Fig:Feltre23} the nuclear regions of other local Seyfert galaxies (namely, type-2s NGC\,7319, NGC\,6552, and NGC\,7172, and the type-1.5 NGC\,7469) and LINERs (namely, NGC\,1052 and NGC\,4594) observed with JWST MIRI/MRS data \citep{PereiraSantaella2022,Alvarez2023,Armus2023,HM2024b,Goold2024}. These Seyfert galaxies present ratios consistent with AGN photoionisation, that are very different from those derived with NGC\,6240. Contrarily, both nuclei lie in a similar region of the diagram to the two LINERs (particularly the N nucleus), indicating that SF and shocks have a dominant role in the ionisation of the nuclear region emission. This is despite the detection of high excitation lines associated to AGN photoionisation. This implies that, as previously proposed \citep{Tecza2000,Armus2006,Engel2010}, the AGN are not the main source of ionisation for the ISM in the mid-IR.

If instead we estimate log([Ne\,III]/[Ne\,II]) for the two nuclei of NGC\,6240 by separating the line profiles into different Gaussian components (see Appendix~\ref{Appendix1} and Fig.~\ref{Fig:Appendix_FitIntLinesParam}), we get a similar ratio for both nuclei and both the narrowest, primary component and the broadest, secondary component (N nucleus $-0.48\pm0.03$ vs $-0.39\pm0.02$ in log, respectively; S nucleus $-0.66\pm0.02$ and $-0.57\pm0.02$ in log, respectively). Nevertheless, in both cases the ratios fall in the same region in the mid-IR diagrams by \cite{Feltre2023} as for the ratios obtained with the integrated fluxes (see Fig.~\ref{Fig:Feltre23}). 

\subsubsection{SF emission}
\label{Subsubsect4:Disc_LineRat_SFR}

Various works from the literature detected significant enhanced emission in between both nuclei (the bridge region, see Fig.~\ref{Fig:SummaryRegions}), associating it to a nuclear starburst \citep[see e.g.][]{vanderWerf1993,Lutz2003,MullerSanchez2018}, while ground-based spectroscopy detected the brightest 11.3$\mu$m PAH emission around the S nucleus \citep{AH2014}. From the maps derived for the PAH feature at 6.2\,$\mu$m (see Fig.~\ref{Fig:PAHmap} and Sect.~\ref{SubSect3:Results_fluxes}), we detect the maximum of the emission at the S nucleus, $\sim 6$ times more intense than in the N nucleus and $\sim 9$ times more than the bridge region. As mentioned in Sect.~\ref{SubSect3:Results_fluxes}, the same applies for emission lines such as [Ne\,II] and other low excitation lines (see Figs.~\ref{Fig:KinMapsLowExcit} and~\ref{FigAp:KinMaps_extralines}), which are mostly ionised by SF processes. In fact, from the [Fe\,II]/[Ar\,II] map (bottom panel in Fig.~\ref{Fig:RatioLines}) the nuclei show the lowest ratios, indicating that SF is stronger there.
Finally, from the [Ne\,III]/[Ne\,II] spatially-resolved line ratios in Fig.~\ref{Fig:RatioLines}, we detected the lowest ratios for the region around the S nucleus, which may also indicate that the relative contribution from the SF processes is larger than in other regions, such as the bridge between both nuclei.
To quantify this, we used the relations from Eq.~12 by \cite{Shipley2016}, that relates the star formation rate (SFR) to the luminosity of the different PAH features. Using the fluxes measured in the integrated spectra with the tool by \cite{Donnan2024} (see Sect.~\ref{Subsect3:HighIonLines}), we found that the SFR of the S nucleus is $\sim 4.1$ times larger than that of the N nucleus. 
Thus, although there is possibly SF occurring in the bridge and in the N nucleus, the peak of the nuclear starburst is detected in the S nucleus.

\subsubsection{Shocked emission}
\label{Subsubsect4:Disc_LineRat_Shocks}

Another way to evaluate the presence of shocks is by means of the [Fe\,II] lines, that are a tracer of both shocks and star-forming processes \citep[see e.g.][]{Allen2008}. In Fig.~\ref{Fig:KinMapsLowExcit} we show the kinematic maps for the [Fe\,II] at 5.34$\mu$m, as it is the iron line detected with the best S/N (see spectra in Fig.~\ref{Fig1:IntSpec}). We note, however, that it covers the smallest FoV as it is located in ch1-short (see Sect.~\ref{Sect2:Data}). There is [Fe\,II] emission arising in the nuclei as well as the bridge region, with similar velocities and velocity dispersion to other low excitation lines, with $\sigma$ particularly enhanced west from the S nucleus. Given the complex profiles detected in that region for both ionised and the warm molecular gas (Hermosa Mu{\~n}oz et al. in prep.), it is likely that shocks are simultaneously affecting all the gas phases. This is consistent with the picture drawn from the near-infrared line ratios, particularly that of H$_{2}$ 1$-$0 S(1) over Pa$\beta$ \citep{Medling2021}. This ratio is high when the molecular gas is excited by shocks, and in fact it shows large values all over the system probably due to the interaction of the galaxies \citep[see also][]{vanderWerf1993}. It is especially enhanced in the region in between nuclei, west of the S nucleus, and along the direction of the detected [O\,III] outflow \citep[see][]{MullerSanchez2018}. This enhanced, bridge region (see Fig.~\ref{Fig:SummaryRegions}) was proposed to be shocked gas as a consequence of the interaction between the galaxies \citep[see e.g.][]{Tecza2000, Max2005}. From the [Fe\,II]/[Ar\,II] ratio (bottom panel in Fig.~\ref{Fig:RatioLines}), the largest values are also in the bridge region, which supports this hypothesis. 

Based on the optical line ratios reported by \cite{Medling2021}, despite the significant dust obscuration (with H$\alpha$/H$\beta \sim 20$, larger in the region between the two nuclei), both nuclear regions fall within the AGN regime. Most of the other regions are consistent with either AGN+shocks or LINER+shocks, which aligns with the mid-IR ratios observed in this work.
Moreover, as mentioned in Sect.~\ref{SubSect3:Results_fluxes}, there is a spatial coincidence between the features traced with the soft X-rays and the ionised gas \citep[see also][]{Nardini2013,Yoshida2016,Paggi2022}. The C2 region partly coincides with the largest "outflow ridge" region defined in Fig.~5 in \cite{Paggi2022}, that was consistent with shock excitation. The bubble region, which forms part of the largest H$\alpha$ bubble detected in the optical \citep{MullerSanchez2018} and the X-ray loop \citep{Paggi2022}, is however associated with a starburst-driven outflow.

\subsection{Detection of high excitation lines in deeply embedded nuclei with mid-IR data}
\label{SubSect4:Disc_HighIonLines}

As mentioned in Sect.~\ref{Sect1:Introduction}, the S nucleus has approximately three times larger bolometric luminosity than the N nucleus \citep{MullerSanchez2018}, and is brighter by a factor $\sim$2.6 in both soft and hard X-rays (i.e 10-40\,keV luminosity of 7.1$\times 10^{43}$\,erg\,s$^{-1}$ and 2.7$\times 10^{43}$\,erg\,s$^{-1}$, for the S and N nucleus respectively, \citealt{Puccetti2016}). The high X-ray luminosities classify both nuclei as AGN \citep{Puccetti2016,Paggi2022}. As seen in Sect.~\ref{Subsect3:HighIonLines}, the mid-IR continuum emission is brighter for the S nucleus, and it is impacting on the detection of high excitation lines in our data. However, although we have detected [Ne\,VI] and [Mg\,V] for both nuclei, the [Ne\,V] line is still seen with low S/N for the S nucleus. 

One possibility proposed to explain the lack of high excitation lines in some U/LIRGs is the presence of high extinction. \cite{Yamada2024} proposed that U/LIRGs with high column densities (N$_{\rm H}$ from $10^{22}$ to $10^{24}$\,cm$^{-1}$) may not show [O\,IV] and [Ne\,V] emission. For the particular case of NGC\,6240, \cite{Komossa2003} derived a column density of 1-2$\times 10^{24}$\,cm$^{-2}$ with Chandra data for both nuclei, which suggest that they are highly absorbed, although there are other determinations ($\sim 10^{22}$\,cm$^{-2}$ in \citealt{Paggi2022}).  
From the H$_{2}$ molecular lines we derived the column densities for both nuclei and found values below $10^{22}$\,cm$^{-2}$ (Hermosa Mu{\~n}oz et al. in prep.). Although indeed the S nucleus shows larger extinction than the N nucleus. We estimated $\tau_{9.8}\sim$2 vs 1.8, respectively, for the H$_{2}$ emission using the tool by \citealt{Donnan2024}, see Sect.~\ref{Subsect3:HighIonLines}), we do not detect a significant difference between both nuclei. 

Another plausible explanation is that the stronger dust emission, in the form of continuum and PAH features, in the S nucleus hinders the detection of the lines. From our results presented in Sect.~\ref{Subsect3:HighIonLines}, this seems to be likely the case for NGC\,6240. The PAH features hamper the measurements of the emission lines for both nuclei. Even in spectral regions with no PAH emission (e.g. close to the [Ne\,V] lines), the continuum is strong. From the [Ne\,V] and [Cl\,II] complex (see lower right panels in Fig.~\ref{Fig:fitPAH}), the relative contribution of [Cl\,II] with respect to the [Ne\,V] changes between nuclei (see Sect.~\ref{Subsect3:HighIonLines}). Particularly, for the S nucleus the brightest [Cl\,II] indicates that SF is stronger than for the N nucleus, consistently with the derived SFRs in Sect.~\ref{Subsubsect4:Disc_LineRat_SFR}. Additionally, the lines for the S nucleus are broader than for the N nucleus, which contributes to dilute the [Ne\,V] line if the [Cl\,II] is also broad and brighter. Thus for the S nucleus, the low contrast of the high excitation emission lines is likely a combination of a strong mid-IR continuum, very broad lines, and intense SF in the nucleus.

In other local U/LIRGs even with JWST observations, high excitation lines remain undetected, for instance, in Mrk\,231 \citep{AH2024}, Arp\,220 with NIRSpec \citep{Perna2024} and MRS (Rieke et al. accepted, Goldberg et al. submitted, van der Werf et al. in prep), and II\,Zw96 \citep{GarciaBernete2024}. For Mrk\,231, the non-detection of high excitation lines is likely due to both the intrinsically faint X-ray nature of the source and the strong mid-IR continuum \citep{AH2024}, while for Arp\,220 the main reason is probably the high levels of obscuration (Goldberg et al. submitted) or that this source is not an AGN (van der Werf et al. in prep., Buiten et al. in prep.). Thus the detection of high excitation lines in deeply embedded AGN depends on their exact nature and intrinsic properties.

\subsection{AGN-driven outflow detection and properties}
\label{SubSect4:Disc_Outflow}

\begin{table*}
	\caption{Outflow parameters derived for NGC\,6240.}
	\label{Table:OutParam}
	\centering          
	\begin{tabular}{lccccccc}
		\hline\hline
		Region & v$_{\rm OF,max}$ & $\sigma_{\rm OF}$ & R$_{\rm OF}$ & M$_{\rm OF,ion}$ & $\dot{M}_{\rm OF,ion}$ & E$_{\rm OF,ion}$ & $\dot{E}_{\rm OF,ion}$ \\ 
		& (km s$^{-1}$) & (km s$^{-1}$) & (pc) & (M$_{\sun}$) & (M$_{\sun}$\,yr$^{-1}$) & (erg) & (erg\,s$^{-1}$) \\ 
		(1) & (2) & (3) & (4) & (5) & (6) & (7) & (8) \\
		\hline           
		Bubble    & $674\pm96$ & 216$\pm$38 & 742  & $5.05\pm0.92\times 10^{5}$  & $1.41\pm0.32$  & $2.3\pm0.9\times10^{53}$ & $2.6\pm0.9\times10^{41}$ \\ 
		S nucleus & $625\pm58$ & 261$\pm$20 & 368  & $3.55\pm0.07\times 10^{5}$  & $1.85\pm0.17$  & $2.4\pm0.4\times10^{53}$ & $3.5\pm0.7\times10^{41}$ \\ 
		\hline        
	\end{tabular}\\
	\tablefoot{Columns indicate: (1) name of the region, (2) maximum velocity of the outflow, (3) average velocity dispersion, (4) distance to the maximum velocity of the outflow, (5) mass of the outflow, (6) mass outflow rate, (7) kinetic energy, and (8) kinetic power of the ionised gas outflow. See Sect.~\ref{SubSect4:Disc_Outflow}) for more details on the calculations.}
\end{table*}

The emission detected with a bubble-like shape towards the north-west part of the galaxy is most likely an expanding outflow. In fact, it is identified as an ionisation cone with the NIRSpec data (Ceci et al. submitted). It is coincident with the base of the H$\alpha$ bubble to the west and the [O\,III] cone to the east detected on much larger scales \citep{MullerSanchez2018}. The H$\alpha$ bubble was associated with the most recent starburst episode in the system, given its kinetic energy and the derived time scale of the event \citep{MullerSanchez2018}. However, the kinetic power derived for the [O\,III] outflow was insufficient to be solely attributed to the starburst, so \cite{MullerSanchez2018} concluded that it has to be AGN-driven. They estimated a mass outflow rate, $\dot{M}_{\rm out}$, of 75\,M$_{\sun}$\,yr$^{-1}$ for the [O\,III] cone, and of 10\,M$_{\sun}$\,yr$^{-1}$ for the H$\alpha$ bubble, though they state that these are probably lower limits to the total mass.

We detected an expanding bubble clearly traced by high excitation lines such as [Ne\,V], [Ne\,VI], and [Mg\,V] (see Fig.~\ref{Fig:KinMapsHighIon}), as well as intermediate excitation lines, such as [Ar\,III] and [S\,IV] (see Fig.~\ref{FigAp:KinMaps_extralines}), and also Pf$\alpha$. This feature, extending up to $\sim 5.2\arcsec$ (i.e. projected distance of 2.74\,kpc) from the N nucleus (see Fig.~\ref{Fig:KinMapsHighIon}), appears to be a gas flow expanding at receding velocities, likely associated with the large-scale H$\alpha$ bubble.
However the complicated kinematics, with broad line widths and wings detected in both nuclei (see Sect.~\ref{SubSect3:Results_kin}) suggest the presence of additional non rotational motions. In particular, the S nucleus exhibits a line component of $\sigma \geq 600$\,km\,s$^{-1}$ (see Fig.~\ref{Fig:Appendix_FitIntLinesParam}), that could only be explained by the existence of strong gas motions, likely driven by complex interactions within the system, and associated with the large scale outflows. 
Due to the contamination from the SF emission, we cannot spatially resolve the outflow located in and/or around the S nucleus using the intermediate or low excitation emission lines. Thus, here we focus on the integrated properties of the southern nuclear outflow, while the N nucleus is already included within the bubble region.

To estimate the properties of the ionised gas outflows, H$\alpha$ (or H$\beta$) and [O\,III] are typically used. Although the NW bubble is better traced by the high excitation lines, and thus we could use the [Ne\,V] line to estimate the outflow properties \citep[see e.g.][]{Zhang2024}, we use the Pf$\alpha$ line, which is detected in both nuclei, for consistency in the measurements. This hydrogen recombination line is detected in the NW bubble and in both nuclei with sufficient S/N ($>3$, see Sect.~\ref{Sect2:Methodology}). Moreover, it traces a similar emission to other hydrogen transitions (e.g. Br$\gamma$ line in \citealt{MullerSanchez2018}), allowing us to avoid additional assumptions on the element abundances when estimating the total ionised mass of the outflow, unlike methods involving [O\,III] \citep[see e.g.][]{Carniani2015,Baron2019}.

We follow the method described by \cite{Davies2020} and \cite{Baron2019} to estimate the total ionised gas outflowing mass through the equation: 
\begin{equation}
	M_{\rm ion} = \frac{\mu \times m_{\rm H} \times L_{\rm line}}{\gamma_{\rm line} \times n_{e}},
\end{equation}
where $m_{\rm H}$ is the hydrogen mass, $\mu$ is the mass per hydrogen atom, assumed to be 1.4 in \cite{Baron2019}, $\gamma_{\rm line}$ is the effective line emissivity, L$_{\rm line}$ is the luminosity of the selected line, and n$_{\rm e}$ is the electron density of the gas. 

For estimating the electron density, we followed \cite{HM2024b} and used the module \textsc{pyneb} in \textsc{python} \citep{Luridiana2015}. We measured the fluxes of the two [Ne\,V] lines in the integrated spectra (R\,$\sim 0.7$\arcsec) of the N nucleus, obtaining a ratio f$_{\rm [Ne\,V]14}$/f$_{\rm [Ne\,V]24} \sim$ 1.3 (see Table~\ref{Table:1}), which gives a density of $\sim 1800$\,cm$^{-3}$. This value is similar to those found in other local Seyfert galaxies \citep[see e.g.][]{Pereira2010,Alvarez2023,HM2024b,Zhang2024}. The electron density assumed by \cite{MullerSanchez2018} was 50\,cm$^{-3}$, that is approximately two orders of magnitude smaller than our determination. 
We estimated the luminosity of the Pf$\alpha$ line for both nuclei from the fluxes measured in the PAH subtracted, integrated spectra (r$\sim$0.7\arcsec, see Sect.~\ref{SubSect4:Disc_HighIonLines} and Fig.~\ref{Fig:fitPAH}) after subtracting a local continuum. As mentioned in Sect.~\ref{SubSect4:Disc_HighIonLines}, the two nuclei are still obscured in the mid-IR, with a derived extinction $\tau_{9.8} \sim 2$. However, we did not correct the fluxes for Pf$\alpha$ and the [Ne\,V] lines from extinction, as they fall in a minimum in the mid-IR extinction curves \citep[see Fig.~4 in][]{HernanCaballero2020}. The same applies for the [Ne\,V] emission lines when used to estimate the electron density \citep[see also][]{Chiar2006,HM2024b}. 
As for $\gamma_{\rm line}$, in \cite{Baron2019} \citep[see also][]{Davies2020} there is a detailed explanation on how to estimate this parameter. Based on the optical line ratios presented in \cite{Medling2021}, for the nuclear region (i.e. the bridge) they estimated a log([O\,III]/H$\beta$) of 0 and log([N\,II]/H$\alpha$) of 0.1 that, following Eq.~2 in \cite{Baron2019}, leads to an ionisation parameter log$U \sim -3.8$. Using Table~5 in \cite{Davies2020}, this corresponds to a $\gamma_{\rm H\alpha} = 2.26\times 10^{-25}$\,cm$^{3}$\,erg\,s$^{-1}$. Following \cite{Draine2011}, assuming a temperature of 10$^4$\,K, j$_{\rm Pf\alpha}$/j$_{H\alpha} \sim 0.009$, thus applying this factor, $\gamma_{\rm Pf\alpha} \sim 2 \times 10^{-27}$\,cm$^{3}$\,erg\,s$^{-1}$. We cross-checked this value by using \textsc{pyneb}, with the corresponding density and temperature, which leads to $\sim 3.1 \times 10^{-27}$\,cm$^{3}$\,erg\,s$^{-1}$, approximately the same value. 

All the parameters estimated for the outflows are summarised in Table~\ref{Table:OutParam}. The derived mass of the ionised gas outflow for the S integrated nucleus is 3.55\,$\pm$\,0.07\,$\times 10^{5}$\,M$_{\sun}$, and the bubble-like structure is 5.1\,$\pm$\,0.9\,$\times 10^{5}$\,M$_{\sun}$. 
From this mass we can estimate the mass outflow rate, $\dot{M}_{\rm ion}$. We assume a bi-conical morphology given that there is a secondary emission seen with the [O\,IV] line (see Fig.~\ref{Fig:KinMapsHighIon}). We follow \cite{Cresci2015} and \cite{Fiore2017}, so that the mass outflow rate is:
\begin{equation}
	\dot{M}_{\rm ion} = 3 \times M_{\rm ion} \times v_{\rm max,OF} / R_{\rm OF}.
\end{equation}

For obtaining this quantity we used the maximum velocity of the outflow, $v_{\rm max,OF}$, and the radius at which this velocity is found, $R_{\rm OF}$. For the S nucleus we take as the radius r$\sim$0.7\arcsec\,(i.e. 368\,pc), and the maximum velocity is estimated as the $|$v$_{\rm line}| + 2\times \sigma_{\rm line}$ \citep[following][]{Rupke2013,Fiore2017}, where $\sigma_{\rm line}$ is the average velocity dispersion of Pf$\alpha$. This average $\sigma$ is similar to that found for the high excitation lines in the N nucleus. We obtained $v_{\rm max,OF} = 625\pm58$\,km\,s$^{-1}$ for the S nucleus. As for the bubble, we find the maximum velocity, $v_{\rm max,OF} = 674\pm96$\,km\,s$^{-1}$, at a distance $R_{\rm OF} \sim 741$\,pc NW of the N nucleus. We derived a mass outflow rate for S nucleus 1.85\,$\pm$\,0.17\,M$_{\sun}$\,yr$^{-1}$, and for the bubble 1.41\,$\pm$\,0.32\,M$_{\sun}$\,yr$^{-1}$. These values are two orders of magnitude lower than the $\dot{M}_{\rm ion}$ derived by \cite{MullerSanchez2018}. The main reason could be the difference in the electron density (50\,cm$^{-3}$ in the optical vs 1800\,cm$^{-3}$ with MIRI), that directly lowers our determination of $M_{\rm ion}$, and consequently that of $\dot{M}_{\rm ion}$. This highlights the importance of n$_{\rm e}$ for estimating the outflow parameters, as was already pointed out by \cite{Baron2019} \citep[see also][]{Davies2020}. These mass outflow rates fall within the \cite{Fiore2017} relation and are consistent with those derived for other Seyfert galaxies in the local Universe. 

With the mass and the mass outflow rate we can estimate the kinetic energy and the kinetic power of the outflow following the expressions described in \cite{Venturi2021} \citep[see also][]{Rose2018,Santoro2020}:
\begin{equation}
	E_{\rm OF} = 0.5 \times M_{\rm OF} \times \sigma^{2}_{\rm OF}.
\end{equation}
\begin{equation}
	\dot{E}_{\rm OF} = 0.5 \times \dot{M}_{\rm OF} \times (3\sigma^{2}_{\rm OF} + v^{2}_{\rm max,OF} ).
\end{equation}
We obtained a similar energy for both the S nucleus and the bubble ($\sim 2\times10^{53}$\,erg, see Table~\ref{Table:OutParam}) and kinetic power ($\dot{E}_{\rm OF} \sim 3\times 10^{41}$\,erg\,s$^{-1}$, see Table~\ref{Table:OutParam}). These values are in contrast to the limits set in \cite{MullerSanchez2018} for the kinetic power of the H$\alpha$ bubble, between 10$^{42}$ and 10$^{44}$\,erg\,s$^{-1}$, and about 15 times larger for the [O\,III] bubble. We note that the $\sigma$ derived with the Pf$\alpha$ line for the bubble (or for [Ne\,V] or [Ne\,VI]) are approximately half the value used for [O\,III] (average $\sigma \sim$505\,km\,s$^{-1}$, \citealt{MullerSanchez2018}). 

Given the AGN bolometric luminosity of $\sim 10^{45}$\,erg\,s$^{-1}$ \citep{Puccetti2016} and the total injection of energy due the star formation of $\sim 7 \times 10^{43}$\,erg\,s$^{-1}$, assuming a SFR of 100\,M$_{\sun}$\,yr$^{-1}$ \citep{MullerSanchez2018}, both the AGN and the SF could trigger the outflows. An additional complication is that both the SF peak and the more luminous AGN in the system are located in the S nucleus (see Sect.~\ref{Subsubsect4:Disc_LineRat_SFR}), which is the brightest AGN. Given that the bubble detected in our MRS observations (see Figs.~\ref{Fig:KinMapsHighIon} and~\ref{Fig:SummaryRegions}) is co-spatial with the base of the starburst-driven H$\alpha$ and soft X-rays outflow \citep{MullerSanchez2018,Paggi2022}, we cannot rule out that it is SF-driven, despite the presence of high excitation lines (see Sect.~\ref{SubSect4:Disc_LineRatios}).

\section{Summary and conclusions}
\label{Sect5:Conclusions}

In this work we presented new mid-IR data for NGC\,6240 covering up to 6.6\arcsec$\times$7.7\arcsec\,(i.e. projected distance of 3.5\,kpc\,$\times 4.1$\,kpc) obtained with MIRI/MRS on board of JWST as part of the GTO program termed MICONIC. We resolved for the first time the full 5-28$\mu$m spectra for the two X-ray-detected nuclei, which are separated by $\sim1.6\arcsec$ (i.e. projected distance of $\sim$840\,pc). The spectra are very rich, with a total 20 different ionised emission lines detected (IP from 7.6 to 187\,eV), two hydrogen recombination lines, and 10 warm molecular emission lines (Hermosa Mu\~noz et al. in prep.). Additionally, there are strong PAH and absorption features, whose in-depth analysis is beyond the scope of this paper. 
We summarise the main results in the following points: 

\begin{itemize}
	\item The AGN nature of the two nuclei: The detection of spatially-resolved high excitation lines (namely [Ne\,V], [Ne\,VI], [Mg\,V], and [Fe\,VIII]) in the N nucleus can only be produced by AGN ionisation. The S nucleus presents a brighter continuum and PAH emission than the N nucleus (factor of $\sim$3 and $\sim$4, respectively). After carefully modelling and subtracting the PAH features, we detected [Mg\,V] and [Ne\,VI] in the integrated spectrum of the S nucleus (see Fig.~\ref{Fig:fitPAH}). All of this together with the large widths and velocities of the ionised emission lines, indicate that the S nucleus contains an AGN. 
	
	\item Extended SF activity: The flux maps of the low excitation lines and the PAH features (see Figs.~\ref{Fig:KinMapsLowExcit},~\ref{Fig:PAHmap}, and~\ref{FigAp:KinMaps_extralines}) show multiple clumps and extended emission, as well as a \textit{bridge} between both nuclei (see Fig.~\ref{Fig:SummaryRegions}). We find that there is SF occurring everywhere over the 3.5\,kpc\,$\times$\,4.1\,kpc mapped region, but the brightest starburst is likely located at and around the S nucleus.
	
	\item Perturbed and complex kinematics: The emission lines have complex profiles (see Fig.~\ref{Fig:Appendix_FitIntLinesParam}) that are a signature of perturbed velocity and $\sigma$ produced by multiple physical effects. Interestingly, the S nucleus has the broadest profiles (in general FWHM\,$> 1500$\,km\,s$^{-1}$, see Figs.~\ref{Fig:LineProfiles} and~\ref{Fig:Appendix_FitIntLinesParam}) with red wings and a shift in velocity for the bulk of the line with respect to the systemic value (z\,$\sim$\,0.02448). The N nucleus has more symmetrical, less broad profiles (on average FWHM\,$\sim$\,700\,km\,s$^{-1}$, see Fig.~\ref{Fig:LineProfiles}), although equally complex with wings towards both the red and blue. 
	
	\item Dominance of shocked regions: From the spatially-resolved maps, there is a clear "V"-like shape in the velocity dispersion (see Figs.~\ref{Fig:KinMapsLowExcit},~\ref{Fig:SummaryRegions}, and~\ref{FigAp:KinMaps_extralines}) reaching values up to $\sim$600\,km\,s$^{-1}$. These high values and the detection of multiple peaks and wings in the lines, both ionised and molecular, are indicative of the presence of shocks throughout the MIRI/MRS FoV. Moreover, the derived mid-IR line ratios are mostly consistent with observations of H\,II and LINER galaxies, and shock models. This is similar to the optical and near infrared, where the most shocked regions are found towards the SW part of our FoV and in between both nuclei, probably due to the interaction of the galaxies. 
	
	\item Ionised gas outflows properties: The bubble extending NW from the N nucleus reaching up to a projected distance of 2.74\,kpc (see Fig.~\ref{Fig:KinMapsHighIon}) coincides with the base of the H$\alpha$-detected outflow. The line ratios within this bubble are consistent with AGN ionisation, and it is located at the base of the large scale H$\alpha$ outflow. Using the Pf$\alpha$ line, we estimate for the bubble a maximum velocity of $\sim$675\,km\,s$^{-1}$ at a distance of $\sim$741\,pc, resulting in an outflow mass of $\sim 3.6\pm0.1\times 10^{5}$\,M$_{\sun}$ and a mass outflow rate of 1.41$\pm$0.31\,$\dot{M}_{\sun}$\,yr$^{-1}$ (see Table~\ref{Table:OutParam}). The large non-rotational motions found in the S nucleus are likely associated with another large-scale outflow. We used the integrated properties for this nucleus to obtain an outflow mass of $3.55\pm0.07\times 10^{5}$\,M$_{\sun}$ and a $\dot{M}_{ion}$ of $1.85\pm 0.17$\,M$_{\sun}$\,yr$^{-1}$ (see Table~\ref{Table:OutParam}). Given the derived kinetic power for both (see Sect.~\ref{SubSect4:Disc_Outflow} and Table~\ref{Table:OutParam}), both the AGN and the starburst could have driven those outflows. 
	
\end{itemize}

\begin{acknowledgements}
	We thank the referee for his/her suggestions that have helped to improve the paper. LHM and AAH acknowledge financial support by the grant PID2021-124665NB-I00 funded by the Spanish Ministry of Science and Innovation and the State Agency of Research MCIN/AEI/10.13039/501100011033 PID2021-124665NB-I00 and ERDF A way of making Europe. LP and MB acknowledge funding from the Belgian Science Policy Office (BELSPO) through the PRODEX project “JWST/MIRI Science exploitation” (C4000142239). LC acknowledges support by grant PIB2021-127718NB-100 from the Spanish Ministry of Science and Innovation/State Agency of Research MCIN/AEI/10.13039/50110001103. IGB is supported by the Programa de Atracci{\'o}n de Talento Investigador "C{\'e}sar Nombela" via grant 2023-T1/TEC-29030 funded by the Community of Madrid. GÖ acknowledges support from the Swedish National Space Agency (SNSA). POL acknowledges financial support by CNES.
	This work is based on observations made with the NASA/ESA/CSA James Webb Space Telescope. The data were obtained from the Mikulski Archive for Space Telescopes at the Space Telescope Science Institute, which is operated by the Association of Universities for Research in Astronomy, Inc., under NASA contract NAS 5-03127 for JWST; and from the European JWST archive (eJWST) operated by the ESDC. These observations are associated with program 1265.
	This research has made use of the NASA/IPAC Extragalactic Database (NED), which is operated by the Jet Propulsion Laboratory, California Institute of Technology, under contract with the National Aeronautics and Space Administration.
	
	\noindent This work has made extensive use of Python (v3.9.12), particularly with \textsc{astropy} \citep[v5.3.3, \nolinkurl{http://www.astropy.org};][]{astropy:2013, astropy:2018}, \textsc{lmfit} (v1.2.2), \textsc{pyneb} \citep[v1.1.18;][]{Luridiana2015}, \textsc{matplotlib} \citep[v3.8.0;][]{Hunter:2007}, and \textsc{numpy} \citep[v1.26.0;][]{Harris2020}.
\end{acknowledgements}

%
\bibliographystyle{aa} 
\bibliography{bibliography.bib} 

%

\begin{appendix}
	
	\section{Modelling of the individual line profiles}
	\label{Appendix1}
	
	We have performed a non-parametric modelling of the ionised emission lines following the lines from \cite{Harrison2014}. In order to select the wavelength range for the modelling, we define the line profile as where the standard deviation of the continuum was larger than 3, similarly to the S/N$>$3 cut for the parametric modelling (see grey, dashed lines in Figs.~\ref{Fig:Appendix_profilesN} and~\ref{Fig:Appendix_profilesS}). When the continuum was noisy, or there were wiggles affecting its shape, we defined the wavelength range as 4 or 5 times the FWHM of the line to obtain proper measurements of the integrated flux. 
	As mentioned in Sect.~\ref{SubSect3:Results_kin}, we excluded from the modelling the lines that were blended, namely [Ne\,V] with [Cl\,II], and [O\,IV] with [Fe\,II] at 25.99$\mu$m.
	
\begin{figure*}
		\centering
		\includegraphics[width=.86\columnwidth]{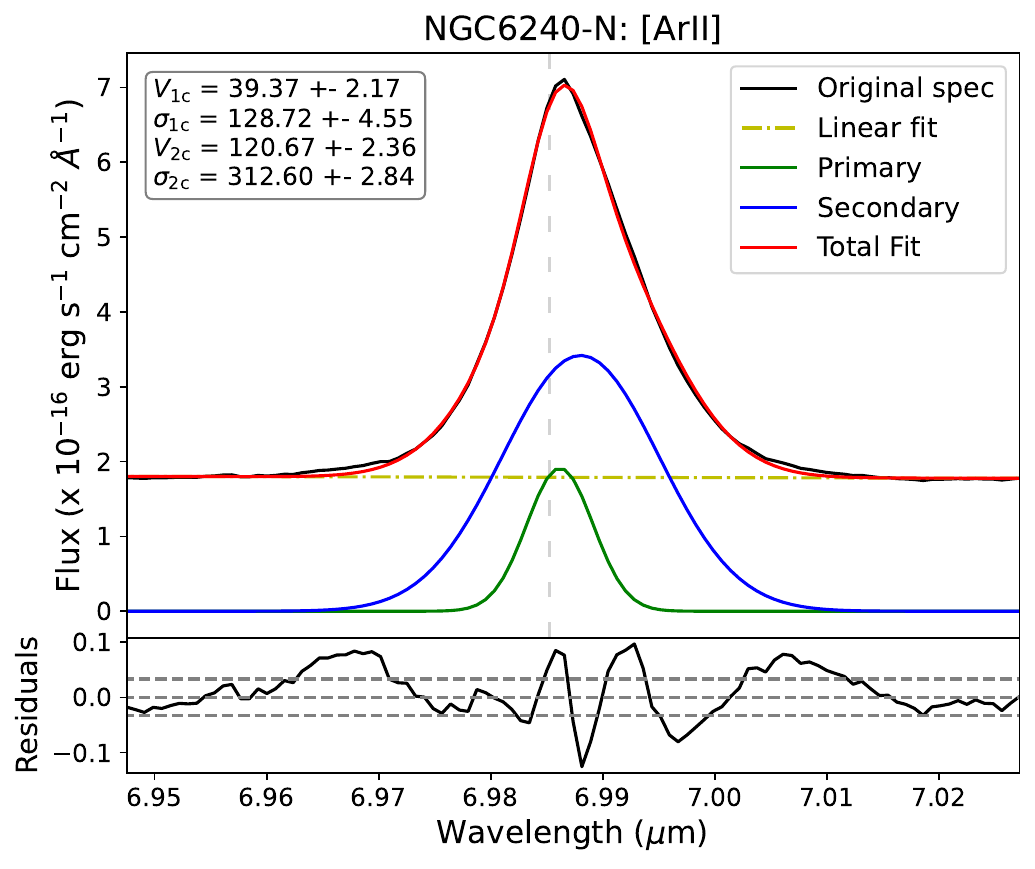}
		\includegraphics[width=.86\columnwidth]{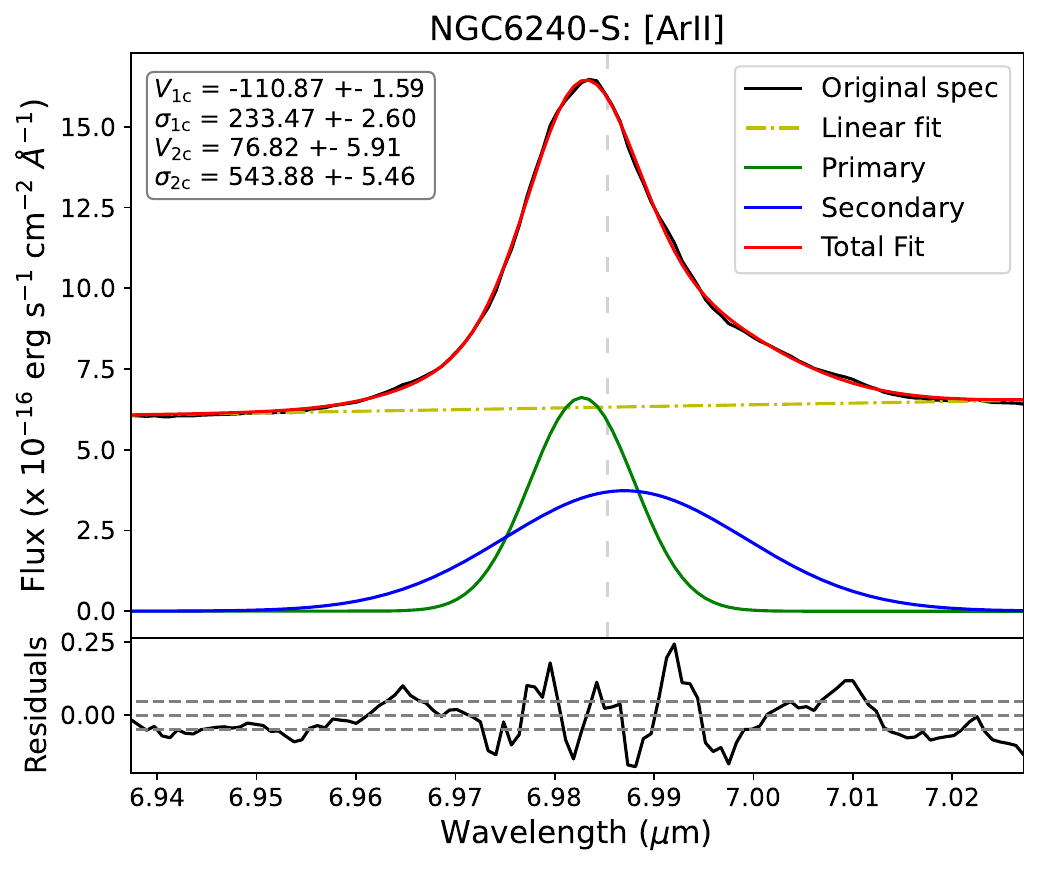}
		\includegraphics[width=.88\columnwidth]{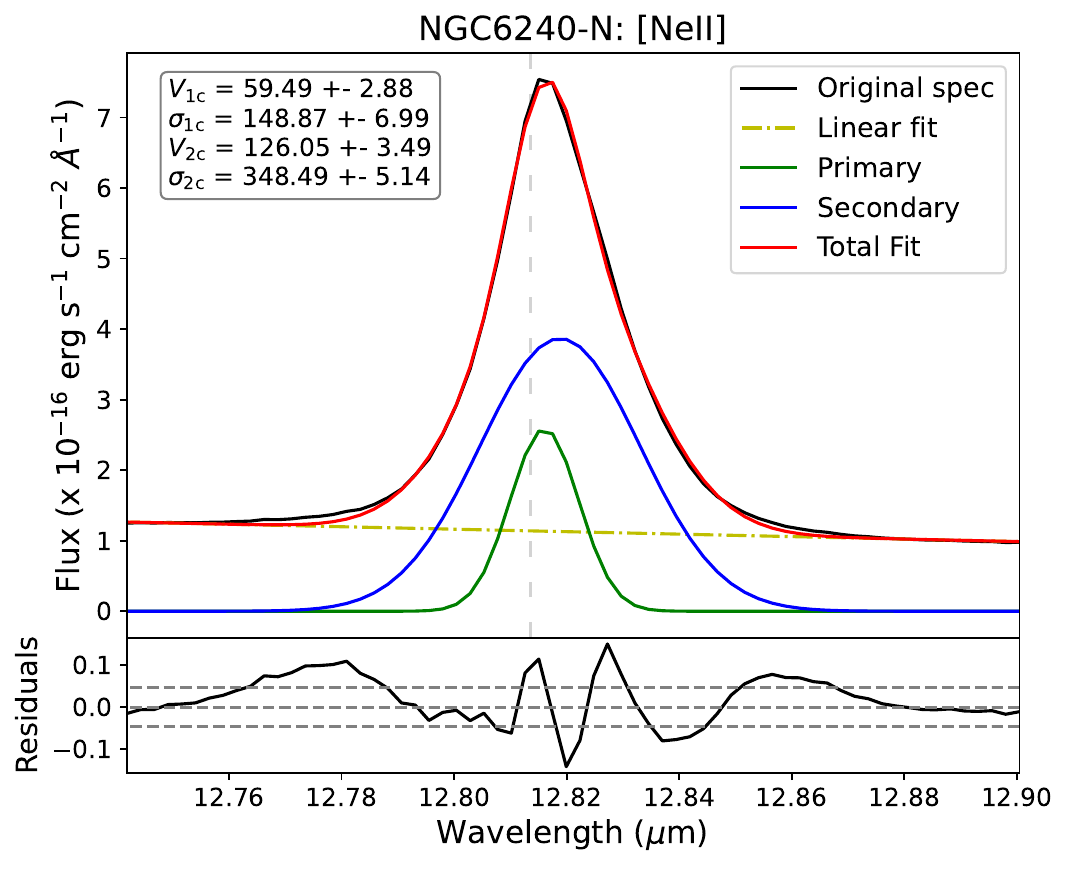}
		\includegraphics[width=.88\columnwidth]{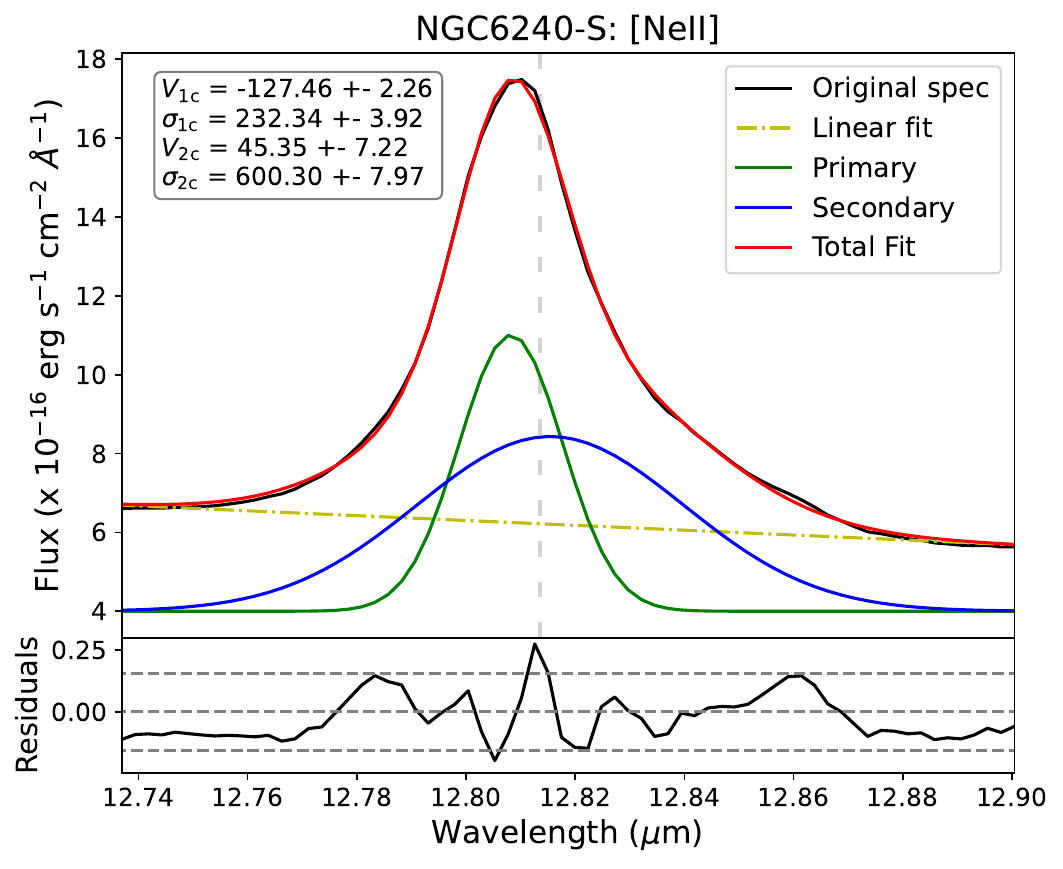}
		\includegraphics[width=.88\columnwidth]{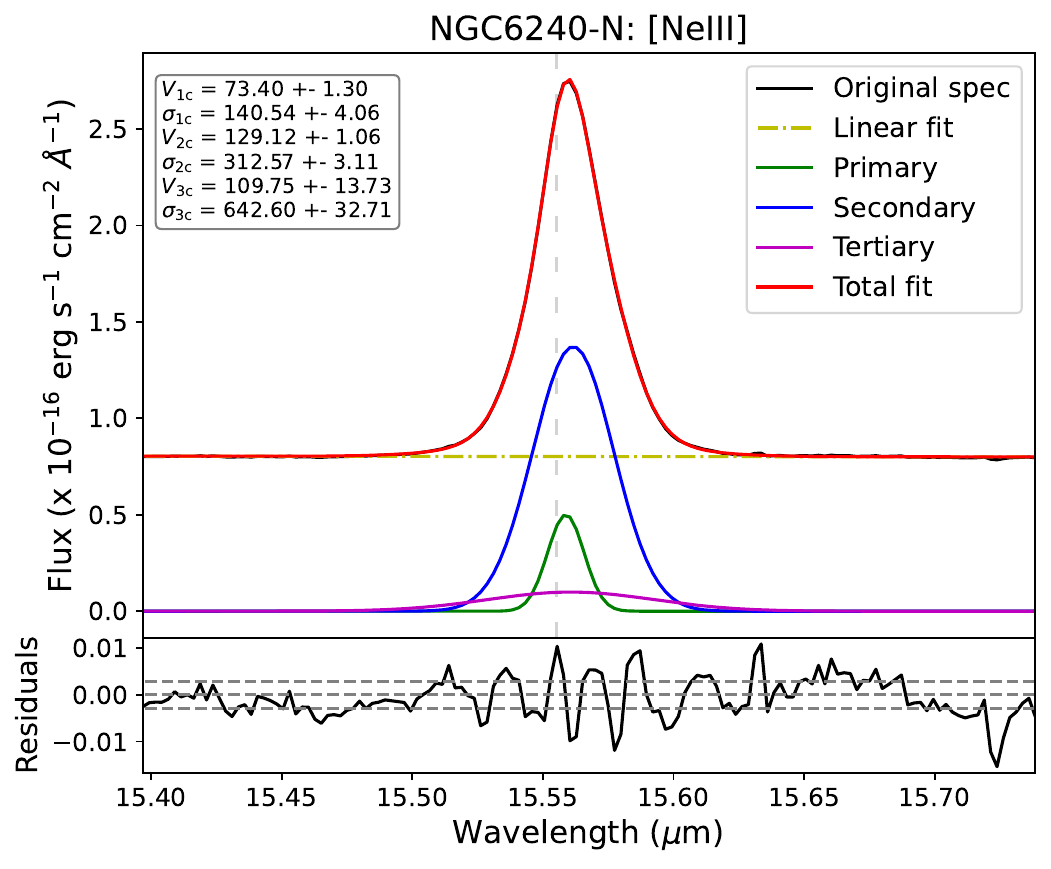}
		\includegraphics[width=.88\columnwidth]{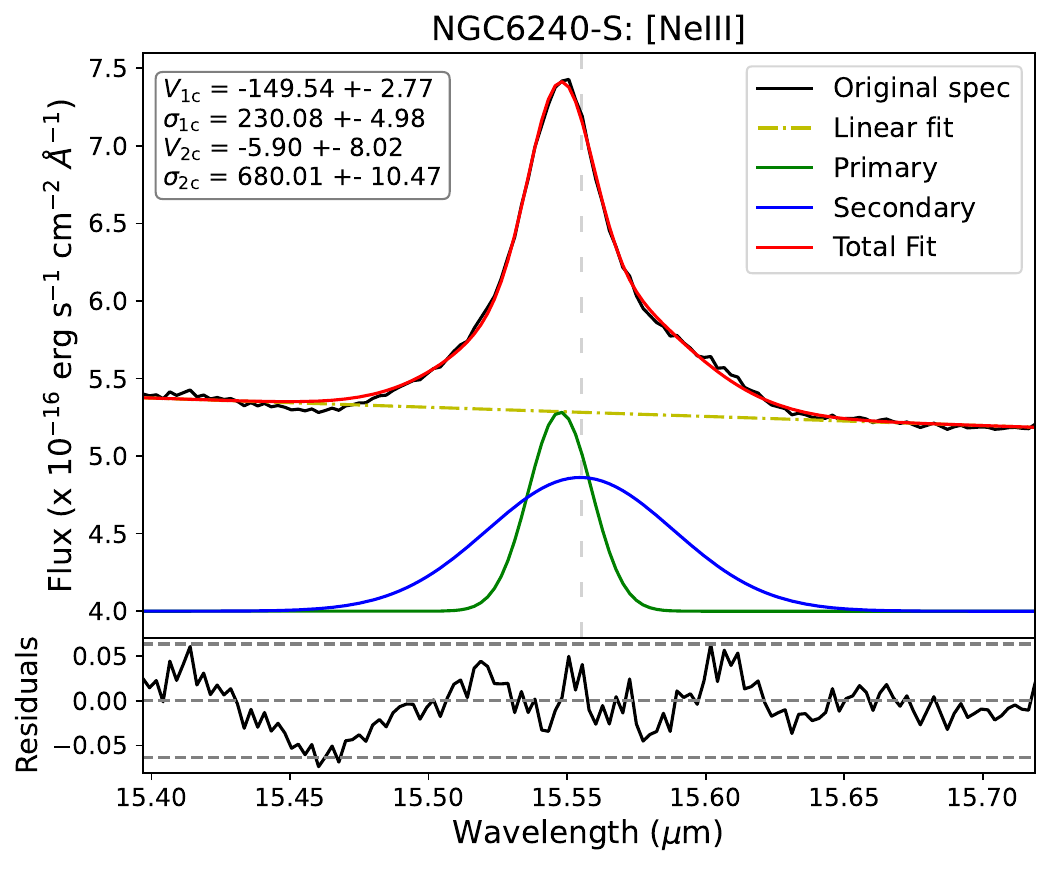}
		\caption{Parametric modelling of the most complex emission lines (from top to bottom, [Ar\,II], [Ne\,II], and [Ne\,III]) for the integrated spectra (R$\sim 0.7$\arcsec) both the N (left panels) and S (right panels) nucleus. The main part of the figure for all panels represent the original spectrum in black, the linear fit as the dashed, yellow line, and the total final modelling as the red line. The different Gaussian components are represented in green (primary), blue (secondary), and magenta (tertiary), when present. The rest-frame position is marked with a dashed, grey, vertical line at the corresponding wavelength in all panels. The bottom part of the figures show the residuals of the fit in black, with the 3$\sigma$ limits as dashed, grey lines (see details in Sect.~\ref{Sect2:Methodology} and \citealt{HM2024}). The velocity (corrected for the systemic value) and the velocity dispersion (corrected for the instrumental value) of every component are marked for each line. }
		\label{Fig:Appendix_FitIntLinesParam}
	\end{figure*}

\begin{figure*}
		\centering
		\includegraphics[width=0.33\textwidth]{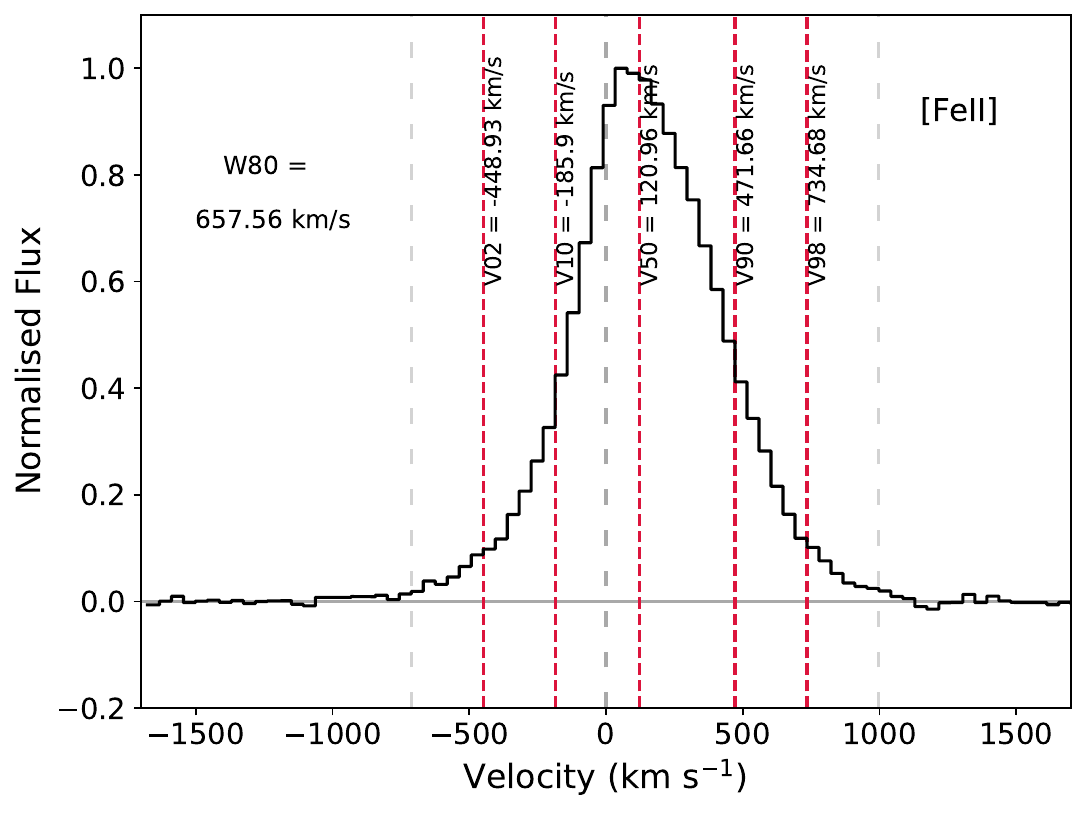}
		\includegraphics[width=0.33\textwidth]{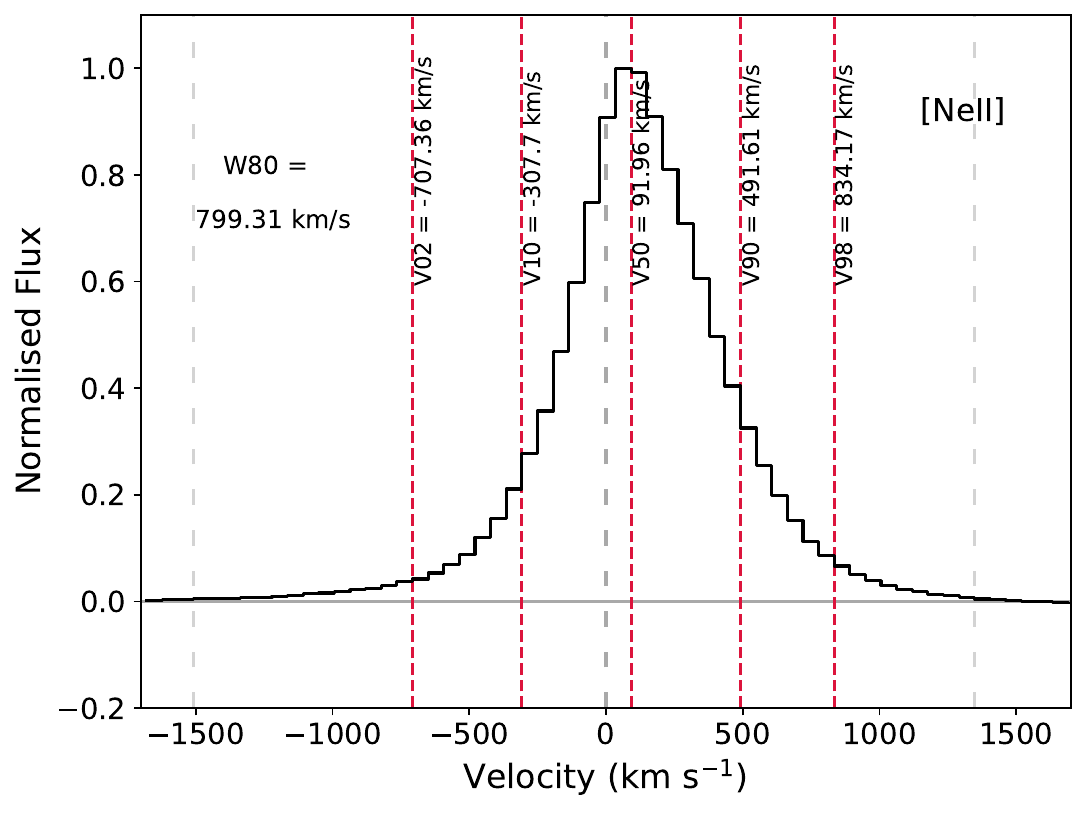} 
		\includegraphics[width=0.33\textwidth]{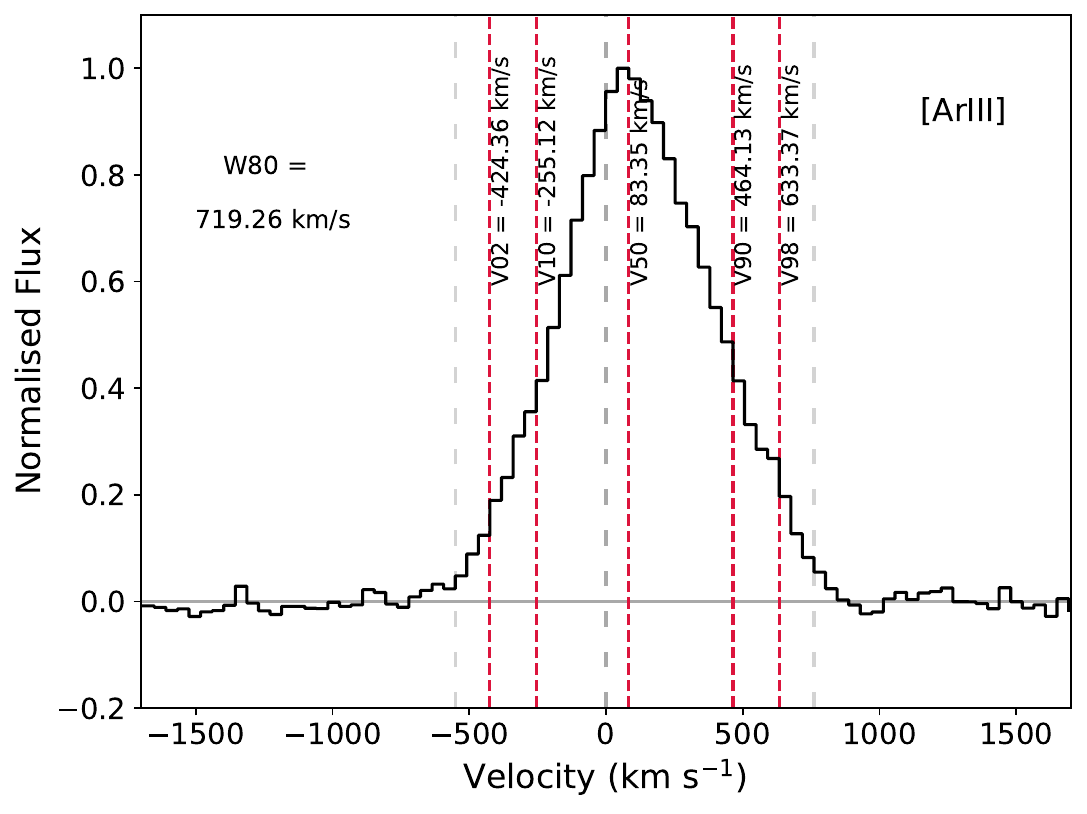} 
		\includegraphics[width=0.33\textwidth]{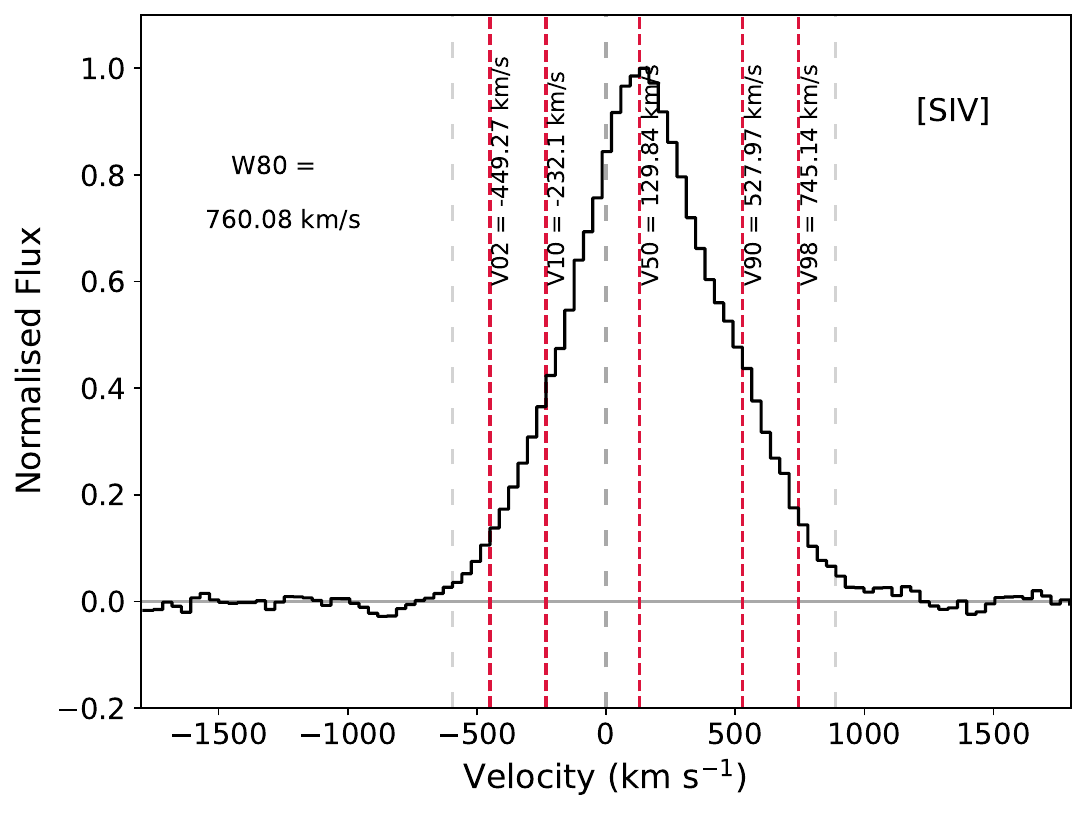}
		\includegraphics[width=0.33\textwidth]{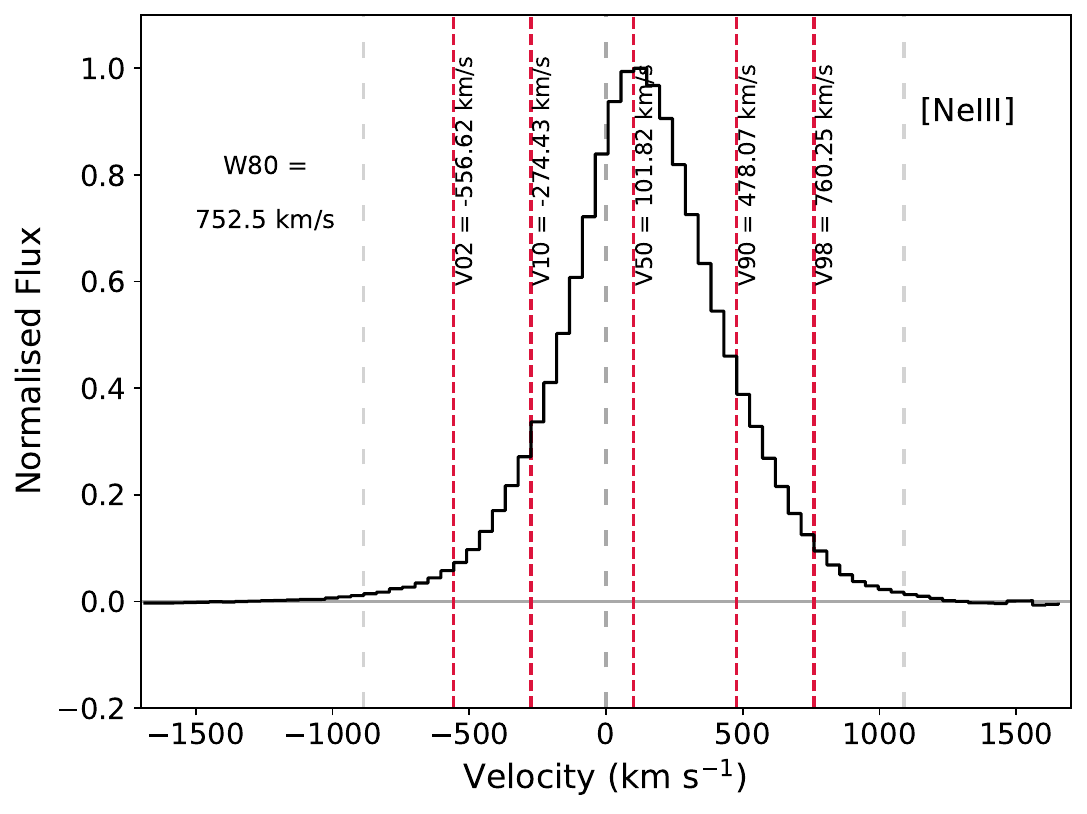} 
		\includegraphics[width=0.33\textwidth]{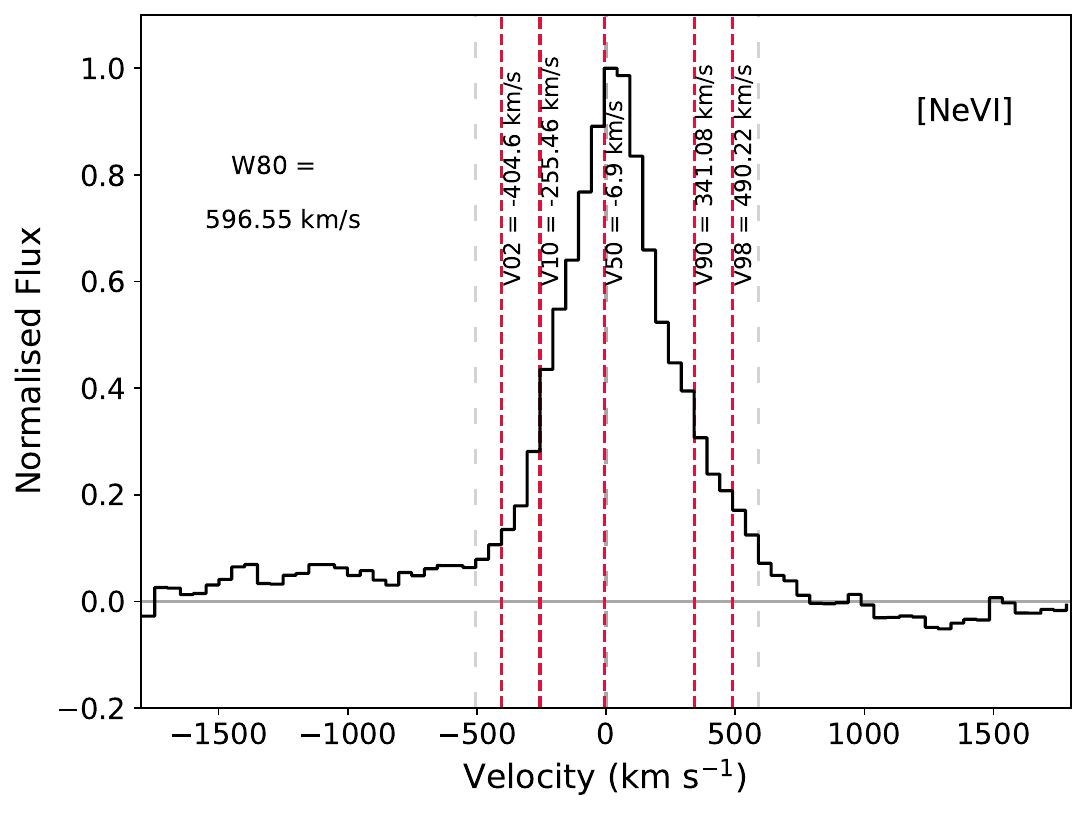}
		\caption{Non-parametric modelling for the integrated profiles of some ionised emission lines at different IP detected for the N nucleus assuming a radius of 0.7\arcsec. The flux has been normalised with the peak of the emission, and the wavelength has been transformed to velocity in km\,s$^{-1}$ after redshift correction (see Sect.~\ref{Sect1:Introduction}). In all panels the grey, dashed lines indicate the selected line limits, and the red, dashed lines mark the V02, V10, V50, V90 and V98 values (see Sect.~\ref{Sect2:Data}). The labels indicate the line in each panel and the W80 parameter (see Sect.~\ref{Sect2:Data}).}
		\label{Fig:Appendix_profilesN}
	\end{figure*}

\begin{figure*}
		\centering
		\includegraphics[width=0.33\textwidth]{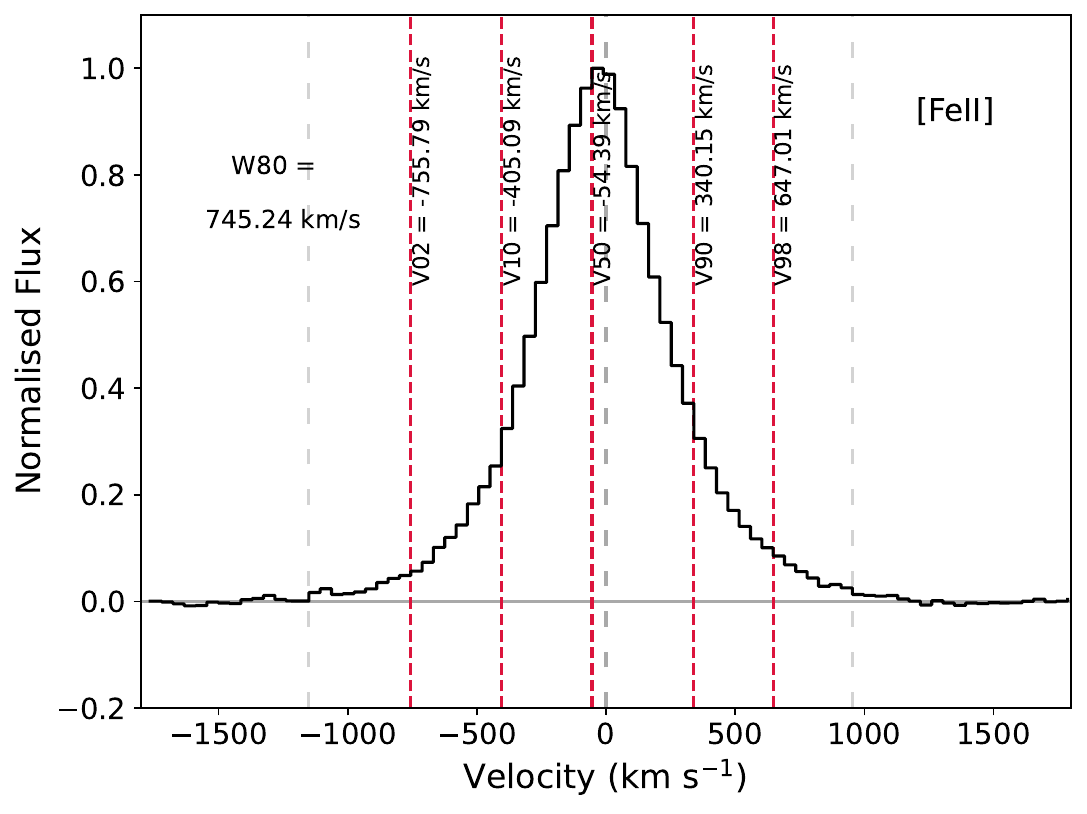}
		\includegraphics[width=0.33\textwidth]{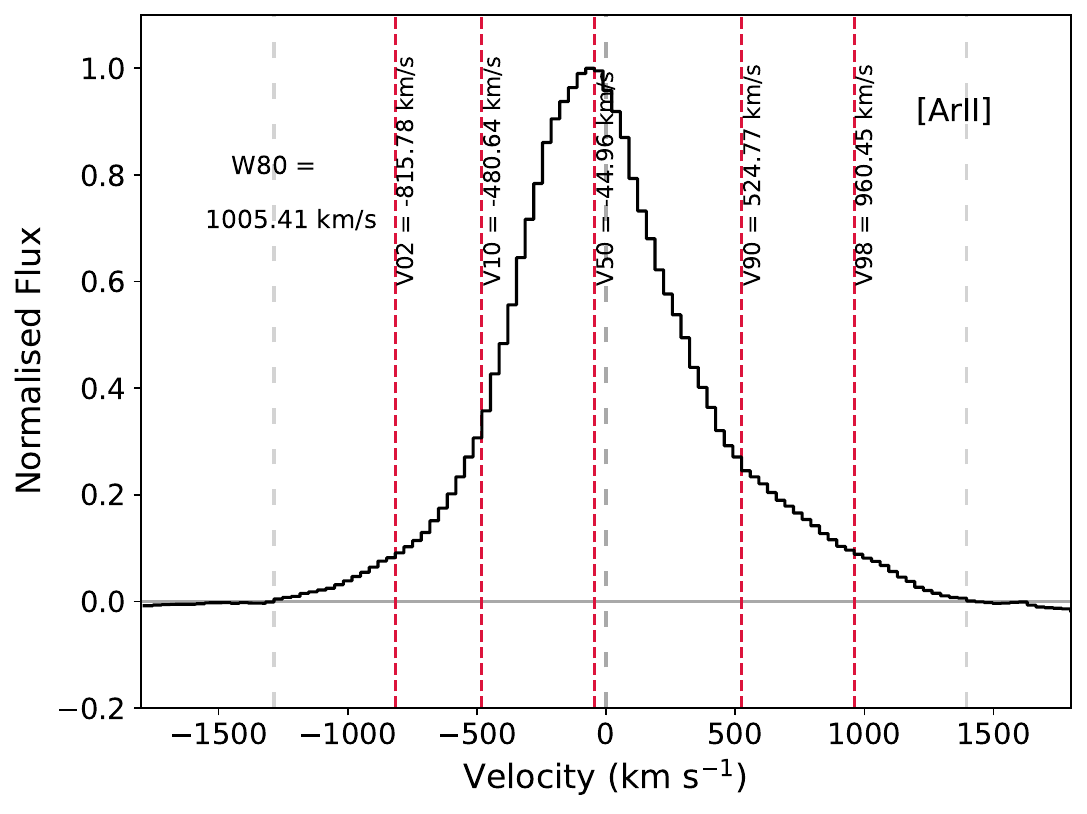} 
		\includegraphics[width=0.33\textwidth]{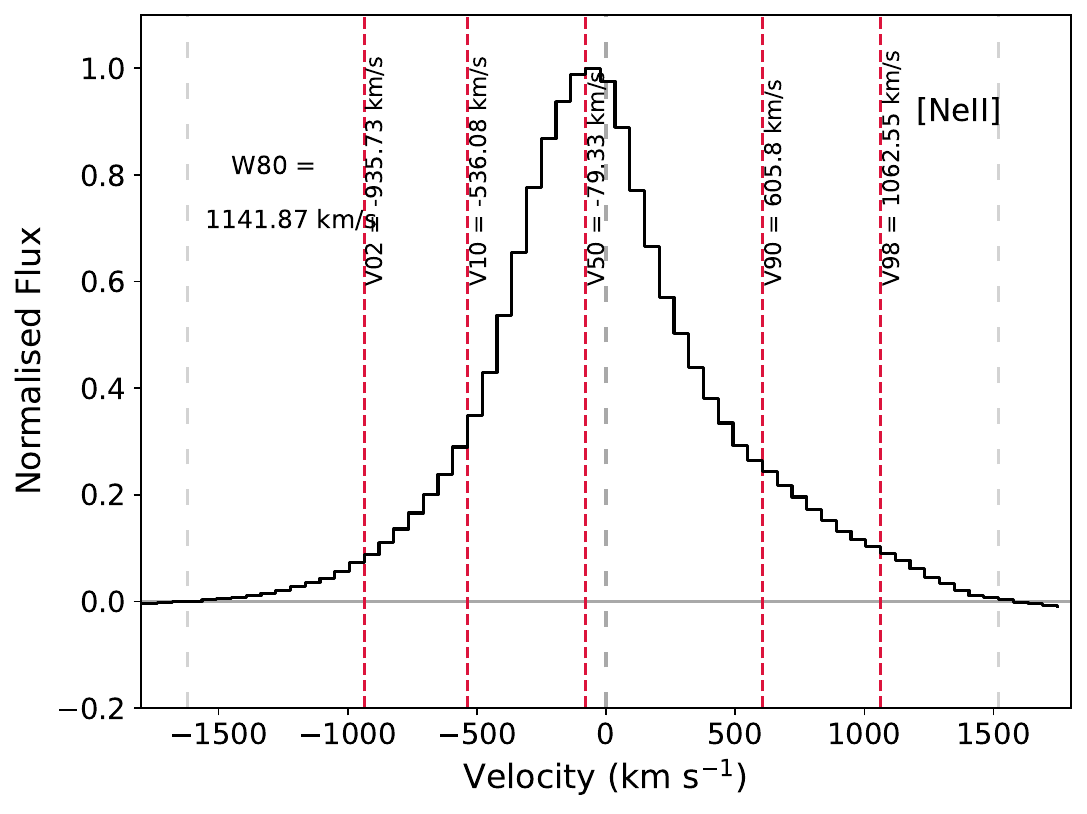}
		\includegraphics[width=0.33\textwidth]{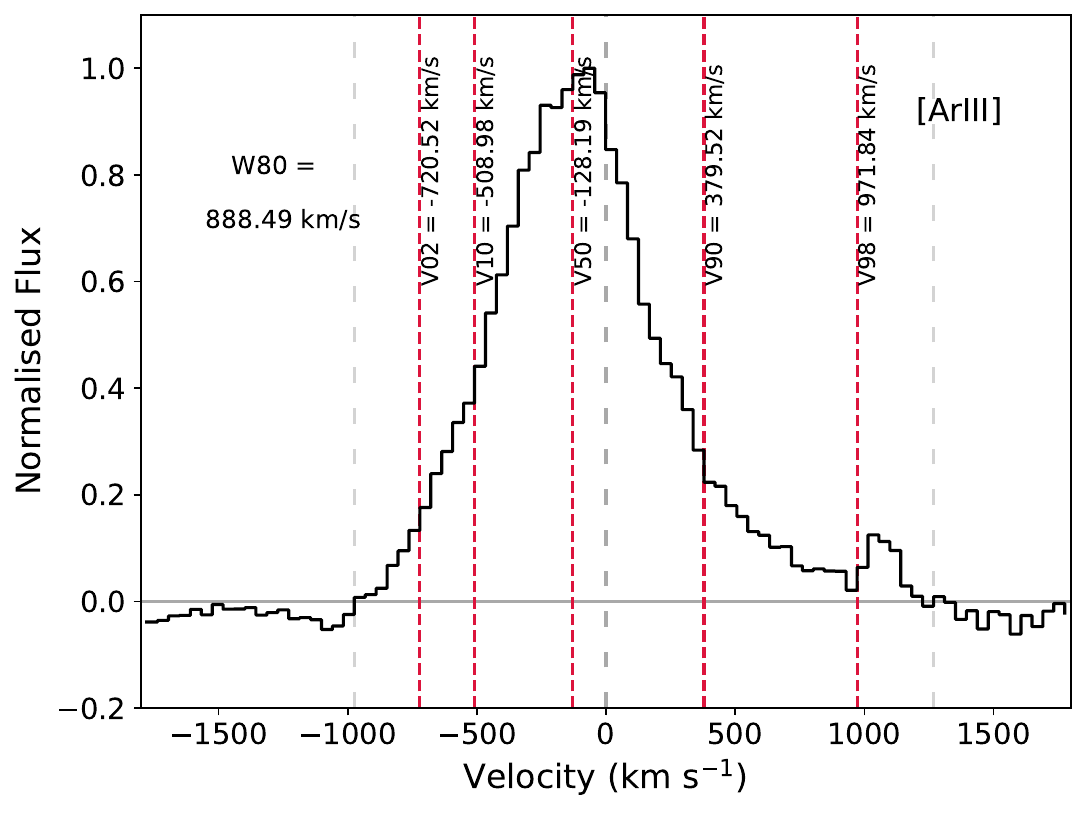} 
		\includegraphics[width=0.33\textwidth]{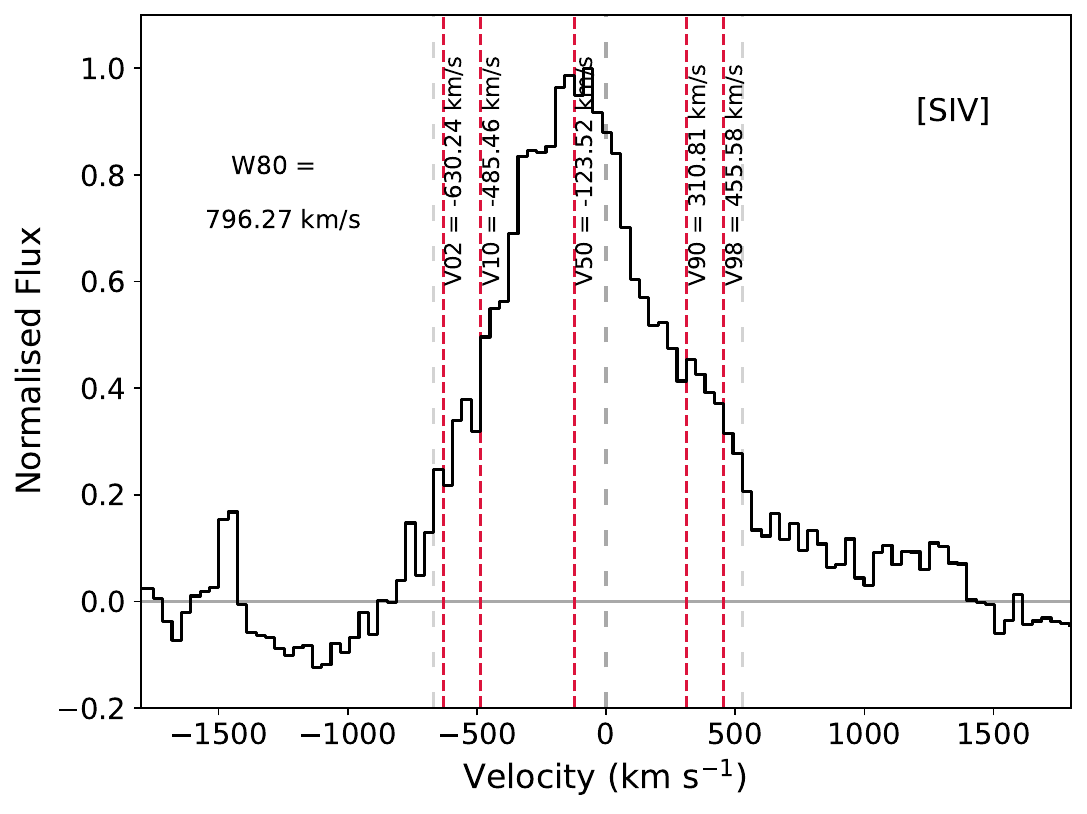}
		\includegraphics[width=0.33\textwidth]{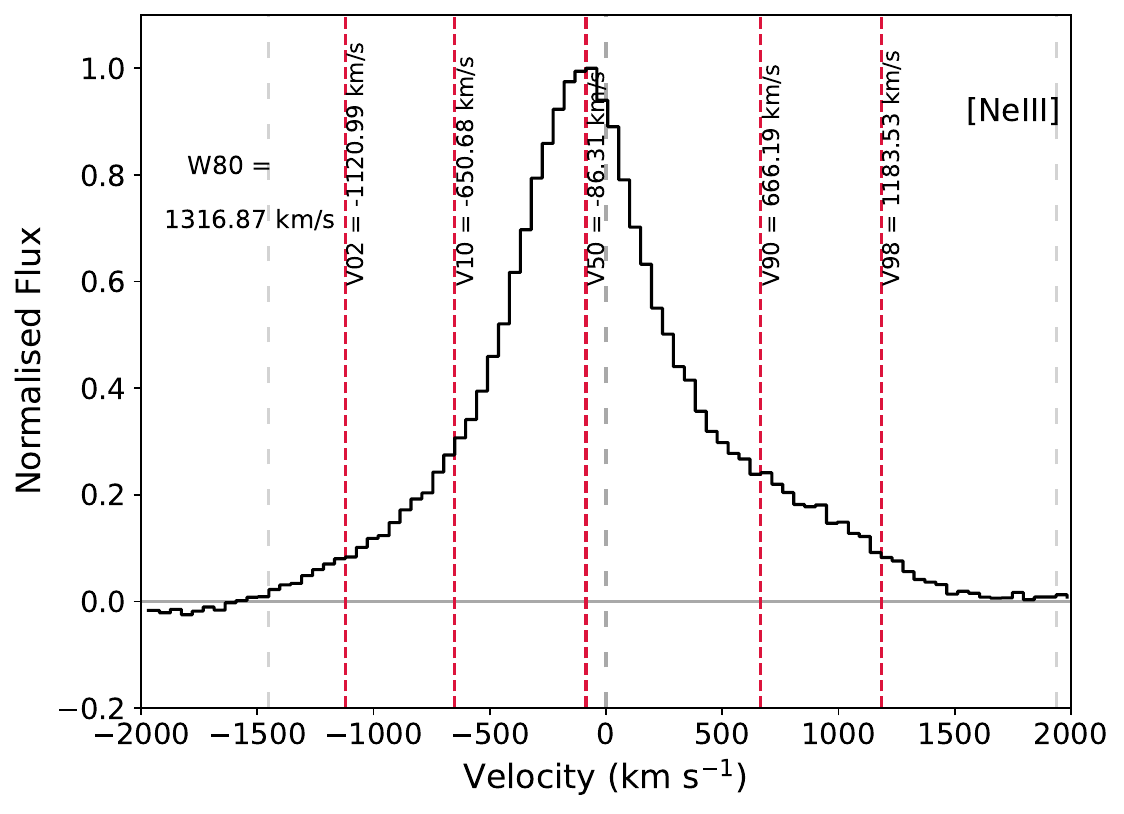}
		\caption{Same as Fig.\ref{Fig:Appendix_profilesN} but for the S nucleus.}
		\label{Fig:Appendix_profilesS}
	\end{figure*}

	We also did a parametric modelling with a multi-Gaussian approach of the integrated profiles (see Sect.~\ref{Sect2:Methodology}). In Fig.~\ref{Fig:Appendix_FitIntLinesParam} we show examples of this modelling for the most complex, broadest emission lines for both nuclei (namely [Ar\,II], [Ne\,II], and [Ne\,III]). We needed up to three Gaussian components for these emission lines to obtain a proper modelling (i.e. the standard deviation of the residuals is lower than 3$\times$ the standard deviation of the continuum; see \citealt{Cazzoli2018,HM2024}). They have been ordered by their $\sigma$, so that the primary is the narrowest Gaussian component. In general for these three lines, as mentioned in Sect.~\ref{SubSect3:Results_kin}, the profiles for the S nucleus have large wings mainly to the red, whereas the profiles for the N nucleus are broad but more symmetrical. For [Ar\,II], that has a better spatial and spectral resolution than [Ne\,II] and [Ne\,III] (located in ch1-long, ch3-short, and ch3-medium, respectively), we modelled it with 2 Gaussian components in both nuclei. The broadest, secondary component has a $\sigma_{\rm 2C,N} \sim 313$\,km\,s$^{-1}$ (i.e. FWHM$_{\rm 2C,N} \sim 736$\,km\,s$^{-1}$) in the N and $\sigma_{\rm 2C,S} \sim 544$\,km\,s$^{-1}$ (i.e. FWHM$_{\rm 2C,N} \sim 1280$\,km\,s$^{-1}$) for the S nucleus. However from the non-parametric modelling, the widths measured with the W80 parameter (see Table~\ref{Table:2_W80}) were $670$ and $971$ for the N and the S nucleus (uncorrected for the instrumental width), respectively, which are much lower than the FWHM with the parametric modelling (corrected for the instrumental width). This proves that the non-parametric modelling is not the correct approach to analyse complex emission lines with non-Gaussian profiles, as the primary (or secondary) component tends to dominate the flux whereas the wing components are not fully captured. 
	
	\section{Additional figures}
	\label{Appendix_figures}
	
\begin{figure*}
		\includegraphics[width=0.34\textwidth]{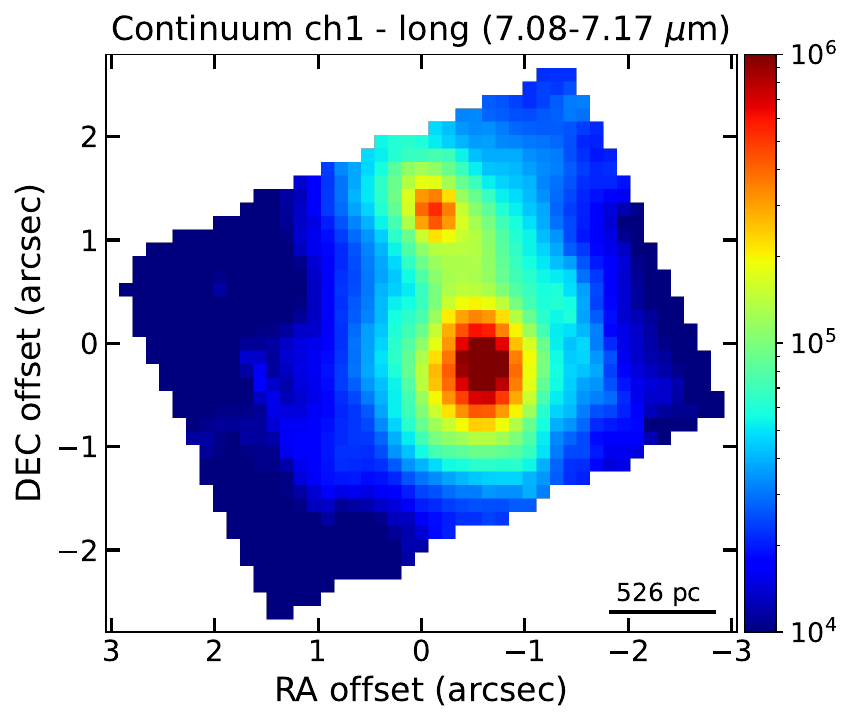}\\
		\includegraphics[width=0.34\textwidth]{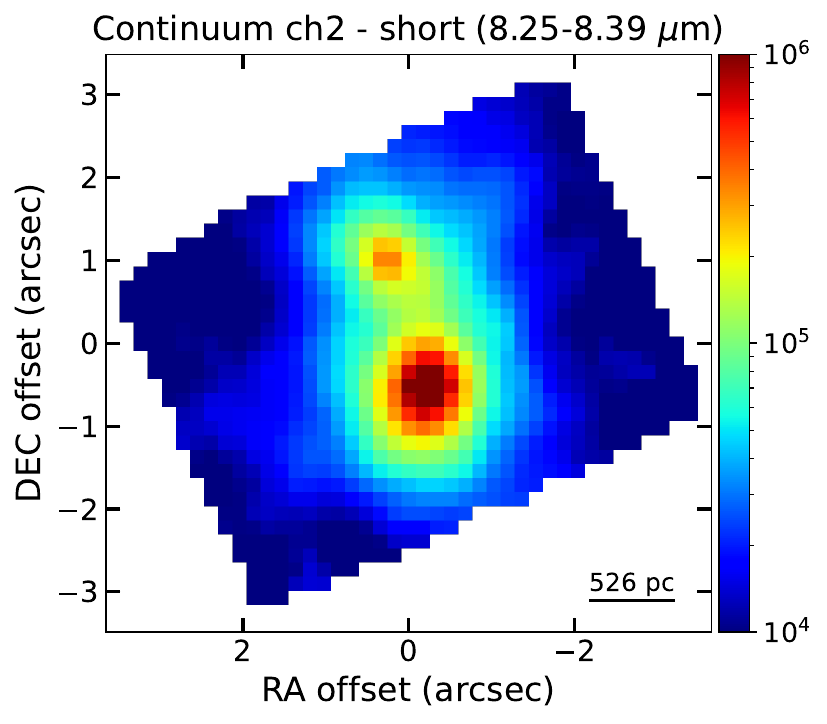}
		\includegraphics[width=0.34\textwidth]{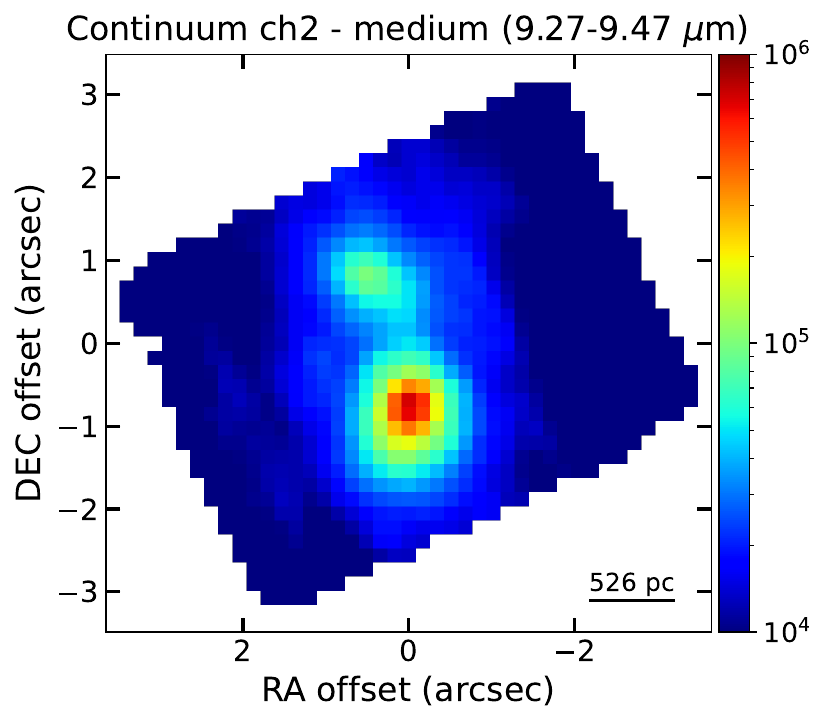} 
		\includegraphics[width=0.34\textwidth]{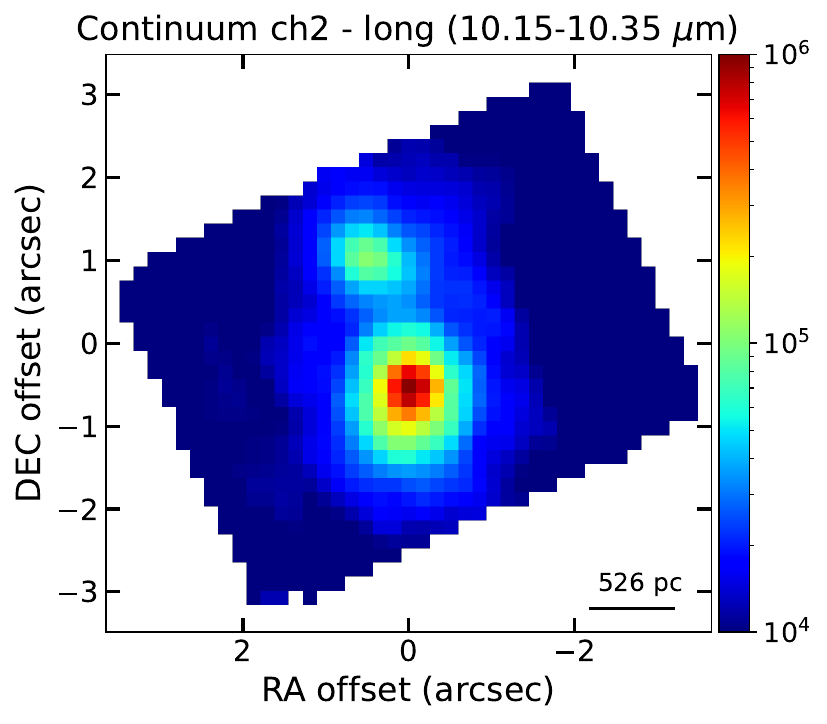}
		\includegraphics[width=0.34\textwidth]{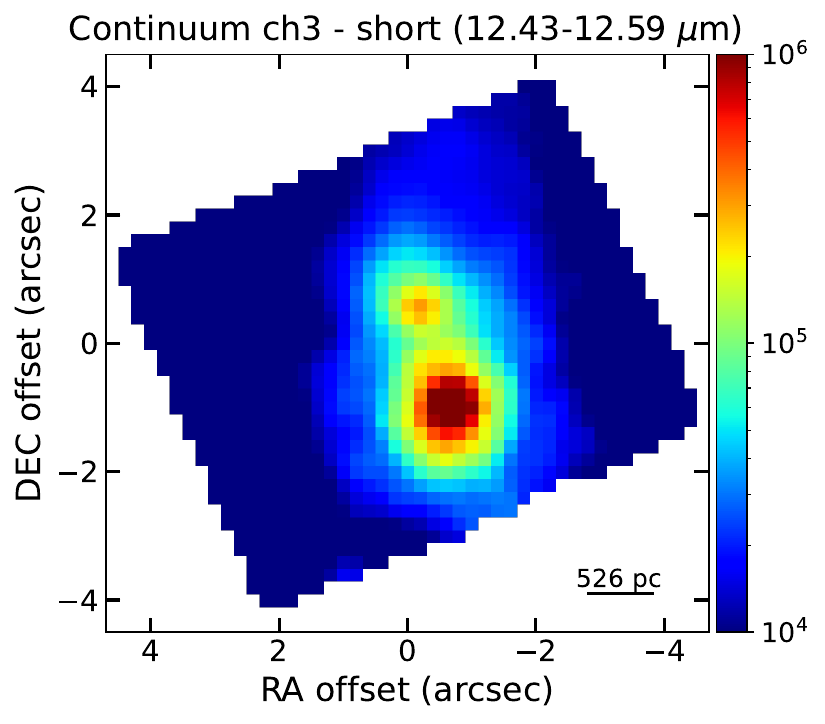} 
		\includegraphics[width=0.34\textwidth]{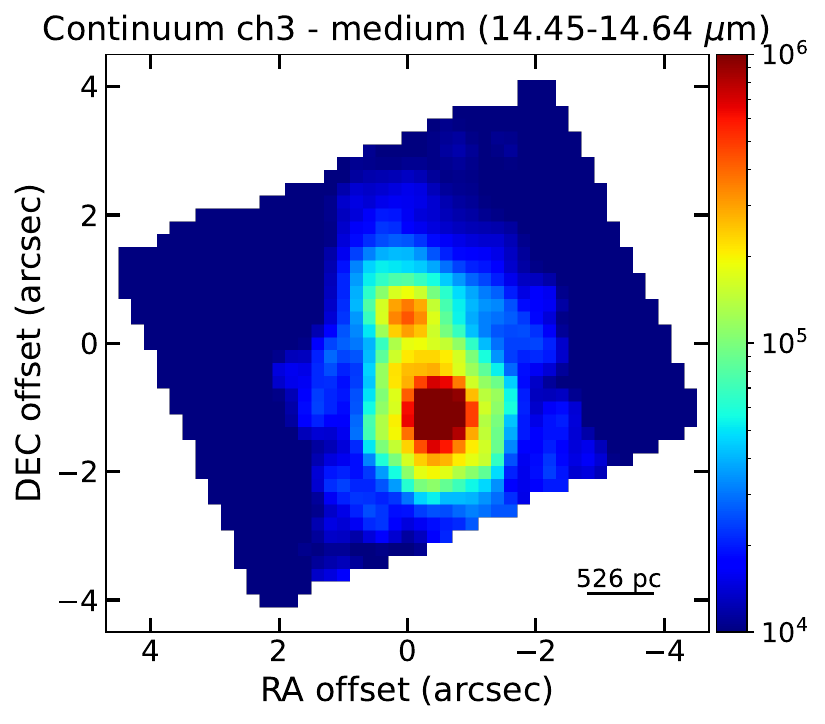}  
		\includegraphics[width=0.34\textwidth]{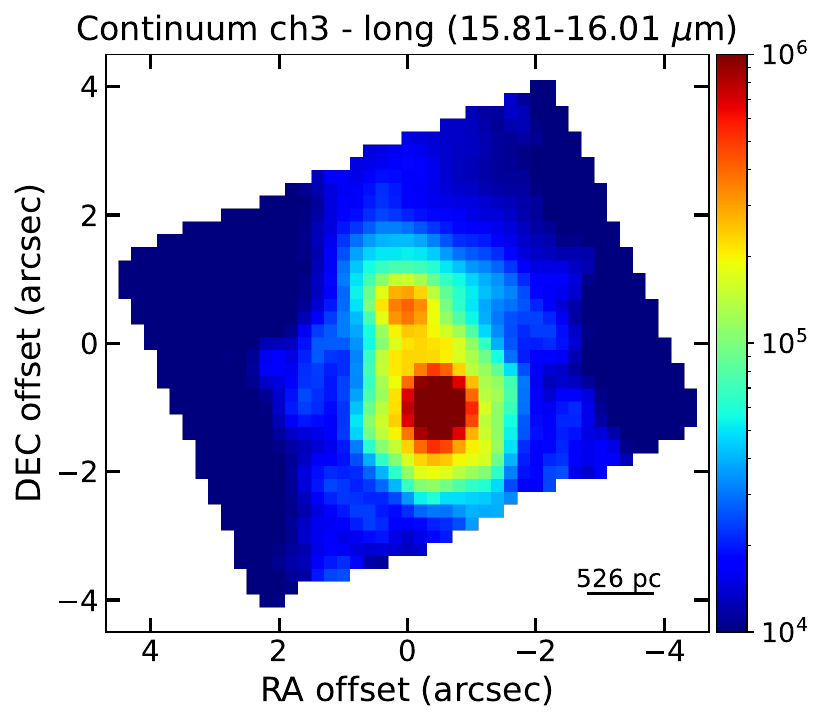}
		\includegraphics[width=0.34\textwidth]{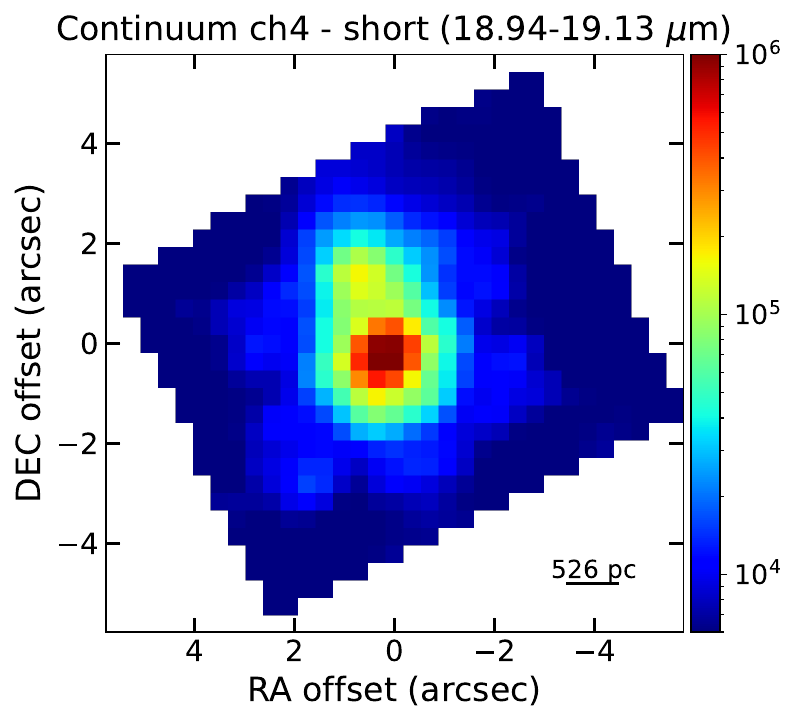}
		\includegraphics[width=0.34\textwidth]{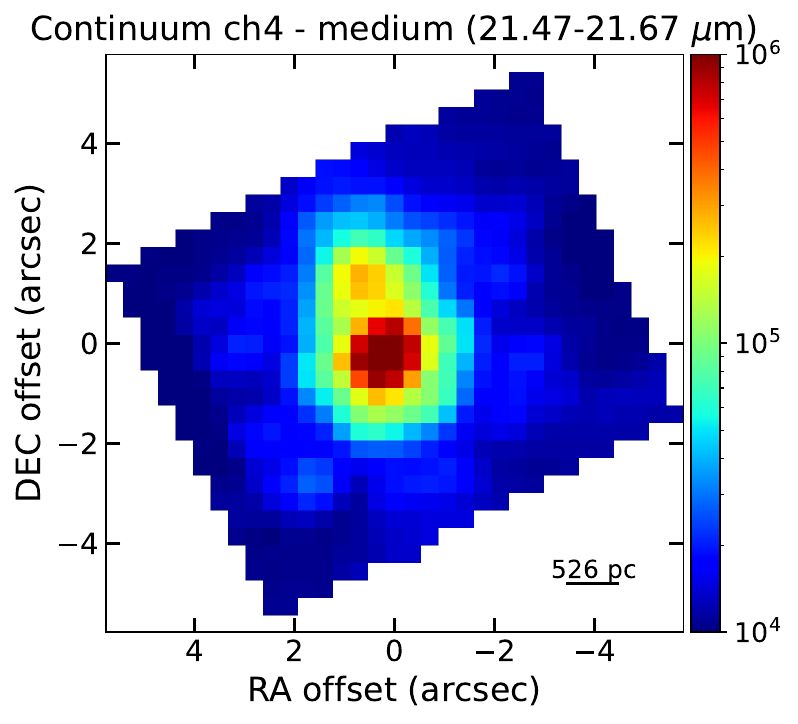}
		\includegraphics[width=0.34\textwidth]{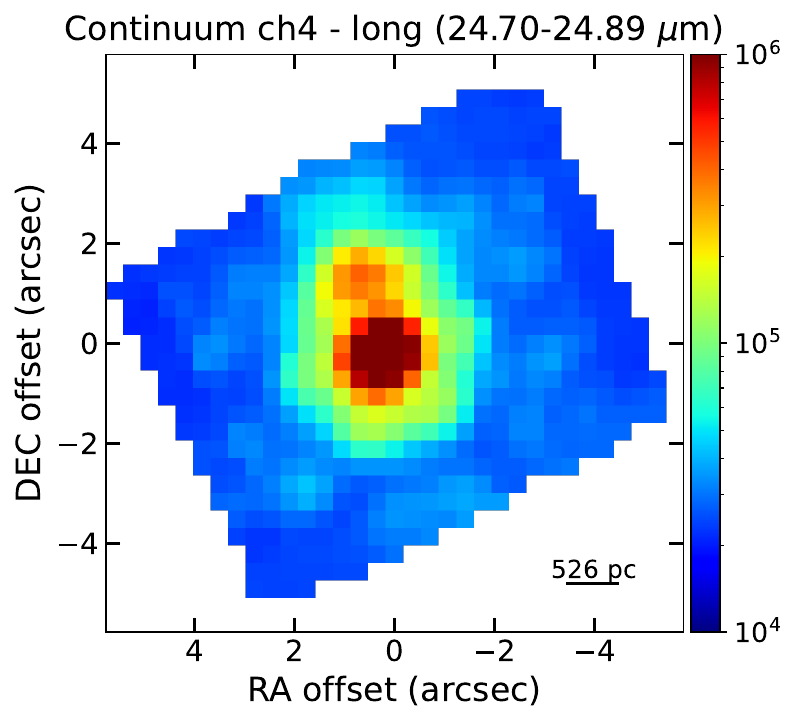}
		\caption{Continuum maps for all the bands in each channel ordered from lower to higher wavelengths. The title of each panel indicates the wavelength range used (in rest-frame) to create the maps. The short and medium band of channel 1 are not shown, as the spectra are dominated by emission and absorption features, thus there is no sufficient featureless range to estimate the continuum.}
		\label{Fig:Appendix_ContMaps}
	\end{figure*}

	In this section we show additional figures from the analysis. Figure~\ref{Fig:Appendix_ContMaps} are the continuum maps obtained for each channel and band of the MIRI/MRS data, except the short and medium bands of ch1, as they presented many lines and emission or absorption features that prevented from selecting a featureless continuum region. We used these maps to estimate the PSF size, by measuring the FWHM of both nuclei. In general the PSF size is below 0.7\arcsec\,in all the wavelength range up to ch4.
	
\begin{figure*}
		\centering
		\includegraphics[width=.85\textwidth]{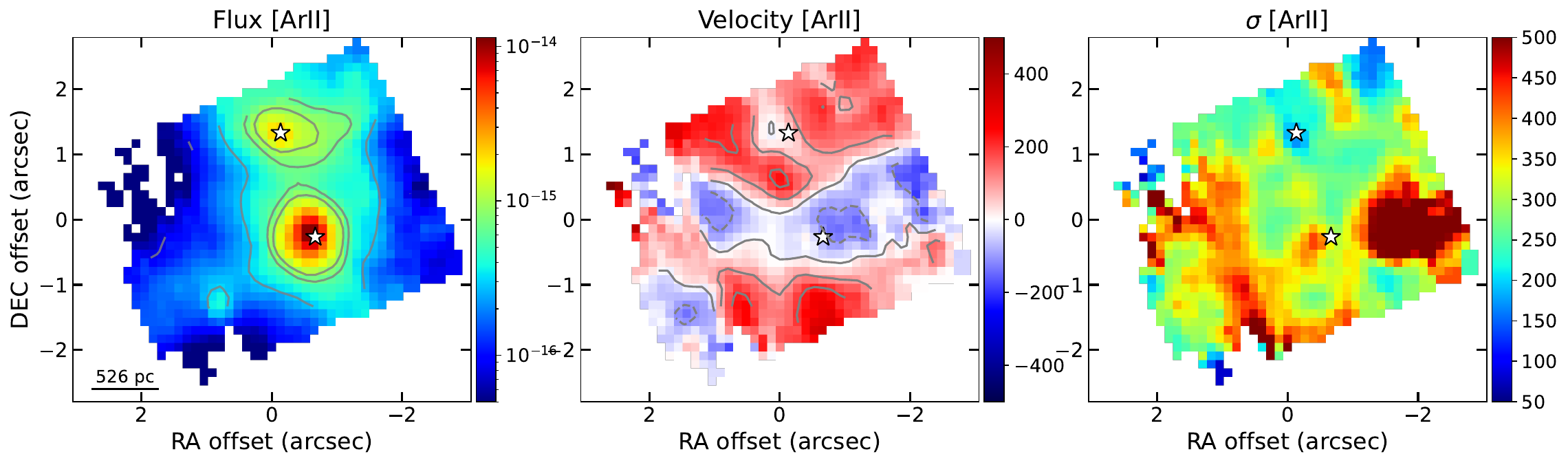} 
		\includegraphics[width=.85\textwidth]{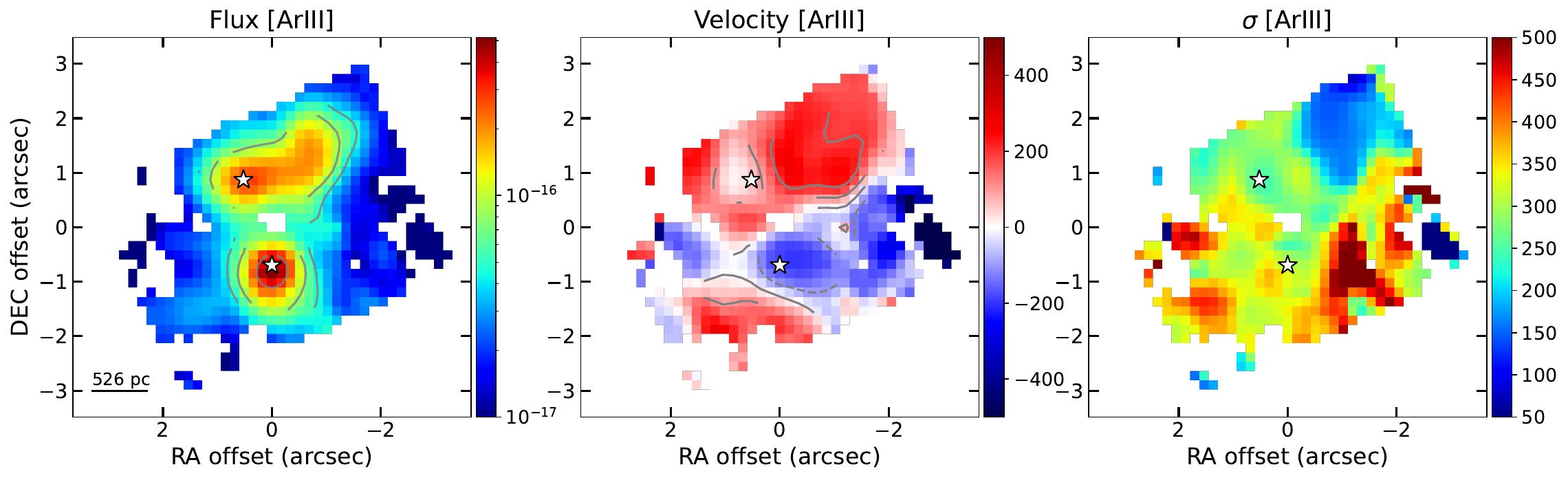}
		\includegraphics[width=.85\textwidth]{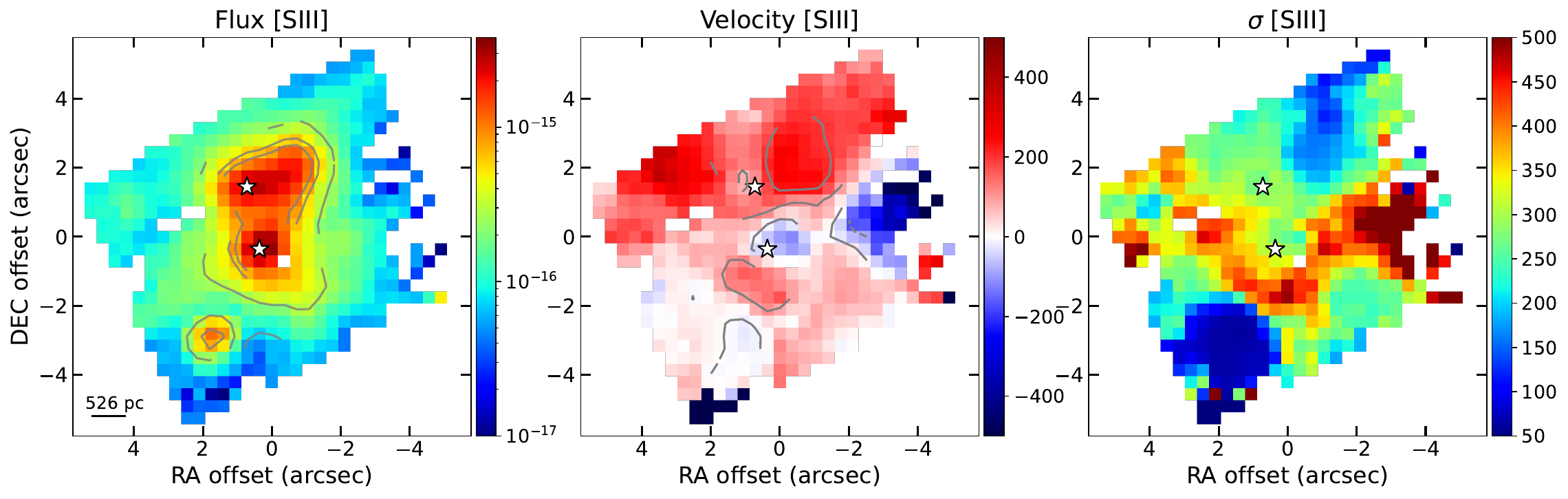}
		\includegraphics[width=.85\textwidth]{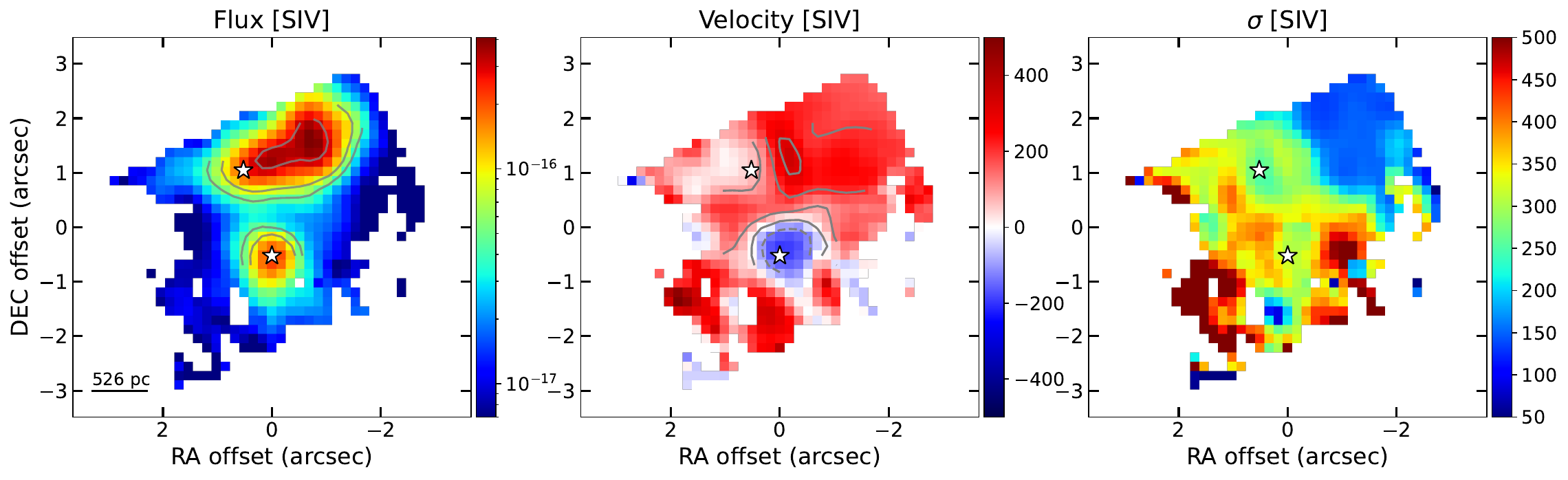}
		\caption{Same as Fig.~\ref{Fig:KinMapsLowExcit} but for [Ar\,II], [Ar\,III], [S\,III], and [S\,IV] lines. These three latter emission lines have been re-binned with a 2$\times$2 box (see Sect.~\ref{Sect2:Methodology}). }
		\label{FigAp:KinMaps_extralines}
	\end{figure*}

	We show in Fig.~\ref{FigAp:KinMaps_extralines} the kinematic maps for the remaining ionised gas emission lines that are spatially resolved with the MRS mode of MIRI. 
	
\end{appendix}

\end{document}